\documentstyle[12pt,psfig]{thesis}

\newcommand{\mysection}{\setcounter{equation}{0}\section}
\renewcommand{\theequation}{\thechapter.\arabic{equation}}
\renewcommand{\thefigure}{\thechapter.\arabic{figure}}

\begin{document}

\pagestyle{prelim}              
%
%

\dissertation

%

\author{Nikolaos Kidonakis}

%

\degree{Doctor of Philosophy}

%

\department{Physics}

%

\title{QCD resummation and heavy quark cross sections}

%

\month{May}
\year{1996}

%

\maketitle

%

%

\begin{approval}
 \member{John Smith\\Professor, Institute for Theoretical Physics}
 \member{George Sterman\\Professor, Institute for Theoretical Physics}
 \member{Robert McCarthy\\Professor, Department of Physics}
 \member{Frank Paige\\ Senior Physicist, Brookhaven National Laboratory }
\end{approval}

%

\begin{abstract}
In this dissertation a detailed analysis of heavy quark production 
is given with an emphasis on the
resummation of soft gluon corrections. 

First we calculate the production cross sections 
for top quark production at the Fermilab Tevatron 
and for bottom quark production at fixed-target $pp$ experiments, 
and in particular HERA-B. 
We consider both the order $\alpha_s^3$ cross 
sections and the resummation of soft gluon corrections in all orders of
QCD perturbation theory.

Then we calculate the inclusive transverse momentum and 
rapidity distributions for top quark production at the Fermilab 
\linebreak
Tevatron and bottom quark production at HERA-B.
We give both $\alpha_s^3$ and resummed results.

Finally, we discuss the resummation of distributions that are singular
at the elastic limit of partonic phase space
(partonic threshold) in
QCD hard-scattering cross sections, such as  heavy quark production.
We show how nonleading soft logarithms exponentiate 
in a manner that depends on the
color structure within the underlying hard scattering.  This result
generalizes the resummation of threshold singularities for the
Drell-Yan process, in which the hard scattering proceeds through
color-singlet annihilation. 
We illustrate our results for the
case of heavy quark production by light quark annihilation
and gluon fusion, and also for light quark 
\linebreak
production through gluon fusion.
 
\end{abstract}

%

\begin{dedication}
For Natalia and my parents
\end{dedication}

%

\tableofcontents
\listoffigures

%


%

\begin{acknowledgements}
I would like to thank my advisor, Professor John Smith, for the guidance
and support that he has offered me throughout my stay at the Institute
for Theoretical Physics at Stony Brook. I would also like to thank 
Professor George Sterman for his help and guidance during my last year
at the Institute. Thanks are also due to Professor Hwa-Tung Nieh for
his assistance and advice during my first semester at the ITP.

Several students in the ITP and elsewhere 
were helpful to me during my studies in many ways.
I would like to thank Brian Harris, Lyndon Alvero, and Sergio Mendoza
for sharing with me their expertise in Fortran programming and
their computer knowledge; Eric Laenen for useful physics
discussions; and most of all Anastasios Liakos and 
Peter Pfeifenschneider for their valuable friendship and help.
  
Finally, I would like to give special thanks to my parents, Ioannis and 
Dimitra, to my sister, Marianna, and to my wife, Natalia, for their strong 
support, encouragement, and love. 

{\hfill Nikolaos Kidonakis}

\end{acknowledgements}

\pagestyle{body}                

\chapter{Introduction}
In this dissertation we will be concerned with the calculation of heavy quark
cross sections and differential distributions in transverse momentum and
rapidity. In particular we will examine top quark production at the
Fermilab Tevatron and bottom quark production at fixed target $pp$ 
experiments and HERA-B.   

The Standard Model of Particle Physics is at present the most successful
model for the description of the interactions of elementary particles.
In this model all the fundamental interactions derive from the principle
of local gauge invariance.
The Standard Model is described by the gauge group 
$SU(3)_C \otimes SU(2)_L \otimes U(1)_Y$,
which incorporates the Glashow-Weinberg-Salam theory of electroweak
processes and Quantum Chromodynamics (QCD). 
QCD is the local $SU(3)_C$ gauge field theory of strong interactions 
and in perturbative QCD we make 
physical predictions by expanding in powers of the 
strong coupling constant $\alpha_s$.

In the Standard Model there are three families of quarks and leptons.
The heaviest family of quarks consists of the top and the bottom quarks.
The latter was discovered in the seventies
but the top was elusive until recently due to its very high mass.
The search for the top has gone on for more than a decade with experimental
groups continually establishing ever higher limits for the top quark mass.
The recent discovery of the top quark by the two experimental groups 
CDF \cite{ObsCDF1} and D0 \cite{ObsD01} at
the Fermilab Tevatron has been the biggest discovery in particle physics
in the last decade and has given us one more confirmation of the Standard
Model. The existence of the top was expected on theoretical grounds for
the cancellation of anomalies. CDF announced a top quark mass of 
$176 \pm 8 \pm 10$ GeV/$c^2$
and a $t \bar t$ production cross section of $6.8^{+3.6}_{-2.4}$ pb.
D0 measured the top quark mass to be $199^{+19}_{-21} \pm 22$ GeV/$c^2$
and its production cross section to be $6.4 \pm 2.2$ pb.

Perturbative QCD (pQCD) is applicable only at high energies since it is
only in this domain that $\alpha_s$ is small according to the principle
of asymptotic freedom \cite{Politzer}. 
The expansion of $\alpha_s$ in the renormalization scale $\mu$ is given by
\begin{equation}
\alpha_s(\mu^2,n_f)=\frac{12 \pi}{(33-2 n_f)\ln(\mu^2/\Lambda^2)}
\left[1-\frac{6(153-19 n_f)}{(33-2 n_f)^2}\frac{\ln(\ln(\mu^2/\Lambda^2))}
{\ln(\mu^2/\Lambda^2)}\right]+\cdots
\end{equation}
where $n_f$ is the number of quarks with mass less than $\mu$ and
$\Lambda$ is the QCD scale parameter. 
\begin{figure}
\centerline{\psfig{file=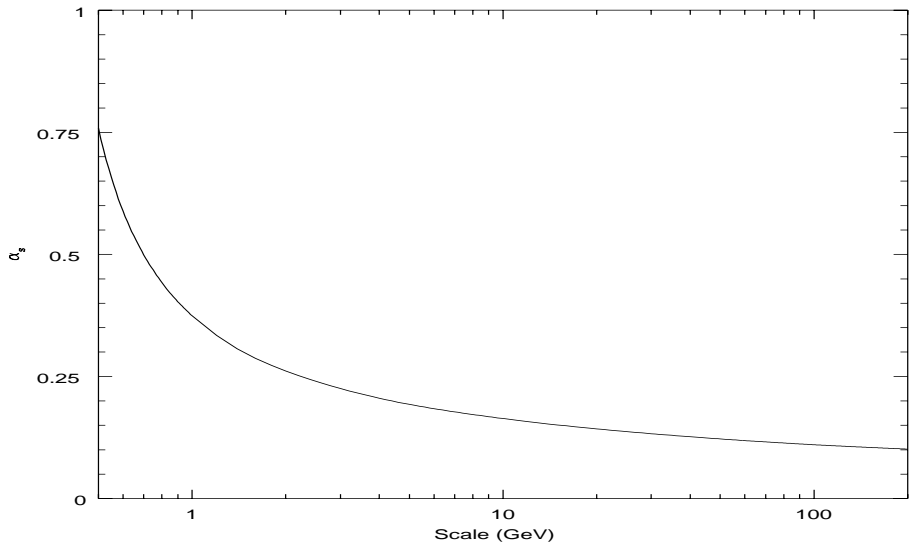,height=3.05in,width=5.05in,clip=}}
\caption[The strong coupling constant $\alpha_s$ versus scale]
{The strong coupling constant $\alpha_s$ versus scale}
\label{fig.1}
\end{figure}

In fig. 1 we plot $\alpha_s$ as a function of scale.
We see that at the scale relevant to top quark production (i.e. 
the top quark mass, which we take to be 175 GeV/$c^2$) 
the value of the coupling is about 0.1. This number is
small enough to allow perturbative QCD calculations to be reliable.
For bottom quark production the relevant scale is the bottom quark mass
(which we take to be 4.75 GeV/$c^2$).
There the value of the coupling is about 0.2, which is still small but
not as much as for the top quark case. 

The calculation of production cross sections for heavy particles 
in QCD is made by invoking the factorization theorem \cite{css1} and 
expanding the contributions to the amplitude in powers of the coupling 
constant $\alpha_s(\mu^2)$. 
One has to perform both renormalization of ultraviolet divergences
and mass factorization of collinear divergences. 
In top and bottom quark production
the heavy quark-antiquark pairs are created in parton-parton 
collisions. For the Born $O(\alpha_s^2)$ cross section the two relevant 
processes are quark-antiquark annihilation
\begin{equation}
q+{\bar q}\rightarrow Q+{\bar Q},
\end{equation}
and gluon-gluon fusion
\begin{equation}
g+g\rightarrow Q+{\bar Q}.
\end{equation}
In $O(\alpha_s^3)$ the relevant parton-parton processes are
\begin{eqnarray} 
q+{\bar q}\rightarrow Q+{\bar Q}+g,\\
g+g\rightarrow Q+{\bar Q}+g,\\ 
q+g\rightarrow Q+{\bar Q}+q,\\
{\bar q}+g\rightarrow Q+{\bar Q}+{\bar q}, 
\end{eqnarray}
together with the virtual corrections to (1.2) and (1.3). 
Recent investigations have
shown that near threshold there are large logarithms in the perturbation
expansion which have to be resummed to make more reliable theoretical
predictions. The standard process is fixed target Drell-Yan production,
which has been the subject of many papers over
the past few years \cite{gs1}. 
The same ideas on resummation were applied to the calculation of 
the top-quark cross section at the Fermilab Tevatron in \cite{LSN1} and
\cite{lsn21}. 
What is relevant in these reactions 
is the existence of a class of logarithms of the type
$(\ln(1-z))^{i}/(1-z)$, where $i$ is the order of the
perturbation expansion, and where one must integrate over the variable
$z$ up to a limit $z=1$. These
terms are not actually singular at $z = 1$ due to the presence
of terms in $\delta(1-z)$. However the remainder
can be quite large. In general one writes such terms 
as ``plus'' distributions, which are then convoluted with
regular test functions (the parton densities).

In perturbation theory with a hard scale we can use the standard
\linebreak
expression for the order-by order cross section
in QCD, namely
\begin{eqnarray}
\sigma(S,m^2)&=& \int_{\frac{4m^2}{S}}^1 dx_1 
\int_{\frac{4m^2}{x_1 S}}^1 dx_2 \sum_{ij}
f_i(x_1,\mu^2) f_j(x_2,\mu^2) \sigma_{ij}(s = x_1 x_2 S, m^2,\mu^2) \, ,
\nonumber \\ &&
\end{eqnarray}
where the $f_i(x,\mu^2)$ are the parton densities at the factorization
scale $\mu^2$ and the $\sigma_{ij}$ are the partonic cross sections. 
The numerical results for the hadronic cross sections
depend on the choice of the
parton densities, which involves the mass factorization
scale $\mu^2$; the choice of the running coupling constant, which
involves the renormalization scale (also normally chosen to be $\mu^2$); 
and the choice for the actual mass of the heavy quark. 
In lowest order or Born approximation the
actual numbers for the cross section show a large sensitivity to
these parameters.  In chapter 2 we will show plots of 
the top and bottom quark production cross sections in
leading order (LO), i.e. $O(\alpha_s^2)$, 
and next-to-leading order (NLO), i.e. $O(\alpha_s^3)$.
The NLO results follow from the work of the 
two groups \cite{betal1} and \cite{nde11}.  
However even including the NLO
corrections does not completely fix the cross section. The sensitivity
to our lack of knowledge of even higher terms in the QCD expansion is usually
demonstrated by varying the scale choice up and down by factors
of two. In general it is impossible to make more precise
predictions given the absence of a calculation in next-to-next-to-leading 
order (NNLO). However in specific kinematical regions we can do 
so.

The threshold region is one of these regions.
In this region one finds that there are large logarithms which arise 
from an imperfect cancellation
of the soft-plus-virtual (S+V) terms in the perturbation expansion.
These logarithms are exactly of the same type mentioned above.
In \cite{LSN1}  the dominant logarithms
from initial state gluon bremsstrahlung (ISGB), 
which are the cause of the large corrections
near threshold, were carefully examined. 
Such logarithms have been studied previously
in Drell-Yan (DY) \cite{gs1} production at fixed target energies 
(again near threshold) where they are responsible for
correspondingly large corrections.   The 
analogy between DY and heavy quark production cross sections was 
exploited in \cite{LSN1} and
a formula to resum the leading logarithms
in pQCD to all orders was proposed. Since the contributions due to these
logarithms are positive (when all scales $\mu$ are set equal
to the heavy quark mass $m$),
the effect of summing the higher order corrections
increases the top and bottom quark production cross sections over those
predicted in $O(\alpha_s^3)$. 
This sum, which will be identified as $\sigma_{\rm res}$, depends
on a nonperturbative parameter $\mu_0$. The reason that
a new parameter has to be introduced is that
the resummation is sensitive to the scale at which
pQCD breaks down. As we approach the threshold region
other, nonperturbative, physics plays a role (higher twist,
bound states, etc.) indicated by a dramatic increase in 
$\alpha_s$ and in the resummed cross section. 
This is commonly called the effect of the infrared renormalon
or Landau pole \cite{Mueller1}. 
We choose to simply cut off the resummation
at a specific scale $\mu_0$ where 
$\Lambda << \mu_0 << m$  since it is not
obvious how to incorporate the nonperturbative effects.
Note that our resummed corrections diverge for small $\mu_0$ 
but this is {\em not} physical since they should be
joined smoothly onto some nonperturbative prescription
and the total cross section will be finite. 
Another way to make it finite would be to avoid the infrared
renormalon by a specific continuation around it, i.e. the
principal value resummation method \cite{ConAlv1}.
However, at the moment
our total resummed corrections depend on the parameter
$\mu_0$ for which we can only make a rough estimate.

We will see in chapter 2 that the gluon-gluon channel 
is the dominant channel for the production
of $b$-quarks near threshold in a fixed-target $pp$ experiment.  
This is not the case for the production of the top quark at the
Fermilab Tevatron, which is a proton-antiproton collider, and where 
the dominant channel is the quark-antiquark one. That was fortunate
as the exponentiation of the soft-plus-virtual terms in \cite{LSN1}
is on a much more solid footing in the $q \bar q$ channel, due to all
the past work which has been done on the Drell-Yan reaction \cite{gs1}. 
We will examine all ``large''
corrections near threshold, including both Coulomb-like 
and large constant terms. We will do that
in chapter 2 where we will present all the relevant formulae at
the partonic level.
In addition we will present subleading S+V terms and discuss their     
contribution to the total S+V cross section.
Chapter 2 also contains the analysis of the hadron-hadron cross section
which is relevant for top quark production at the Fermilab Tevatron
as well as for bottom quark production for the HERA-B experiment and 
for fixed-target
$pp$ experiments in general. We give
results in LO, in NLO and after resummation.
Most of our results in chapter 2 have appeared  
in \cite{KSb1} and \cite{KSh1}.

In chapter 3 we will present 
results for the heavy quark differential distributions
in transverse momentum $p_T$ and rapidity $Y$. After showing the relevant
formulas at the partonic level and discussing resummation we will
give results for top and bottom quark production at the Fermilab Tevatron
and HERA-B, respectively. We will show that resummation produces an
\linebreak
enhancement~of the NLO results with little change in shape.  
Our results in this chapter have appeared in \cite{KSb1} and \cite{NKJS1}.

In chapter 4 we will present a new resummation formula that takes into
account subleading logarithms. A strong motivation for this research
is the inadequacy of the leading-log approximation for the $gg$ channel
as will be discussed in chapter 2. 
We will show how nonleading soft logarithms exponentiate in a manner that
depends on the color structure within the underlying hard scattering.
We will derive the anomalous dimension
matrices and exhibit the exponentiation of nonleading soft logarithms
for the production of heavy quarks 
through both light quark-antiquark annihilation and
gluon-gluon fusion.
Some of our results in chapter 4 have been presented in \cite{NKGS1}.

\renewcommand{\theequation}{\thesection.\arabic{equation}}
\chapter{Top and bottom quark production cross sections} 

In this chapter we present
general formulas for heavy quark cross sections.
First we discuss our results at the partonic level and then we give
the calculations for
the cross sections for top quark production at the Fermilab Tevatron
and for bottom quark production at fixed-target energies and HERA-B. 
We present both the 
order $\alpha_s^3$ cross sections and the resummation of 
soft gluon corrections in all orders of QCD perturbation theory.

\mysection{Results for parton-parton reactions}
The partonic processes that we examine are
\begin{equation}
i(k_1) + j(k_2) \rightarrow Q(p_1) + \bar Q(p_2),
\end{equation}
where $i,j = g, q, \bar q$.
The square of the parton-parton center-of-mass (c.m.) energy is 
$s=(k_1+k_2)^2$. 

We begin with an analysis of heavy quark production in the $q \bar q$ channel.
The Born cross section in this channel is given by
\begin{equation}
\sigma_{q \bar q}^{(0)}(s,m^2)=\frac{2 \pi}{3} \alpha_s^2(\mu^2)
 K_{q \bar q} N C_F \frac{1}{s} \beta \left(1+\frac{2m^2}{s}\right)
\end{equation}
where $C_F=(N^2-1)/(2N)$ is the Casimir invariant for the fundamental
representation of $SU(N)$,  
$K_{q \bar q}=N^{-2}$ is a color average factor, $\mu$ denotes the 
renormalization scale, and $\beta=\sqrt{1-4m^2/s}$.
Also $N=3$ for the $SU(3)$ color group in QCD.
The threshold behavior ($s \rightarrow 4m^2$) of this expression is given by
\begin{equation}
\sigma_{q \bar q ,\, \rm thres}^{(0)}(s,m^2)=\pi\alpha_s^2(\mu^2) 
K_{q \bar q} N C_F \frac{1}{s} \beta.
\end{equation}
Complete analytic results are not available for the NLO cross section
as some integrals are too complicated to do by hand.
However in \cite{bnmss2} analytic results are given for
the soft-plus-virtual (S+V) contributions to the cross section, and for the
approximation to the cross section near threshold. 
Simple formulas which yield reasonable approximations to the exact 
$O(\alpha_s^3)$ results have been constructed in \cite{MSSN2}. From these
results one can derive that the Coulomb terms to first order in the 
$q \bar q$ channel are given by
\begin{equation}
\sigma^{(\pi^2)}_{q \bar q}(s,m^2)=\sigma^{(0)}_{q \bar q}(s,m^2)
\frac{\pi \alpha_s(\mu^2)}{2\beta}\left(C_F-\frac{C_A}{2}\right)
\end{equation}
in the $\overline {\rm MS}$ scheme,
where $C_A=N$ is the Casimir invariant for the adjoint representation
of $SU(N)$. 
These terms are distinguished by their typical
$\beta^{-1}$ behavior near threshold which, after multiplication by the Born
cross section, yield finite cross sections at threshold in NLO.
We note that $C_F-C_A/2=-1/6$ is negative for $SU(3)$ 
and that the first-order
Coulomb correction is negative (the interaction is repulsive).

In the DIS scheme, in addition to the Coulomb terms, we also have 
a large constant contribution so that the 
first order result near threshold is
\begin{equation}
\sigma^{(\pi^2)}_{q \bar q}(s,m^2)=\sigma^{(0)}_{q \bar q}(s,m^2) 
\left[\frac{\pi\alpha_s(\mu^2)}{2\beta}\left(C_F-\frac{C_A}{2}\right)
+\frac{\alpha_s(\mu^2)}{\pi}C_F\left(\frac{9}{2}+\frac{\pi^2}{3}\right)\right].
\end{equation}
We have included the constant terms to see their effect at
larger values of $\beta$.

Since the total parton-parton cross sections only depend on the variables
$s$ and $m^2$ they can be expressed in terms of scaling functions
as follows
\begin{eqnarray}
\sigma_{ij}(s,m^2)&=&\sum_{k=0}^{\infty} \sigma_{ij}^{(k)}(s,m^2)
\nonumber \\ & =&
\frac{\alpha_s^2(\mu^2)}{m^2} \sum_{k=0}^{\infty}
(4\pi\alpha_s(\mu^2))^k \sum_{l=0}^{k} f_{ij}^{(k,l)}(\eta)
\ln^l\frac{\mu ^2}{m^2},
\end{eqnarray}
where we denote by $\sigma^{(k)}$ the $O(\alpha_s^{k+2})$ contribution
to the cross section.
The scaling functions $f_{ij}^{(k,l)}(\eta)$ depend on the scaling variable
$\eta=s/4m^2-1=s\beta^2/4m^2$.

\begin{figure}
\centerline{\psfig{file=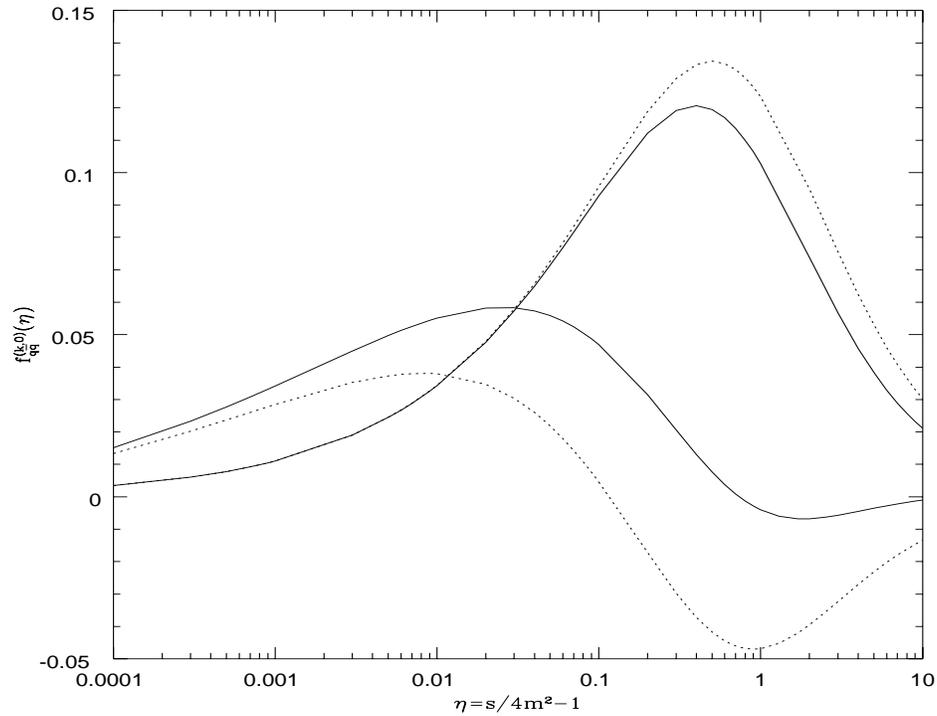,height=4.05in,width=5.05in,clip=}}
\caption[The scaling functions $f_{q\bar q}^{(k,0)}$ 
in the $\overline {\rm MS}$ scheme]
{\baselineskip=12pt
{The scaling functions $f_{q\bar q}^{(k,0)}$ in the $\overline {\rm MS}$
scheme. Plotted are $f_{q\bar q}^{(0,0)}$ (exact, upper solid line at large
$\eta$; threshold approximation, upper dotted line at large $\eta$),
$f_{q\bar q}^{(1,0)}$ (exact, lower solid line at large $\eta$; 
threshold approximation, lower dotted line at large $\eta$).}}
\label{fig. 2.1}
\end{figure}

In fig. 2.1 we plot $f_{q \bar q}^{(k,0)}(\eta)$ for $k=0,1$ for the
exact and threshold  expressions (from \cite{bnmss2}) in the 
$\overline {\rm MS}$ scheme. 
We see that the threshold Born approximation is excellent for small
$\eta$ and reasonable for the entire range of $\eta$ shown. 
We also note that the theshold first-order approximation is good
only very near to threshold.
\begin{figure}
\centerline{\psfig{file=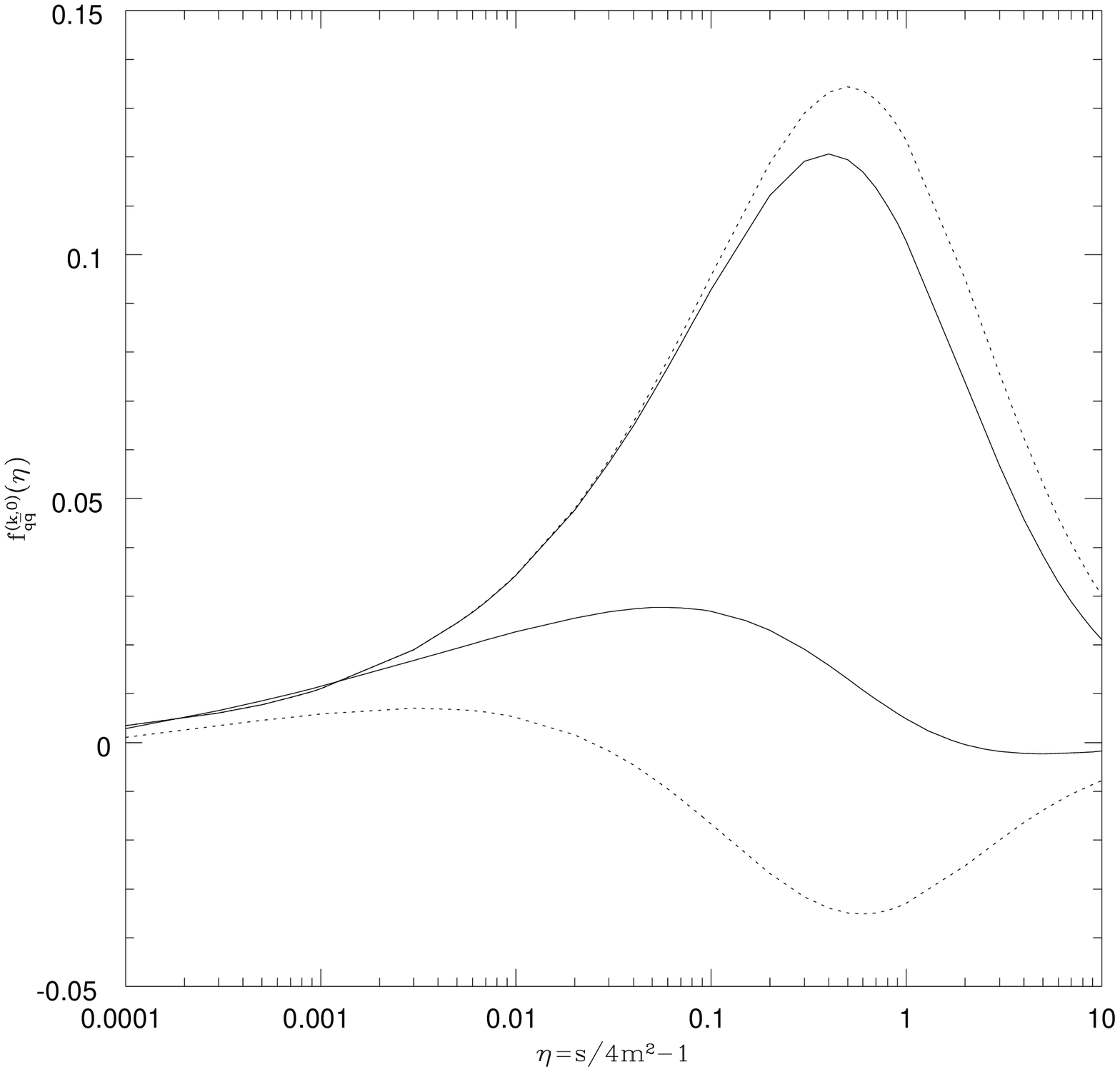,height=4.05in,width=5.05in,clip=}}
\caption[The scaling functions $f_{q\bar q}^{(k,0)}$ 
in the DIS scheme]
{Same as fig. 2.1 but now for the DIS scheme.}
\label{fig. 2.2}
\end{figure}
In fig. 2.2 we plot the corresponding functions for the DIS scheme.
Here the first-order corrections are smaller than in the 
$\overline {\rm MS}$ scheme. Again the threshold first-order approximation
is good only very close to threshold. 

The analysis of the contributions to the gluon-gluon channel in NLO
is much more complicated. First of all there are three Born diagrams
each with a different color structure.  Therefore only few terms
near threshold are proportional to the Born cross section.
The exact Born term in the $gg$ channel is 
\begin{eqnarray}
\sigma_{gg}^{(0)}(s,m^2)&=&4\pi\alpha_s^2(\mu^2) K_{gg}N C_F \frac{1}{s}
\left\{C_F\left[-\left(1+\frac{4m^2}{s}\right)\beta\right.\right.
  \nonumber \\ && \quad \quad \quad \quad
+\left. \left(1+\frac{4m^2}{s}-\frac{8m^4}{s^2}\right)
 \ln\frac{1+\beta}{1-\beta}\right] 
\nonumber \\ &&  
+\left.C_A\left[-\left(\frac{1}{3}+\frac{5}{3}\frac{m^2}{s}\right)\beta
+\frac{4m^4}{s^2}\ln\frac{1+\beta}{1-\beta}\right]\right\},
\end{eqnarray}
where $K_{gg}=(N^2-1)^{-2}$ is a color average factor.
The threshold behavior ($s \rightarrow 4m^2$) of this expression is given by
\begin{equation}
\sigma_{gg,\, \rm thres}^{(0)}(s,m^2)= \pi \alpha_s^2(\mu^2) K_{gg}
\frac{1}{s} N C_F [4 C_F-C_A] \beta.
\end{equation}
Again, the complete NLO expression for the cross section in the $gg$ channel 
is unavailable but analytic results are given for the S+V terms in 
\cite{betal2}. These were used in \cite{MSSN2} to analyze the magnitude of the
cross section near threshold. 
From the approximate expressions given in \cite{MSSN2} 
one can extract the $\pi^2$ terms to first order in the $gg$ channel. 
These are 
\begin{eqnarray}
\sigma_{gg}^{(\pi^2)}(s,m^2)&=&\alpha_s^3(\mu^2) N C_K K_{gg} \frac{\pi^2}{s}
\left[\frac{5}{8}+\frac{1}{24}\beta^2+16\frac{m^6}{s^3} \right.
\nonumber \\ && \quad \quad \quad
+\left.\left(32\frac{m^8}{s^4}-10\frac{m^4}{s^2}\right)\frac{1}{\beta}
\ln\frac{1+\beta}{1-\beta}\right] \nonumber \\ &&  
+\alpha_s^3 C_{\rm QED} K_{gg} \frac{\pi^2}{s}
\left[-\frac{1}{4}-16\frac{m^6}{s^3} \right. \nonumber \\ && \quad \quad \quad
\left.+\left(-32\frac{m^8}{s^4}+8\frac{m^4}{s^2}\right)\frac{1}{\beta}
\ln\frac{1+\beta}{1-\beta}\right],
\end{eqnarray}
where $C_K=(N^2-1)/N=2N C_F C_A-4N C_F^2$, and $C_{\rm QED}=(N^4-1)/N^2
=-4C_F^2+4C_A C_F$.
These are not proportional to the Born term.
The threshold behavior of (2.1.9) is given by
\begin{equation}
\sigma_{gg,\, \rm thres}^{(\pi^2)}(s,m^2)=\alpha_s^3(\mu^2) K_{gg}
\frac{\pi^2}{4}\frac{1}{s}\left[\frac{-N C_K}{2}+C_{\rm QED}\right],
\end{equation} 
which is proportional to the 
threshold Born term.
Therefore the threshold approximation for the $\pi^2$ terms
in NLO can be written as
\begin{equation}
\sigma_{gg,\, \rm thres}^{(0)+(\pi^2)}(s,m^2)=
\sigma_{gg,\, \rm thres}^{(0)}(s,m^2) \left[1+\frac{\pi\alpha_s(\mu^2)}
{4\beta}\left(\frac{-NC_K/2+C_{\rm QED}}{(4C_F-C_A)N C_F}\right)\right],
\end{equation}
or, writing the color factors in terms of $N$, as
\begin{equation}
\sigma_{gg,\, \rm thres}^{(0)+(\pi^2)}(s,m^2)=
\sigma_{gg,\, \rm thres}^{(0)}(s,m^2) \left[1+\frac{\pi\alpha_s(\mu^2)}
{4\beta}\frac{N^2+2}{N(N^2-2)}\right].
\end{equation}

\begin{figure}
\centerline{\psfig{file=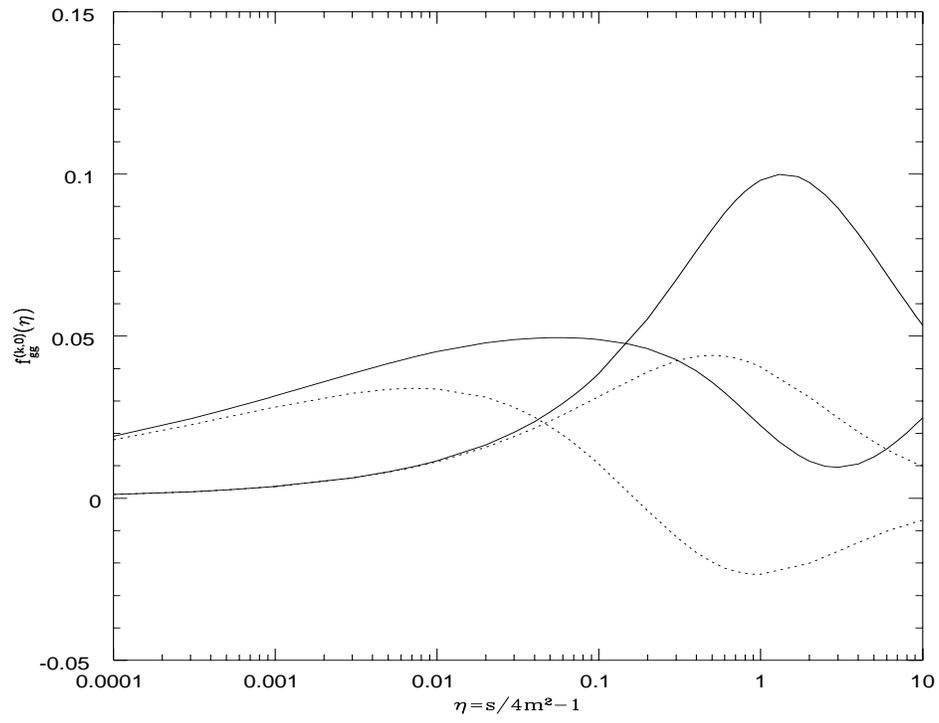,height=4.05in,width=5.05in,clip=}}
\caption[The scaling functions $f_{gg}^{(k,0)}$ 
in the $\overline {\rm MS}$ scheme]
{\renewcommand{\baselinestretch}{1}
{Same as fig. 2.1 but now for $f_{gg}^{(k,0)}$ in the 
$\overline {\rm MS}$ scheme.}}
\label{fig. 2.3}
\end{figure}

In fig. 2.3 we plot the scaling functions $f_{gg}^{(k,0)}(\eta)$ with 
$k=0,1$ in the $\overline {\rm MS}$ scheme for the exact and threshold
expressions (from \cite{betal2}). 
We see that the Born and first-order threshold
approximations are good only very close to threshold. 

In \cite{LSN2} an approximation was given for the 
NLO soft-plus-virtual contributions
and the analogy with the Drell-Yan process was exploited
to resum them to all orders of perturbation theory. The
S+V approximation is adequate in the region 
$0.1<\eta<1$ (which is the kinematical region of interest
as we will see in sections 2.2 and 2.3) for the $q \bar q$ channel,
but not as good for the $gg$ channel in the $\overline {\rm MS}$ scheme.
Therefore we  reexamined the approximate formulae given in \cite{MSSN2}
for the initial state gluon bremsstrahlung (ISGB) mechanism 
to see if there are subleading terms that will improve the
S+V approximation. Let us see the structure of these terms. 
We are discussing partonic reactions of the type 
$i(k_1)+j(k_2) \rightarrow Q(p_1) + \bar Q(p_2)+g(k_3)$,
and we introduce the kinematic variables $t_1=(k_2-p_2)^2-m^2$, 
$u_1=(k_1-p_2)^2-m^2$, and $s_4 = s+t_1+u_1$. The variable $s_4$ depends
on the four-momentum of the extra partons  emitted in the reaction.   
We write the differential
cross section in order $\alpha_s^k(\mu^2)$ as follows
\begin{eqnarray}
s^2\frac{d^2\sigma_{ij}^{(k)}(s,t_1,u_1)}{dt_1 \: du_1} &=&
\alpha_s^k(\mu^2) \sum_{l=0}^{2k-1} \Big[\frac{1}{s_4}a_l(\mu^2) 
\ln^l\Big(\frac{s_4}{m^2}\Big)\theta(s_4 - \Delta)
  \nonumber \\ &&  
+ \frac{1}{l+1} a_l(\mu^2) \ln^{l+1}\Big(\frac{\Delta}{m^2}\Big) \delta(s_4)
\Big] \sigma^B_{ij}(s,t_1,u_1) \,.
\nonumber \\ 
\end{eqnarray}
Here a small parameter $\Delta$ has been introduced 
to allow us to distinguish between the
soft ($s_4 < \Delta$) and the hard ($s_4> \Delta$)
regions in phase space. The quantities $a_l(\mu^2)$ contain 
terms involving the QCD $\beta$-functions and color factors.
The variables $t_1$ and $u_1$ are then mapped onto the variables $s_4$ and
$\cos\theta$, where $\theta$ is the parton-parton
c.m. scattering angle:
\begin{eqnarray}
t_1&=&-\frac{1}{2}\left\{s-s_4-[(s-s_4)^2-4sm^2]^{1/2} \cos \theta\right\},
\\ 
u_1&=&-\frac{1}{2}\left\{s-s_4+[(s-s_4)^2-4sm^2]^{1/2} \cos \theta\right\}.
\end{eqnarray}
After explicit integration over the angle 
$\theta$, the series becomes
\begin{eqnarray}
\sigma_{ij}^{(k)}(s,m^2)&=&\alpha_s^{(k)}(\mu^2)\sum_{l=0}^{2k-1} a_l(\mu^2)
\left\{\int_0^{s-2m s^{1/2}}ds_4 \frac{1}{s_4} \ln^l\frac{s_4}{m^2}\right.
\nonumber \\ &&
\times[{\bar{\sigma}}_{ij}^{(0)}(s,s_4,m^2)
-{\bar{\sigma}}_{ij}^{(0)}(s,0,m^2)]
\nonumber \\ &&
\left. +\frac{1}{l+1}\ln^{l+1}\left(\frac{s-2ms^{1/2}}{m^2}\right)
{\bar{\sigma}}_{ij}^{(0)}(s,0,m^2)\right\},
\end{eqnarray}
where
\begin{equation}
{\bar{\sigma}}_{ij}^{(0)}(s,s_4,m^2)=\frac{1}{2s^2}
[(s-s_4)^2-4sm^2]^{1/2}\int_{-1}^1d\cos \theta \; \;  
\sigma_{ij}^B(s,s_4,\cos \theta).
\end{equation} 

The Born approximation differential cross sections can be expressed by
\begin{equation}
s^2\frac{d^2\sigma^{(0)}_{ij}(s,t_1,u_1)}{dt_1 \: du_1} = \delta
(s+t_1+u_1) \, \sigma^B_{ij}(s,t_1,u_1)\,,
\end{equation}
with 
\begin{equation}
\sigma^B_{q\bar q}(s,t_1,u_1) = \pi \alpha_s^2(\mu^2) K_{q\bar{q}}
NC_F \Big[ \frac{t_1^2 + u_1^2}{s^2} + \frac{2m^2}{s}\Big]\,,
\end{equation}
and
\begin{eqnarray}
\sigma^B_{gg}(s,t_1,u_1)& = & 2\pi \alpha_s^2(\mu^2) K_{gg}
NC_F \Big[C_F - C_A \frac{t_1u_1}{s^2}\Big] \nonumber \\ &&
\times\Big[ \frac{t_1}{u_1} + \frac{u_1}{t_1} + \frac{4m^2s}{t_1u_1}
\Big(1 - \frac{m^2s}{t_1u_1}\Big) \Big] \,.
\end{eqnarray}

The first-order approximate S+V result for the
$q \bar {q}$ channel in the $\overline{\rm MS}$ scheme is 
\begin{eqnarray}
s^2\frac{d^2\sigma^{(1)}_{q \bar q}(s,t_1,u_1)}{dt_1 du_1}&=&
\sigma_{q \bar q}^B(s,t_1,u_1)\frac{2C_F}{\pi} \alpha_s(\mu^2)
\nonumber \\ &&
\times
\left\{\left[\frac{1}{s_4}\left(2\ln\frac{s_4}{m^2}+\ln\frac{m^2}{\mu^2}
\right)\theta(s_4-\Delta) \right.\right.\nonumber \\ &&
\left.+\left(\ln^2\frac{\Delta}{m^2}
+\ln\frac{\Delta}{m^2}
\ln\frac{m^2}{\mu^2}\right) \delta(s_4)\right]
  \nonumber \\ &&
+\left.\left[-\frac{C_A}{2C_F}\frac{1}{s_4}\theta(s_4-\Delta)
-\frac{C_A}{2C_F}\ln\frac{\Delta}{m^2}
\delta(s_4)\right]\right\} \, . 
\nonumber \\
\end{eqnarray} 
The terms in the first pair of square brackets in (2.1.21) are the 
leading
S+V terms given in \cite{LSN2} and those in the second pair of square
brackets are subleading terms that we want to examine.
\begin{figure}
\centerline{\psfig{file=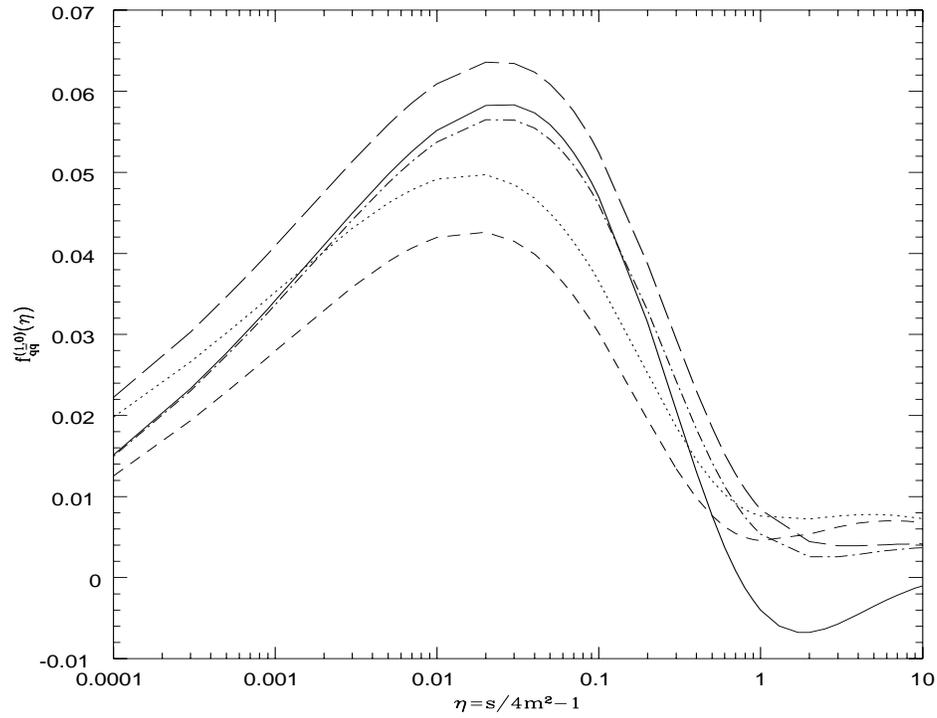,height=4.05in,width=5.05in,clip=}}
\caption[The scaling functions $f_{q \bar q}^{(1,0)}$ 
in the $\overline {\rm MS}$ scheme]
{\renewcommand{\baselinestretch}{1}
{The scaling functions $f_{q\bar q}^{(1,0)}$ in the $\overline {\rm MS}$
scheme. Plotted are the exact result (solid line), the leading S+V 
approximation (dotted line),
the leading S+V approximation plus Coulomb terms (short-dashed line), 
the S+V approximation with leading plus subleading terms (long dashed line), 
and the S+V approximation with leading plus subleading terms
plus Coulomb terms (dash-dotted line).}}
\label{fig. 2.4}
\end{figure}
In fig. 2.4 we plot the scaling functions $f_{q \bar q}^{(1,0)}$
for the exact result, the leading approximate S+V result, 
and the approximate S+V result with
both leading and subleading logarithms. The leading S+V result is
a reasonable approximation to the exact result in our region of interest
$0.1<\eta<1$. 
The addition of Coulomb terms
worsens the leading S+V result. We also see that when we include the
subleading terms our approximation does not improve much in the region of 
interest. However, when we add both the first order Coulomb terms and
subleading terms to the leading 
S+V result we get a very good agreement with the exact result.
In the DIS scheme the analogous results are
\begin{eqnarray}
s^2\frac{d^2\sigma^{(1)}_{q \bar q}}{dt_1 du_1}(s,t_1,u_1)&=&
\sigma_{q \bar q}^B(s,t_1,u_1)
\frac{2 C_F}{\pi}\alpha_s(\mu^2)
 \nonumber \\ &&
\times
\left\{\left[\frac{1}{s_4}\left(\ln\frac{s_4}{m^2}+\ln\frac{m^2}{\mu^2}
\right)\theta(s_4-\Delta) \right.\right. \nonumber \\ &&
\left.+\left(\frac{1}{2}\ln^2\frac{\Delta}{m^2}+\ln\frac{\Delta}{m^2}
\ln\frac{m^2}{\mu^2}\right) \delta(s_4)\right]
 \nonumber \\ &&
+\left[\left(\frac{3}{4}+\ln2-\frac{C_A}{2C_F}\right)\frac{1}{s_4}
\theta(s_4-\Delta)\right.\nonumber \\ &&
\left.\left.+\left(\frac{3}{4}+\ln2-\frac{C_A}{2C_F}\right)
\ln\frac{\Delta}{m^2}\delta(s_4)\right]\right\}.
\end{eqnarray} 
\begin{figure}
\centerline{\psfig{file=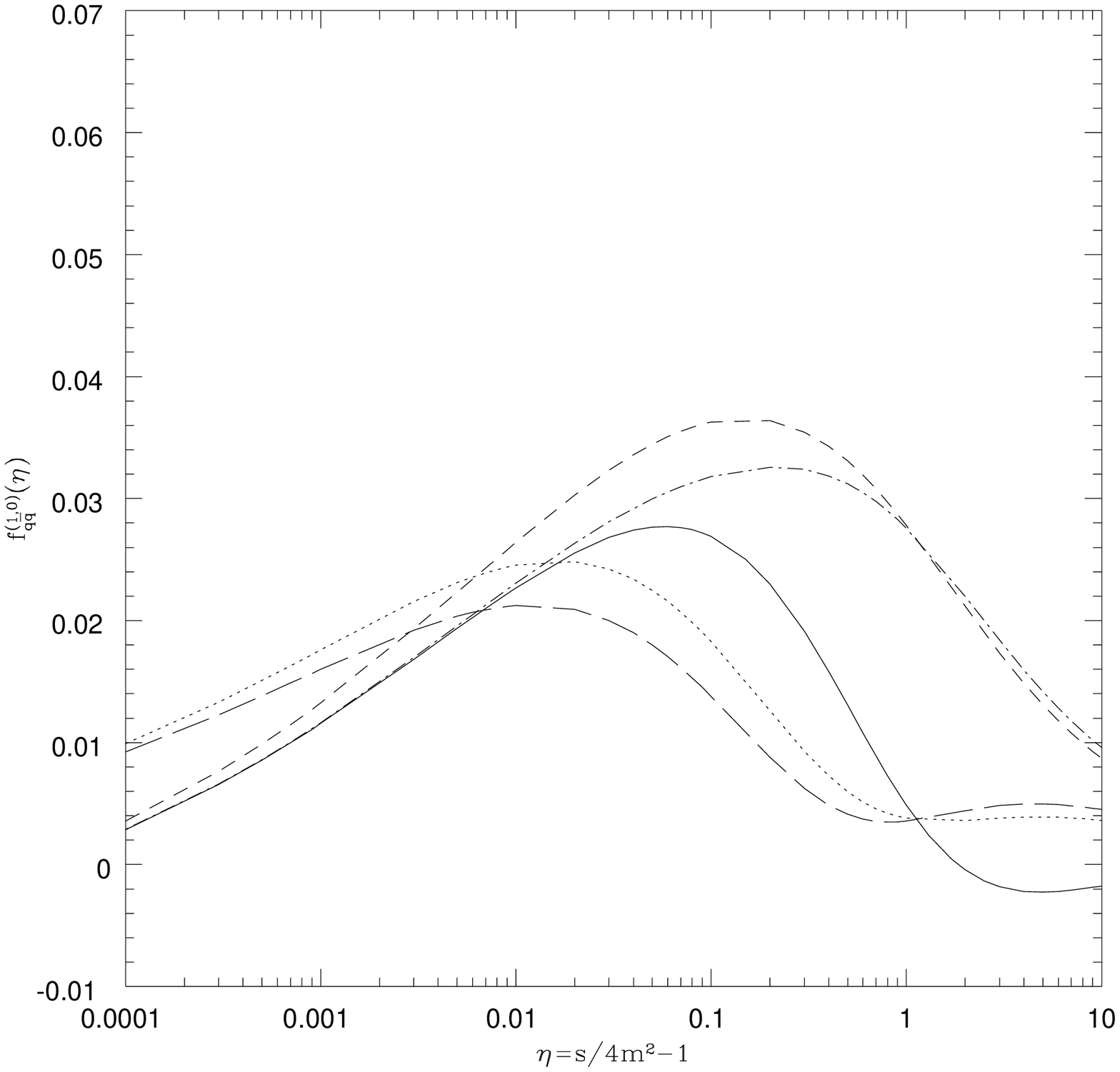,height=4.05in,width=5.05in,clip=}}
\caption[The scaling functions $f_{q \bar q}^{(1,0)}$ 
in the DIS scheme]
{Same af fig. 2.4 but now for the DIS scheme.} 
\label{fig. 2.5}
\end{figure}
In fig. 2.5 we plot the corresponding scaling functions. Here 
the addition of subleading terms worsens the leading S+V
approximation. The addition of Coulomb terms and large constants
enhances the first-order approximate results considerably. 

The resummation of the leading S+V terms has been given in \cite{LSN2}.
The result is 
\begin{eqnarray}
s^2\frac{d^2\sigma^{\rm res}_{q \bar q}
(s,t_1,u_1)}{dt_1 du_1}&=&\sigma_{q \bar q}^B(s,t_1,u_1)
\left[\frac{df(s_4/m^2,m^2/\mu^2)}{ds_4}\theta(s_4-\Delta)\right.
\nonumber \\ && \quad \quad \quad \quad \quad \quad 
+\left.f(\frac{\Delta}{m^2},\frac{m^2}{\mu^2})\delta(s_4)\right],
\end{eqnarray} 
where 
\begin{equation}
f\left(\frac{s_4}{m^2},\frac{m^2}{\mu^2}\right)=
\exp\left[A\frac{C_F}{\pi}\bar\alpha_s\left(\frac{s_4}{m^2},m^2\right)
\ln^2\frac{s_4}{m^2}\right]
\frac{[s_4/m^2]^{\eta}}{\Gamma(1+\eta)}\exp(-\eta \gamma_E).
\end{equation}
The straightforward expansion of the exponential plus the change
of the argument in $\bar\alpha_s$ via the renormalization group equations
generates the corresponding leading logarithmic terms.
The scheme dependent $A$ and $\bar\alpha_s$ in the above expression
are given by
\begin{equation}
A=2, \; \; \; \; \; \bar\alpha_s(y,\mu^2)=\alpha_s(y^{2/3}\mu^2)
=\frac{4\pi}{\beta_0 \ln (y^{2/3}\mu^2/\Lambda^2)}\,,
\end{equation}
in the $\overline{\rm MS}$ scheme, and
\begin{equation}
\! \! \! \! \! \! \! \! \! \! \! \! A=1, \; \; \; \; \; 
\bar\alpha_s(y,\mu^2)=\alpha_s(y\mu^2)
=\frac{4\pi}{\beta_0 \ln (y\mu^2/\Lambda^2)}\,,
\end{equation}
in the DIS scheme, 
where $\beta_0=11/3 \; C_A-2/3 \; n_f$ 
is the lowest order coefficient of the QCD $\beta$-function.
The color factors $C_{ij}$ are defined by $C_{q\bar{q}}=C_F$ and
$C_{gg}=C_A$, and $\gamma_E$ is the Euler constant.
The quantity $\eta$ is given by
\begin{equation}
\eta=\frac{8C_{ij}}{\beta_0}\ln\left(1+\beta_0
\frac{\alpha_s(\mu^2)}{4\pi}\ln\frac{m^2}{\mu^2}\right)\,.
\end{equation}
As the NNLO cross section is not known exactly 
we do not how to resum the subleading terms. 
In chapter 4, however, we will derive their exponentiation
using a different formalism.

Now let us see the analogous results for the $gg$ channel in the
$\overline{\rm MS}$ scheme. We have
\begin{eqnarray}
s^2\frac{d^2\sigma^{(1)}_{gg}(s,t_1,u_1)}{dt_1 du_1}
&=&\sigma_{gg}^B(s,t_1,u_1)
\frac{2 C_A}{\pi}\alpha_s(\mu^2)
\nonumber \\ &&
\times\left\{\left[\frac{1}{s_4}
\left(2\ln\frac{s_4}{m^2}+\ln\frac{m^2}{\mu^2}\right)\theta(s_4-\Delta)
\right.\right.
\nonumber \\ &&
\left.+\delta(s_4)\left(\ln^2\frac{\Delta}{m^2}+\ln\frac{\Delta}{m^2}
\ln\frac{m^2}{\mu^2}\right)\right]
\nonumber \\ &&
+\left.\left[\frac{3C_A-8C_F}{-2C_A+8C_F}\left(\frac{1}{s_4}\theta(s_4-\Delta)
+\ln\frac{\Delta}{m^2}\delta(s_4)\right)\right]\right\}. \nonumber\\ &&
\end{eqnarray}
Again, the terms in the first pair of square brackets in (2.1.28) are the
leading S+V terms and those in the second pair of square brackets 
are subleading terms.
\begin{figure}
\centerline{\psfig{file=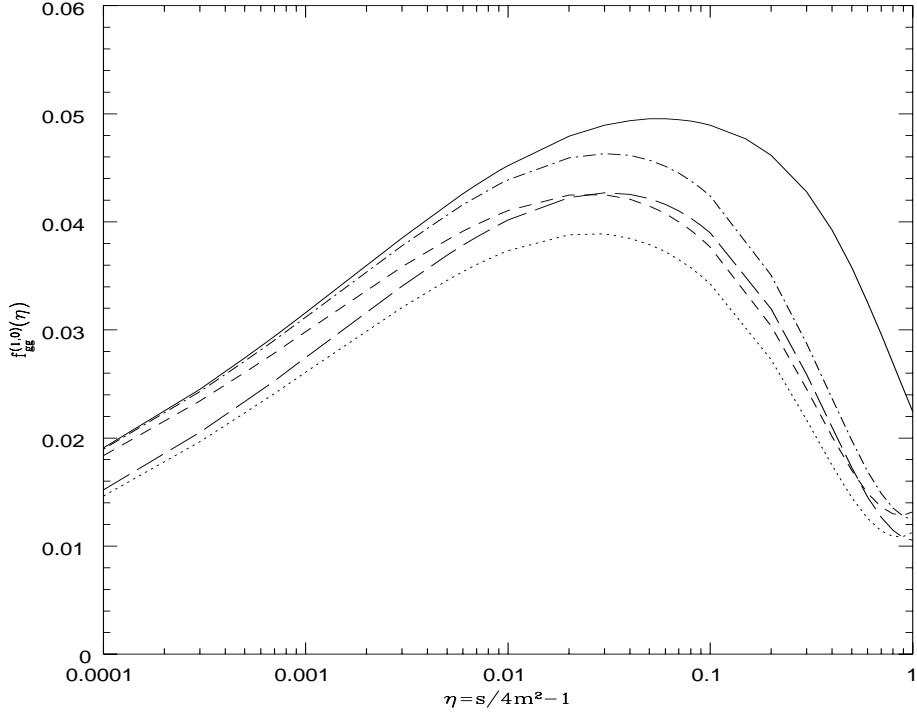,height=4.05in,width=5.05in,clip=}}
\caption[The scaling functions $f_{gg}^{(1,0)}$ 
in the $\overline {\rm MS}$ scheme]
{Same as fig. 2.4 but now for $f_{gg}^{(1,0)}$ in the 
$\overline {\rm MS}$ scheme.}
\label{fig. 2.6}
\end{figure}
In fig. 2.6 we plot the scaling functions $f_{gg}^{(1,0)}$
for the exact result, the leading approximate S+V result, 
and the approximate S+V result with
both leading and subleading terms. We note that the leading
S+V approximate result is significantly smaller than the exact result
and that the addition of subleading terms improves the approximation
considerably. This is important since, as we will see in the next section,
the $gg$ channel is dominant for the production of $b$-quarks at HERA-B.
Also the addition of Coulomb terms further improves the approximation.
The resummation of the leading S+V terms for the $gg$ channel has 
also been given in \cite{LSN2}. The result is
\begin{eqnarray}
s^2\frac{d^2\sigma^{\rm res}_{gg}(s,t_1,u_1)}
{dt_1 du_1}&=&\sigma_{gg}^B(s,t_1,u_1)
\left[\frac{df(s_4/m^2,m^2/\mu^2)}{ds_4}\theta(s_4-\Delta)\right.
\nonumber \\ && \quad \quad \quad \quad \quad \quad
+\left.f\left(\frac{\Delta}{m^2},\frac{m^2}{\mu^2}\right)\delta(s_4)\right],
\end{eqnarray} 
where the function $f$ is the same as for the $q {\bar q}$ channel
in the $\overline {\rm MS}$ scheme (with the substitution 
$C_F \rightarrow C_A$). 
Again, we do not know how to resum the subleading logs, but in chapter 4
we will derive their exponentiation using a different formalism. 

After we map $t_1$ and $u_1$ onto $s_4$ and $\cos \theta$
and we integrate over $\theta$, as we saw earlier,
the resummed cross section for either channel becomes
\begin{equation}
\sigma_{ij}(s,m^2)=-\int_{s_0}^{s-2m s^{1/2}} ds_4 \; f\left(\frac{s_4}{m^2},
\frac{m^2}{\mu^2}\right)\frac{d}{ds_4} 
{\bar{\sigma}}_{ij}^{(0)}(s,s_4,m^2).
\end{equation}
Note that we now have cut off the lower limit of the $s_4$ integration
at $s_4=s_0$ because $\bar\alpha_s$ diverges as $s_4 \rightarrow 0$.
This parameter $s_0$ must satisfy
the conditions $0<s_0<s-2ms^{1/2}$ and $s_0/m^2 << 1$.
It is convenient to rewrite $s_0$ in terms of the scale $\mu$ as
\begin{equation}
\frac{s_0}{m^2}=\left(\frac{\mu_0^2}{\mu^2}\right)^{3/2}
(\overline{\rm MS} \: \rm scheme); 
\end{equation}
\begin{equation}
\frac{s_0}{m^2}=\frac{\mu_0^2}{\mu^2} 
\; \; \; \; \; \; \; \; \; \! (\rm DIS \:scheme).
\end{equation}
Here $\mu_0$ is a nonperturbative parameter  \cite {LSN2}  
satisfying $\Lambda^2<<\mu_0^2<<\mu^2$.

\mysection{Results for top quark production at the Fermilab Tevatron}
In this section we examine the production of top quarks at the Fermilab 
Tevatron.
Following the notation in \cite{LSN2} the total hadron-hadron
cross section in order $\alpha_s^{k}$ is 
\begin{equation}
\sigma^{(k)}_H(S,m^2) = \sum_{ij}\int_{4m^2/S}^1
\,d\tau \,\Phi_{ij}(\tau,\mu^2)\, \sigma_{ij}^{(k)}(\tau S,m^2,\mu^2)\,,
\end{equation}
where $S$ is the square of the hadron-hadron c.m. energy and
$i,j$ run over $q,\bar q$ and $g$.
The parton flux $\Phi_{ij}(\tau,\mu^2)$ is defined via
\begin{equation}
\Phi_{ij}(\tau,\mu^2) = \int_{\tau}^1\, \frac{dx}{x} 
H_{ij}(x,\frac{\tau}{x},\mu^2) \,,
\end{equation}
and $H_{ij}$ is a product of the scale-dependent parton distribution
functions $f^h_i(x,\mu^2)$, where $h$ stands for the hadron which is
the source of the parton~$i$
\begin{equation}
H_{ij}(x_1, x_2, \mu^2) = f_i^{h_1}(x_1, \mu^2) f_j^{h_2}(x_2,\mu^2)\,.
\end{equation}
The mass factorization scale $\mu$ is chosen to be identical with
the renormalization scale in the running coupling constant.

In the case of the all-order 
resummed expression the lower boundary in (2.2.1)  
has to be modified according to the condition
$s_0 < s - 2ms^{1/2}$.
\linebreak
Resumming the soft gluon contributions to all orders we obtain
\begin{equation}
\sigma^{\rm res }_H(S,m^2) = \sum_{ij}\int_{\tau_0}^1
\,d\tau \,\Phi_{ij}(\tau,\mu^2)\, \sigma_{ij}(\tau S,m^2,\mu^2)\,,
\end{equation}
where $\sigma_{ij}$ is given in (2.1.30)  and
\begin{equation}
\tau_0 = \frac{[m+(m^2+s_0)^{1/2}]^2}{S}\,.
\end{equation}
Here $s_0$ (or equivalently $\mu_0$) is the non-perturbative
parameter used to cut off the resummation
since the resummed corrections diverge for small $s_0$.

We now specialize to top quark production at the Fermilab Tevatron
where $\sqrt{S}=1.8$ TeV.
Taking the top quark mass as
$m_t = 175\: {\rm GeV}/c^2$ then the ratio of $m_t/\sqrt{S} \approx 0.1$. 
If we choose the renormalization scale in the running coupling constant
as $m_t$ then $\alpha_s(m_t^2) \approx 0.1$ so 
$\alpha_s(m_t^2) \ln(\sqrt{S}/m_t) \approx  0.2$. This number is
small enough that we expect a reasonably convergent perturbation series.
In the presentation of our results for the exact, approximate,
and resummed hadronic cross sections 
we use the MRSD$\_ ' \:$ parametrization for the parton distributions
\cite{mrs2}.
Note that the hadronic results only involve partonic distribution
functions at moderate and large $x$, where there is little difference
between the various sets of parton densities.
We have used the MRSD$\_ '\:$ set 34 as given in PDFLIB \cite{PDFLIB2} in the 
DIS scheme with the number of active light flavors $n_f=5$ and the QCD
scale $\Lambda_5=0.1559$ GeV. We have used the two-loop corrected 
running coupling constant as given by PDFLIB.   

First, we discuss the NLO contributions to top quark production at
the Tevatron using the results in \cite{nde12} and \cite{{bnmss2},{betal2}}.
Except when explicitly stated otherwise we will take the factorization
scale $\mu=m_t$.
\begin{figure}
\centerline{\psfig{file=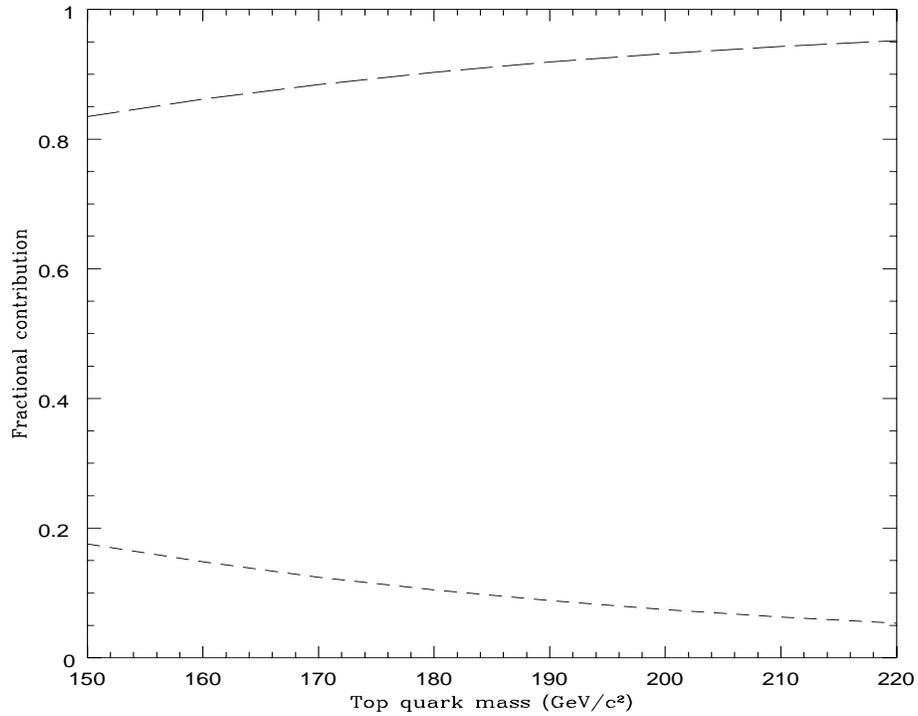,height=4.05in,width=5.05in,clip=}}
\caption[Fractional contributions of the $q \bar q$ and 
$gg$ channels to the total $O(\alpha_s^3)$
top quark production cross section at the Fermilab Tevatron 
versus top quark mass]
{Fractional contributions of the $q \bar q$ (DIS scheme, long-dashed line)
and $gg$ ($\overline{\rm MS}$ scheme, short-dashed line) 
channels to the total $O(\alpha_s^3)$
top quark production cross section at the Fermilab Tevatron 
as a function of the top quark mass.}
\label{fig. 2.7}
\end{figure}
In fig. 2.7 we show the relative contributions 
to the NLO cross section of the $q \bar q$ channel in
the DIS scheme and the $gg$ channel in the $\overline{\rm MS}$ scheme 
as a function of the top quark mass. 
We see  that the $q \bar q$ contribution is the dominant
one and it is about 90\% of the total NLO cross section 
for $m_t=175$ GeV/$c^2$. 
The $gg$ contribution is smaller and makes up the rest of the  
cross section.
The contributions of the $g q$ and the $g \bar q$ channels 
are negligible and are not shown.

\begin{figure}
\centerline{\psfig{file=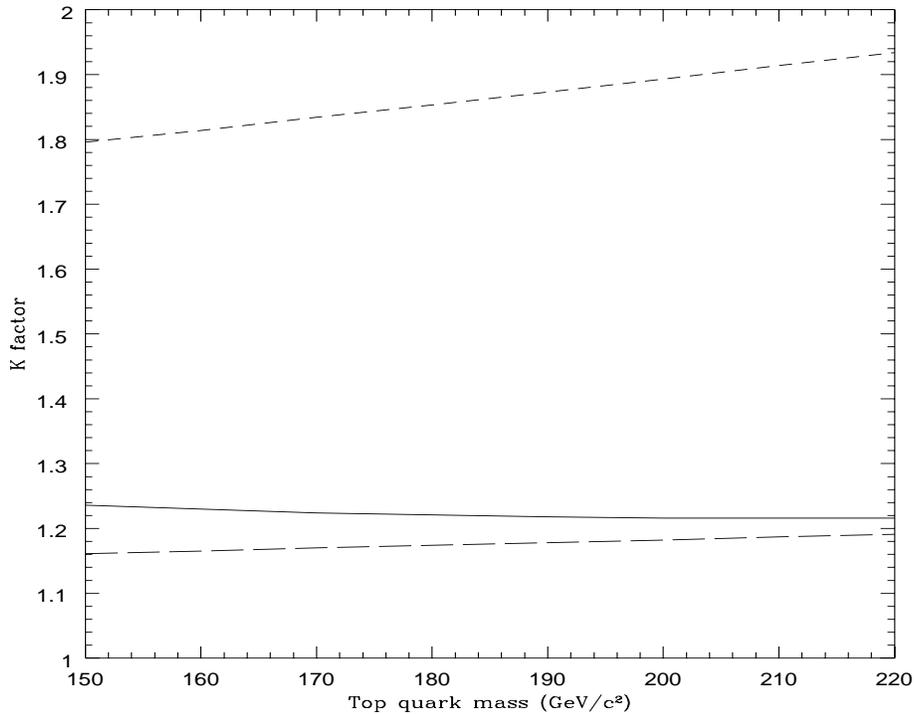,height=4.05in,width=5.05in,clip=}}
\caption[The K factors as a function of top quark mass 
for top quark production at the Fermilab Tevatron]
{The K factors as a function of top quark mass 
for top quark production at the Fermilab Tevatron for the 
$q \bar q$ channel (DIS scheme, long-dashed line),
the $gg$ channel ($\overline{\rm MS}$ scheme, short-dashed line), 
and their sum (solid line).}
\label{fig. 2.8}
\end{figure}
In fig. 2.8 we show the $K$ factors for the $q \bar q$ and $gg$ channels and 
for their sum as a function of the top quark mass.
The $K$ factor is defined by $K=(\sigma^{(0)}
+\sigma^{(1)}\mid _{\rm exact})/\sigma^{(0)}$, where $\sigma^{(0)}$ is the Born
term and $\sigma^{(1)}\mid _{\rm exact}$ is the exact first order correction.
We notice that the $K$ factor is quite large for the $gg$ channel,
which means that higher order effects are more important for this channel
than for $q \bar q$. However, since the $q \bar q$ channel is dominant,
the $K$ factor for the sum of the two channels is only slightly larger
than that for $q \bar q$.
These large corrections come predominantly from the threshold region for 
top quark production where it has been shown that initial state gluon 
bremsstrahlung (ISGB) is responsible for the large corrections at 
NLO \cite{MSSN2}.
\begin{figure}
\centerline{\psfig{file=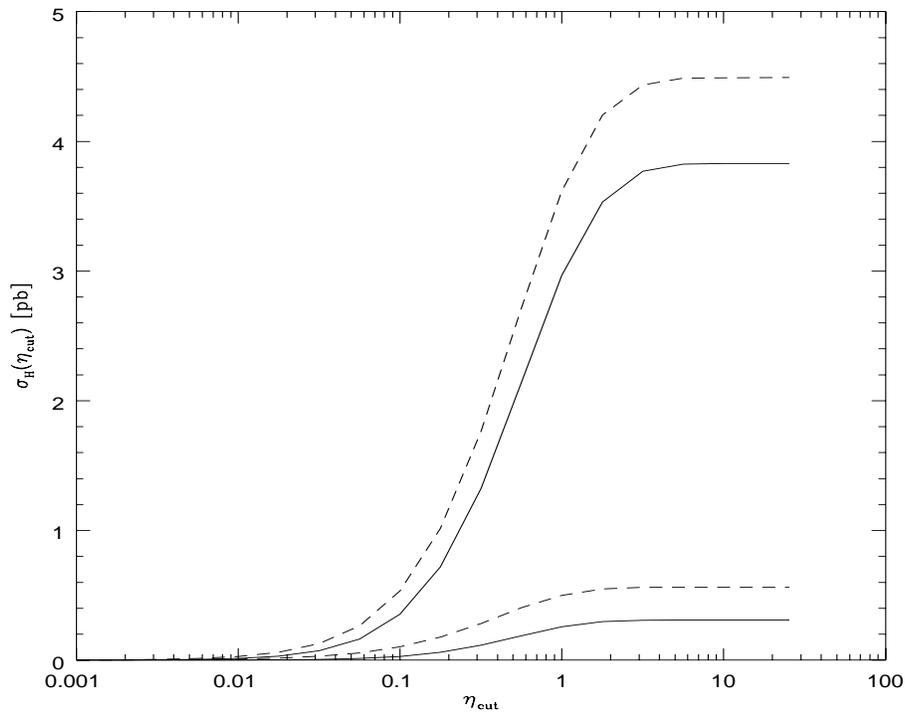,height=4.05in,width=5.05in,clip=}}
\caption[Cross sections for top quark production at the Fermilab Tevatron 
versus $\eta_{\rm cut}$ for the $q \bar q$ and $gg$ channels]
{Cross sections for top quark production at the Fermilab Tevatron 
versus $\eta_{\rm cut}$ with $m_t=175$ GeV$/c^2$ for the $q\bar q$ channel
in the DIS scheme and the $gg$ channel in the $\overline{\rm MS}$ scheme.
Plotted are the Born term ($q \bar q$, upper solid line;
$gg$, lower solid line)
and the $O(\alpha_s^3)$ cross section ($q\bar q$, upper dashed line;  
$gg$, lower dashed line).}
\label{fig. 2.9}
\end{figure}
This can easily be seen in fig. 2.9 where the Born term and the
$O(\alpha_s^3)$ cross section are plotted as functions of 
$\eta_{\rm cut}$ for the 
$q \bar q$ and $gg$ channels, where
$\eta=(s -4m^2)/4m^2$ is the variable into which we have incorporated
the cut in our programs for the cross sections. 
As we increase $\eta_{\rm cut}$ the
cross sections increase. The cross sections rise sharply for values of
$\eta_{\rm cut}$ between 0.1 and 1 and they reach a plateau at higher values 
of $\eta_{\rm cut}$ indicating that the threshold region is very important
and that the region where $s>> 4m^2$ only makes a small contribution 
to the cross sections.  This is the reason why we stressed in section 2.1
that our region of interest for comparison of the various approximations
at the partonic level was $0.1 <\eta <1$. 
Note that in the last figure as well as throughout the 
rest of this section we are assuming that the top quark mass is
$m_t=175$ GeV$/c^2$.

\begin{figure}
\centerline{\psfig{file=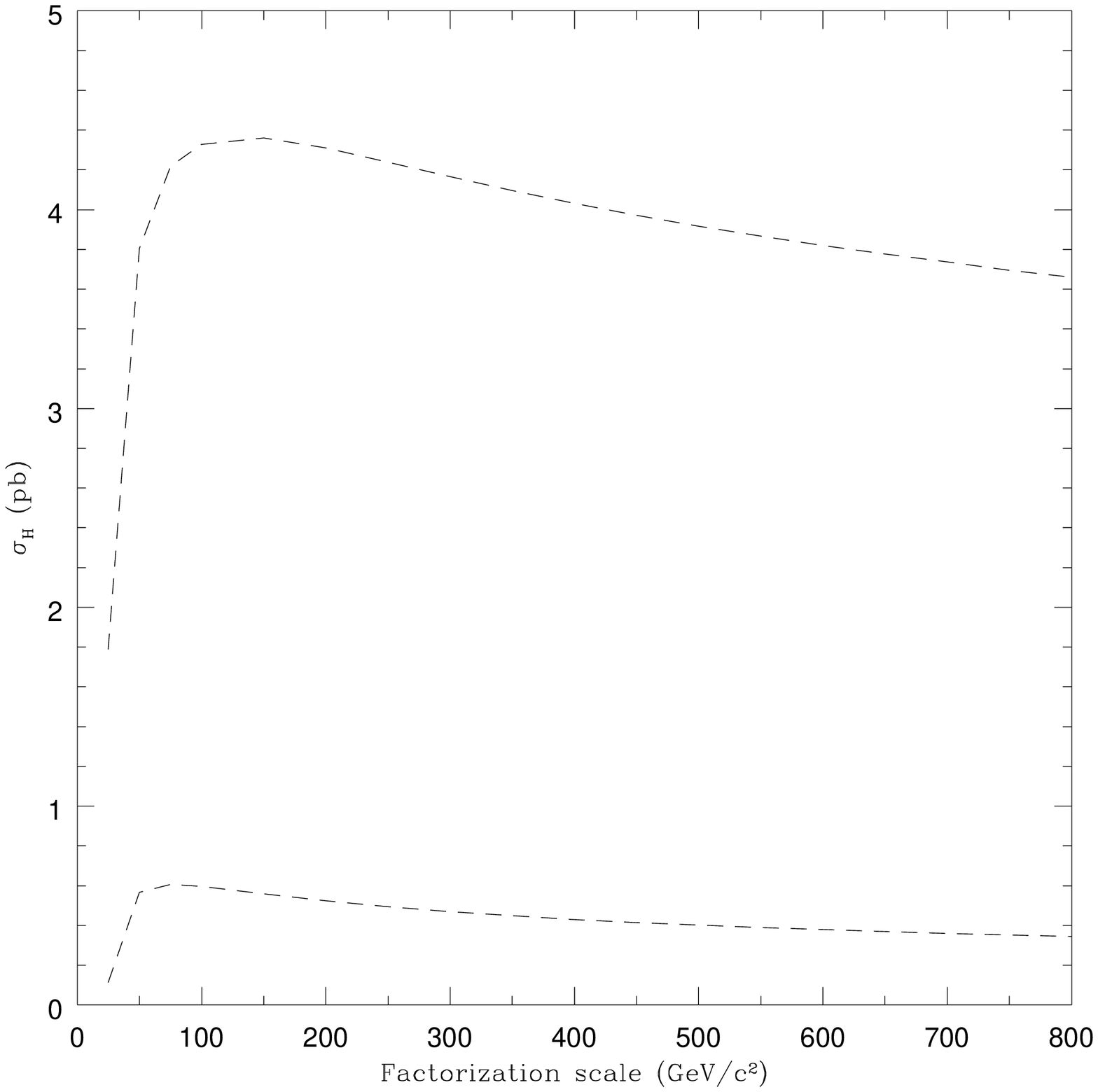,height=4.05in,width=5.05in,clip=}}
\caption[The scale dependence of the cross section for top quark production
at the Fermilab Tevatron]
{The scale dependence of the cross section for top quark production
at the Fermilab Tevatron with $m_t=175$ GeV$/c^2$. 
Plotted are the $O(\alpha_s^3)$ cross section
for the $q \bar q$ channel (upper dashed line)
and for the $gg$ channel (lower dashed line).}
\label{fig. 2.10}
\end{figure}
Next, we discuss the scale dependence of our NLO results. In fig. 2.10 
we show the 
$O(\alpha_s^3)$ cross section as a function of the 
factorization scale for the $q \bar q$ and $gg$ channels.
As the scale decreases, the Born cross section increases without
bound but the exact first order correction decreases faster so that   
the NLO cross section peaks at a scale close to half the mass of the
top quark and then decreases for smaller values of the scale. For both the
$q \bar q$ and $gg$ channels the NLO cross section is relatively flat. 
Thus the variation in
the NLO cross section for scales between $m_t/2$ and $2m_t$ is small.
\begin{figure}
\centerline{\psfig{file=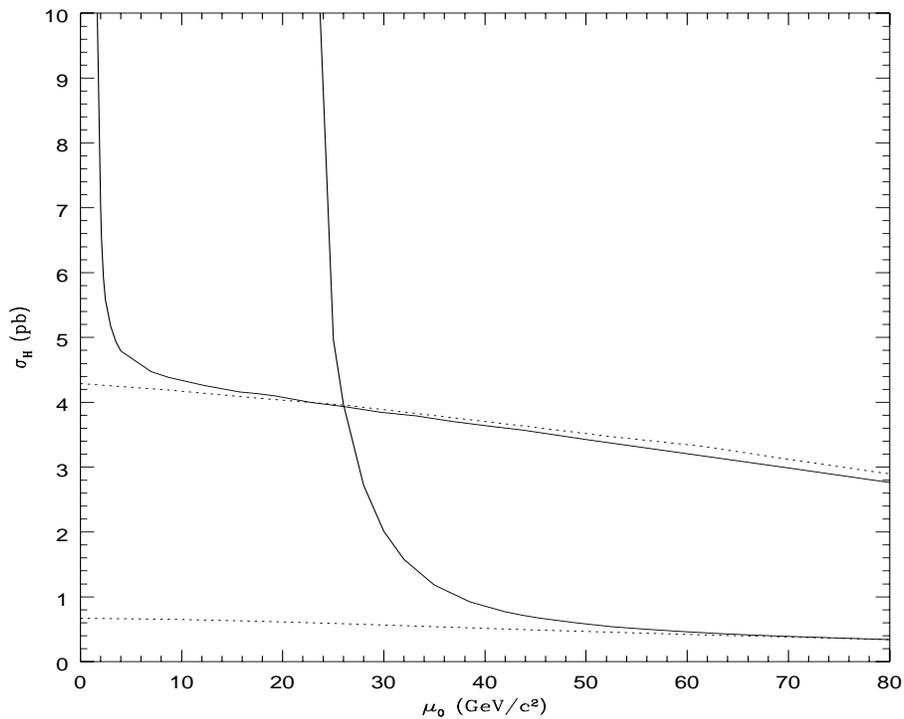,height=4.05in,width=5.05in,clip=}}
\caption[The $\mu_0$ dependence of the resummed cross section for 
top quark production at the Fermilab Tevatron]
{The $\mu_0$ dependence of the resummed cross section for 
top quark production at the Fermilab Tevatron with $m_t=175$ GeV$/c^2$
for the $q \bar q$ channel in the DIS scheme and the $gg$ channel in the
$\overline{\rm MS}$ scheme.
Plotted are $\sigma_{q\bar q}^{\rm res}$ (upper solid line at high $\mu_0$),
$\sigma_{gg}^{\rm res}$ (lower solid line at high $\mu_0$), 
and the sums 
$\sigma_{q \bar q}^{(0)}+\sigma_{q \bar q}^{(1)}\mid _{\rm app}
+\sigma_{q \bar q}^{(2)}\mid _{\rm app}$
(upper dotted line) and
$\sigma_{gg}^{(0)}+\sigma_{gg}^{(1)}\mid _{\rm app}
+\sigma_{gg}^{(2)}\mid _{\rm app}$
(lower dotted line).}
\label{fig. 2.11}
\end{figure}
In fig. 2.11 we examine the $\mu_0$ dependence of the 
resummed cross section for the $q \bar q$ and $gg$ channels.
We also show, for comparison, the $\mu_0$ dependence of
$\sigma^{(0)}+\sigma^{(1)}\mid _{\rm app}+\sigma^{(2)}\mid _{\rm app}$ 
where we have imposed the 
same cut on the phase space of $s_4$ ($s_4>s_0$)
as for the resummed cross section. Here $\sigma^{(1)}\mid _{\rm app}$
and $\sigma^{(2)}\mid _{\rm app}$ denote the approximate first and second
order corrections, respectively, where only soft gluon contributions
are taken into account.
The effect of the resummation shows in the difference between the two curves
for each channel.
At small $\mu_0$, $\sigma^{\rm res}$ diverges, signalling the 
presence of the infrared renormalon.
There is a region for each channel where the higher-order terms 
are numerically important.
At high values of $\mu_0$ the two lines for each channel 
are practically the same.
For the $q \bar q$ channel in the DIS scheme the resummation is successful
in the sense that there is a relatively large region of $\mu_0$ where 
resummation is well behaved before we encounter the divergence. For the 
$gg$ channel, however, the situation is not as good.
From these curves we choose what we think are reasonable values for $\mu_0$.
We choose $\mu_0=8.75$ GeV/$c^2$ ($0.05 \: m_t$)
and $17.5$ GeV/$c^2$ ($0.1 \: m_t$) for the $q \bar q$ channel
and $\mu_0=35$ GeV/$c^2$ ($0.2 \: m_t$)
and $43.75$ GeV/$c^2$ ($0.25 \: m_t$) for the $gg$ channel, which are the 
choices made in \cite{lsn22} corresponding to upper and lower values for
the cross section, respectively. 
Note that $\mu_0$ need not be the same in the $q\bar{q}$ and $gg$ reactions
because the convergence properties of the QCD perturbation series could be
different in these channels and moreover depend on the factorization scheme.

Since we know the exact $O(\alpha_s^3)$ result, we can make an even 
better estimate by calculating
the perturbation theory improved cross sections
defined by
\begin {equation}
\sigma_H^{\rm imp}=\sigma_H^{\rm res}
+\sigma_H^{(1)}\mid _{\rm exact}
-\sigma_H^{(1)}\mid _{\rm app}\,,
\end{equation}
to exploit the fact that $\sigma_H^{(1)}\mid _{\rm exact}$ 
is known and $\sigma_H^{(1)}\mid _{\rm app}$ 
is included in 
$\sigma_H^{\rm res}$.

The value of the NLO cross section for the production
of a top quark with a mass of 175 GeV/$c^2$ at the Fermilab Tevatron with 
$\sqrt{S}=1.8$ TeV is 4.8 pb. The upper and lower values 
of the resummed cross section are 5.6 pb and 4.9~pb, respectively.
The upper and lower values of the improved cross section are 5.8 pb 
and 5.1 pb, respectively.  
Finally, in fig. 2.12 we show the dependence of the top quark production
cross section at the Fermilab Tevatron on the top quark mass. 
Several theoretical curves [8-10] are
compared with recent experimental results from the D0 Collaboration.
\begin{figure}
\centerline{\psfig{file=figz.ps,height=4.05in,width=5.05in,clip=}}
\caption[The  dependence on the top mass of the cross section for 
top quark production at the Fermilab Tevatron]
{The  dependence on the top mass of the cross section for 
top quark production at the Fermilab Tevatron.}
\label{fig. 2.12}
\end{figure}

\mysection{Results for bottom quark production at fixed-target
$pp$ experiments and HERA-B}
In this section we examine the production of $b$-quarks 
in a situation where the presence of large logarithms is 
of importance, namely in a fixed-target experiment to be performed
in the HERA ring at DESY. This actual experiment has the name HERA-B 
\cite{herab12,herab22}
and involves colliding the circulating proton beam against a stationary
copper wire in the beam pipe. The nominal beam energy of the protons
is 820 GeV, so that the square root of the c.m. energy
is $\sqrt{S} = 39.2 $ GeV. Taking the $b$-quark mass as
$m_b = 4.75\: {\rm GeV}/c^2$ then the ratio of $m_b/\sqrt{S} \approx 1/8$. 
If we choose the renormalization scale in the running coupling constant
as $m_b$ then $\alpha_s(m_b^2) \approx 0.2$ so 
$\alpha_s(m_b^2) \ln(\sqrt{S}/m_b) \approx  0.4$. This number is
small enough that we expect a reasonably convergent perturbation series.

In the presentation of our results for the exact, approximate,
and resummed hadronic cross sections 
we use again the same set of MRSD$\_ ' \:$  
parton distributions as for top quark production.
In this case the number of active light flavors is $n_f=4$.

First, we discuss the NLO contributions to bottom quark production at HERA-B.
Except when explicitly stated otherwise we will take the factorization
scale $\mu=m_b$ where $m_b$ is the $b$-quark mass.
\begin{figure}
\centerline{\psfig{file=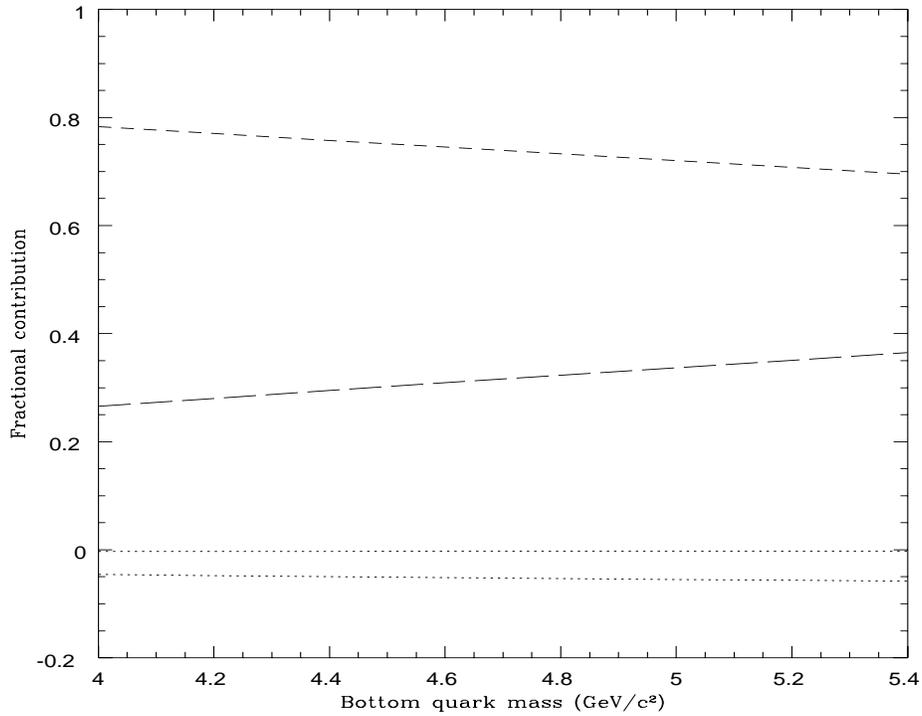,height=4.05in,width=5.05in,clip=}}
\caption[Fractional contributions of the $gg$, $q \bar q$, $qg$, and 
$\bar q g$ channels to the total $O(\alpha_s^3)$
$b$-quark production cross section at HERA-B versus $b$-quark mass]
{Fractional contributions of the $gg$ ($\overline{\rm MS}$ scheme,
short-dashed line), $q \bar q$ (DIS scheme, long-dashed line), 
$qg$ (DIS scheme, lower dotted line), and $\bar q g$ (DIS scheme,
upper dotted line) channels to the total $O(\alpha_s^3)$
$b$-quark production cross section at HERA-B as a function of $b$-quark mass .}
\label{fig. 2.13}
\end{figure}
In fig. 2.13 we show the relative contributions of the $q \bar q$ channel in
the DIS scheme and the $gg$ channel in the $\overline{\rm MS}$ scheme 
as a function
of the bottom quark mass. We see  that the $gg$ contribution is the dominant
one, lying between 70\% and 80\% of the total NLO cross section 
for the range of bottom mass values given.
The $q \bar q$ contribution is smaller and makes up most of the remaining 
cross section.
The relative contributions of the $g q$ and the $g \bar q$ channels in the 
DIS scheme are negative and very small and they are also shown in the plot.
The situation here is the reverse of what we saw in the previous section
for top quark production
at the Fermilab Tevatron where $q \bar q$ is the dominant channel 
with $gg$ making up
the remainder of the cross section, and $g q$ and  $g \bar q$ making an even
smaller relative contribution than is the case for bottom quark production
at HERA-B. The reason for this difference between top quark and bottom quark 
production is that the Tevatron is a $p \bar p$ collider while HERA-B is a
fixed-target $pp$ experiment. Thus, the parton densities involved 
are different and since
sea quark densities are much smaller than valence quark densities, the
$q \bar q$ contribution to the hadronic cross section diminishes for a 
fixed-target $pp$ experiment 
relative to a $p \bar p$ collider for the same partonic cross section.
\begin{figure}
\centerline{\psfig{file=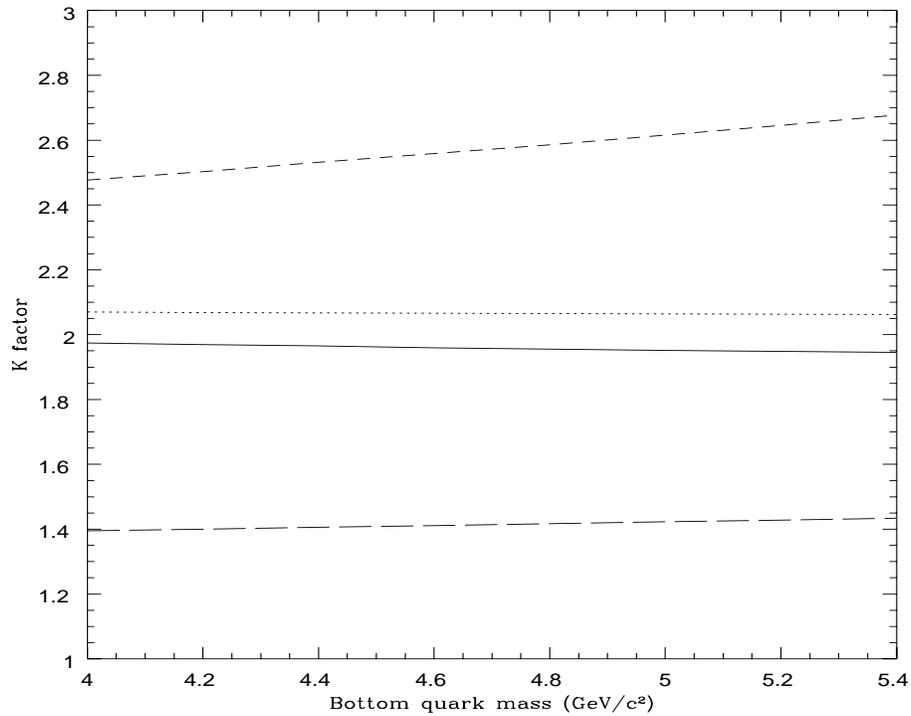,height=4.05in,width=5.05in,clip=}}
\caption[The K factors as a function of $b$-quark mass 
for $b$-quark production at HERA-B]
{The K factors as a function of $b$-quark mass 
for $b$-quark production at HERA-B for the 
$gg$ channel ($\overline{\rm MS}$ scheme,
short-dashed line), the $q \bar q$ channel (DIS scheme, long-dashed line),
the sum of the $gg$ and $q \bar q$ channels (dotted line), and the sum of all
channels (solid line).}
\label{fig. 2.14}
\end{figure}

In fig. 2.14 we show the $K$ factors for the $q \bar q$ and $gg$ channels and 
for their sum as a function of the bottom quark mass.
We notice that the $K$ factor is quite large for the $gg$ channel,
which means that higher order effects are more important for this channel
than for $q \bar q$. Since $gg$ is the more important channel numerically,
the $K$ factor for the sum of the two channels is also quite large.
We also show the $K$ factor for the total which is slightly lower since
we are also taking into account the
negative contributions of the $qg$ and $\bar q g$ channels.   

As in the case of top quark production these large corrections come 
predominantly from the threshold region.
\begin{figure}
\centerline{\psfig{file=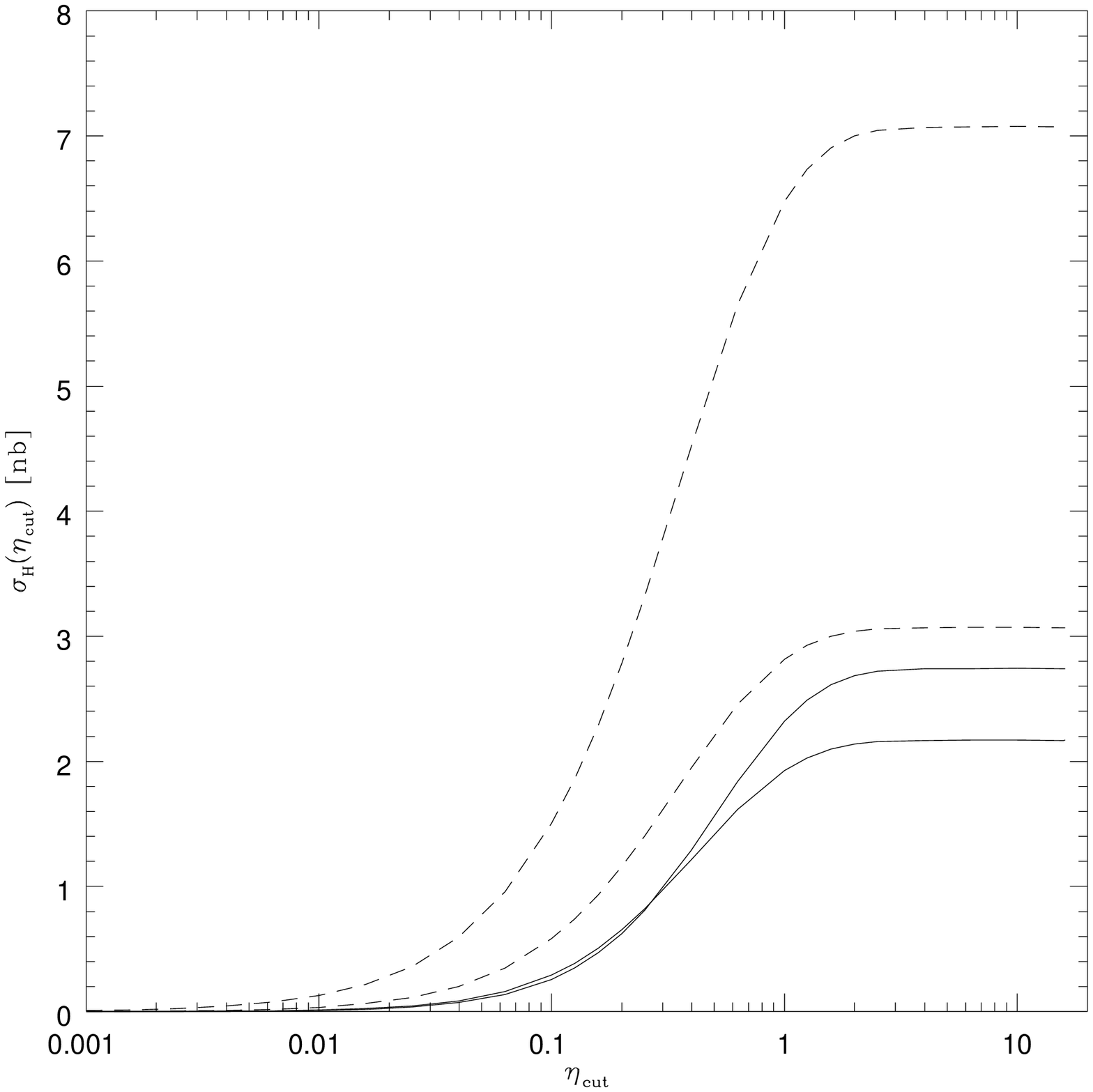,height=4.05in,width=5.05in,clip=}}
\caption[Cross sections for $b$-quark production at HERA-B 
versus $\eta_{\rm cut}$ for the $q \bar q$ and $gg$ channels]
{Cross sections for $b$-quark production at HERA-B 
versus $\eta_{\rm cut}$ with $m_b=4.75$ GeV$/c^2$ for the $q\bar q$ channel
in the DIS scheme and the $gg$ channel in the $\overline{\rm MS}$ scheme.
Plotted are the Born term ($gg$, upper solid line at high $\eta_{\rm cut}$;
$q\bar q$, lower solid line at high $\eta_{\rm cut}$)
and the $O(\alpha_s^3)$ cross section ($gg$, upper dashed line;  
$q\bar q$, lower dashed line).}
\label{fig. 2.15}
\end{figure}
This can easily be seen in fig. 2.15 where the Born term and the
$O(\alpha_s^3)$ cross section are plotted as functions of 
$\eta_{\rm cut}$ for the 
$q \bar q$ and $gg$ channels.
As we increase $\eta_{\rm cut}$ the
cross sections increase. 
As for the top, the cross sections rise sharply for values of
$\eta_{\rm cut}$ between 0.1 and 1 and they reach a plateau at higher values 
of $\eta_{\rm cut}$ indicating that the threshold region is very important
and that the region where $s>> 4m^2$ only makes a small contribution 
to the cross sections.  
Note that in the last figure as well as throughout the 
rest of this section we are assuming that the bottom quark mass is
$m_b=4.75$ GeV$/c^2$.
\begin{figure}
\centerline{\psfig{file=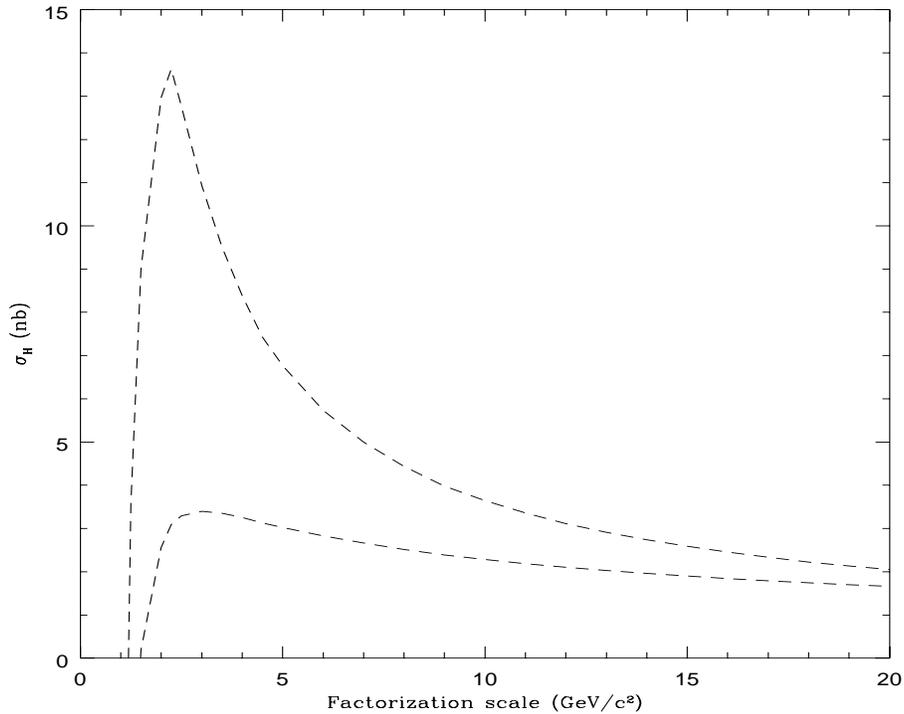,height=4.05in,width=5.05in,clip=}}
\caption[The scale dependence of the cross section for $b$-quark production
at HERA-B]
{The scale dependence of the cross section for $b$-quark production
at HERA-B with $m_b=4.75$ GeV$/c^2$. 
Plotted are the $O(\alpha_s^3)$ cross section
for the $q \bar q$ channel (lower dashed line)
and for the $gg$ channel (upper dashed line).
}
\label{fig. 2.16}
\end{figure}

Next, we discuss the scale dependence of our NLO results. In fig. 2.16  
we show the 
$O(\alpha_s^3)$ cross section as a function of the 
factorization scale for the $q \bar q$ and $gg$ channels.
We see that
the NLO cross section peaks at a scale close to half the mass of the
bottom quark and then decreases for smaller values of the scale. For the
$q \bar q$ channel the NLO cross section is relatively flat. The situation
is much worse for the $gg$ channel, however, since the peak is very sharp and
the scale dependence is much greater. Since the $gg$ channel dominates, this
large scale dependence is also reflected in the total cross section. 
Thus the variation in
the NLO cross section for scales between $m_b/2$ and $2m_b$ is large. For
comparison we note that, as we saw in the previous section, 
the scale dependence for top quark 
production at the Fermilab Tevatron for $m_t=175$ GeV $/c^2$ 
is much smaller.
\begin{figure}
\centerline{\psfig{file=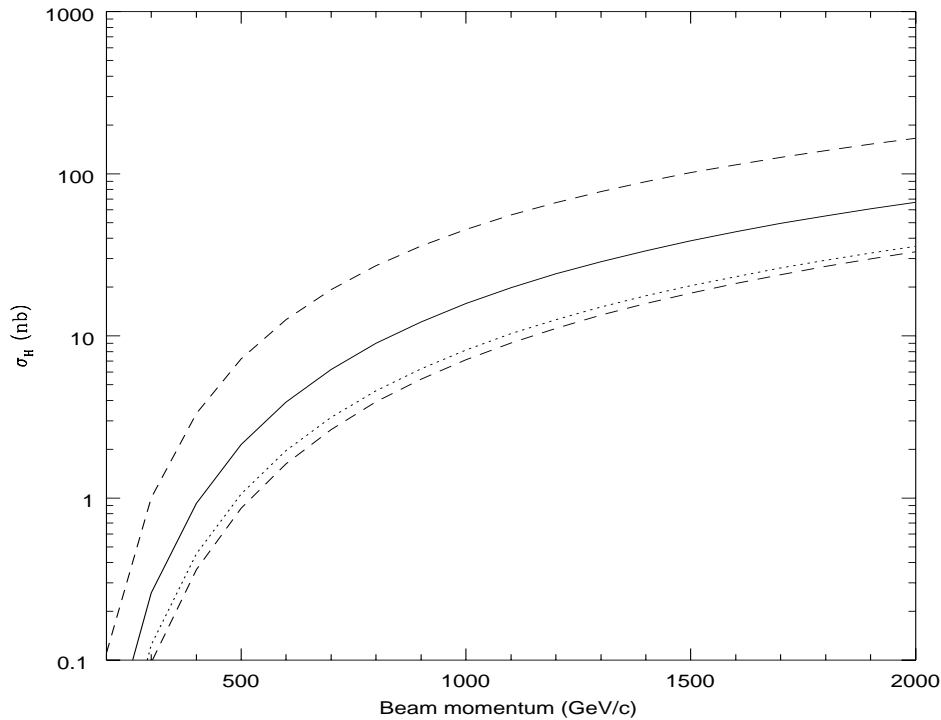,height=4.05in,width=5.05in,clip=}}
\caption[The total Born and $O(\alpha_s^3)$ 
$b$-quark production cross sections at fixed-target $pp$ experiments 
versus beam momentum]
{The total Born (dotted line) and $O(\alpha_s^3)$ 
($\mu=m_b$ solid line, $\mu=m_b/2$ upper dashed line, and $\mu=2m_b$ 
lower dashed line)
$b$-quark production cross sections at fixed-target $pp$ experiments 
versus beam momentum for $m_b=4.75$ GeV$/c^2$.}
\label{fig. 2.17}
\end{figure}

In fig. 2.17 we plot the Born contribution for $\mu=m_b$ and the NLO 
cross section for $\mu=m_b/2$, $m_b$, and $2m_b$,
as a function of the beam momentum for $b$-quark production at fixed-target
$pp$ experiments.
The big width of the 
band reflects the large scale dependence that we discussed above.
We see that the NLO  cross section is almost twice as big
as the Born term for the whole range of beam momenta that we 
are showing, and in particular for 820 GeV$/c$ which is the value
of the beam momentum at HERA-B.
\begin{figure}
\centerline{\psfig{file=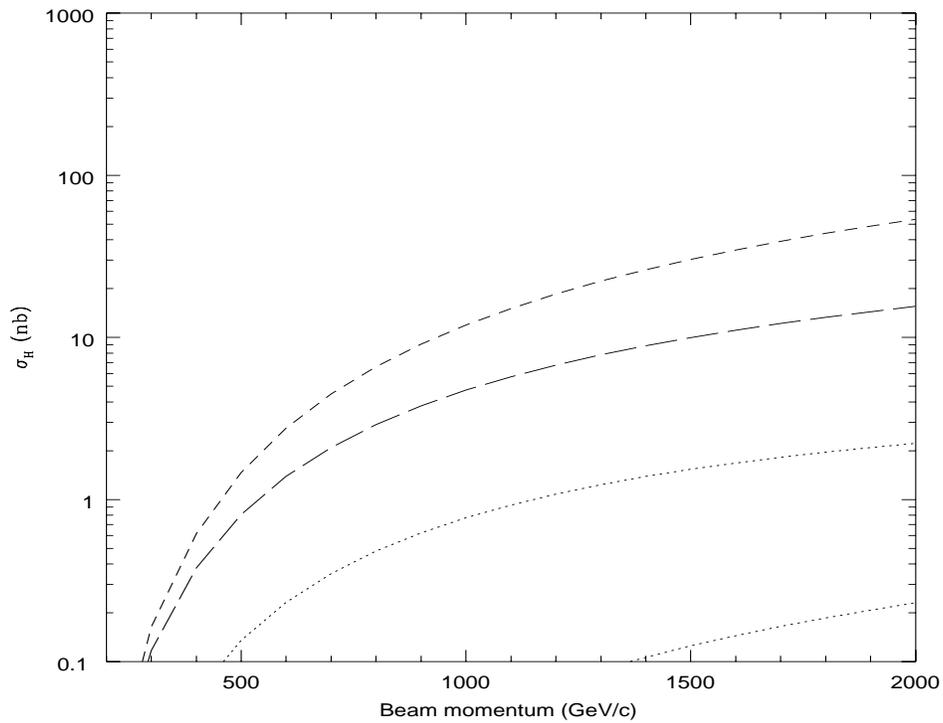,height=4.05in,width=5.05in,clip=}}
\caption[Contributions of individual channels to the total 
$O(\alpha_s^3)$ $b$-quark 
production cross section at fixed-target $pp$ experiments
versus beam momentum]
{Contributions of individual channels to the total 
$O(\alpha_s^3)$ $b$-quark 
production cross section at fixed-target $pp$ experiments
versus beam momentum for $m_b=4.75$ GeV$/c^2$. 
Plotted are the contributions of the $gg$ 
($\overline {\rm MS}$ scheme, short-dashed
line) and $q \bar q$ (DIS scheme, long-dashed line) channels, 
and the absolute value of 
the contributions of the $qg$ (DIS scheme, upper dotted line) and $\bar q g$ 
(DIS scheme, lower dotted line) channels.}
\label{fig. 2.18}
\end{figure}
We also give the NLO results for the individual channels in fig. 2.18
for $\mu=m_b$.  
\begin{figure}
\centerline{\psfig{file=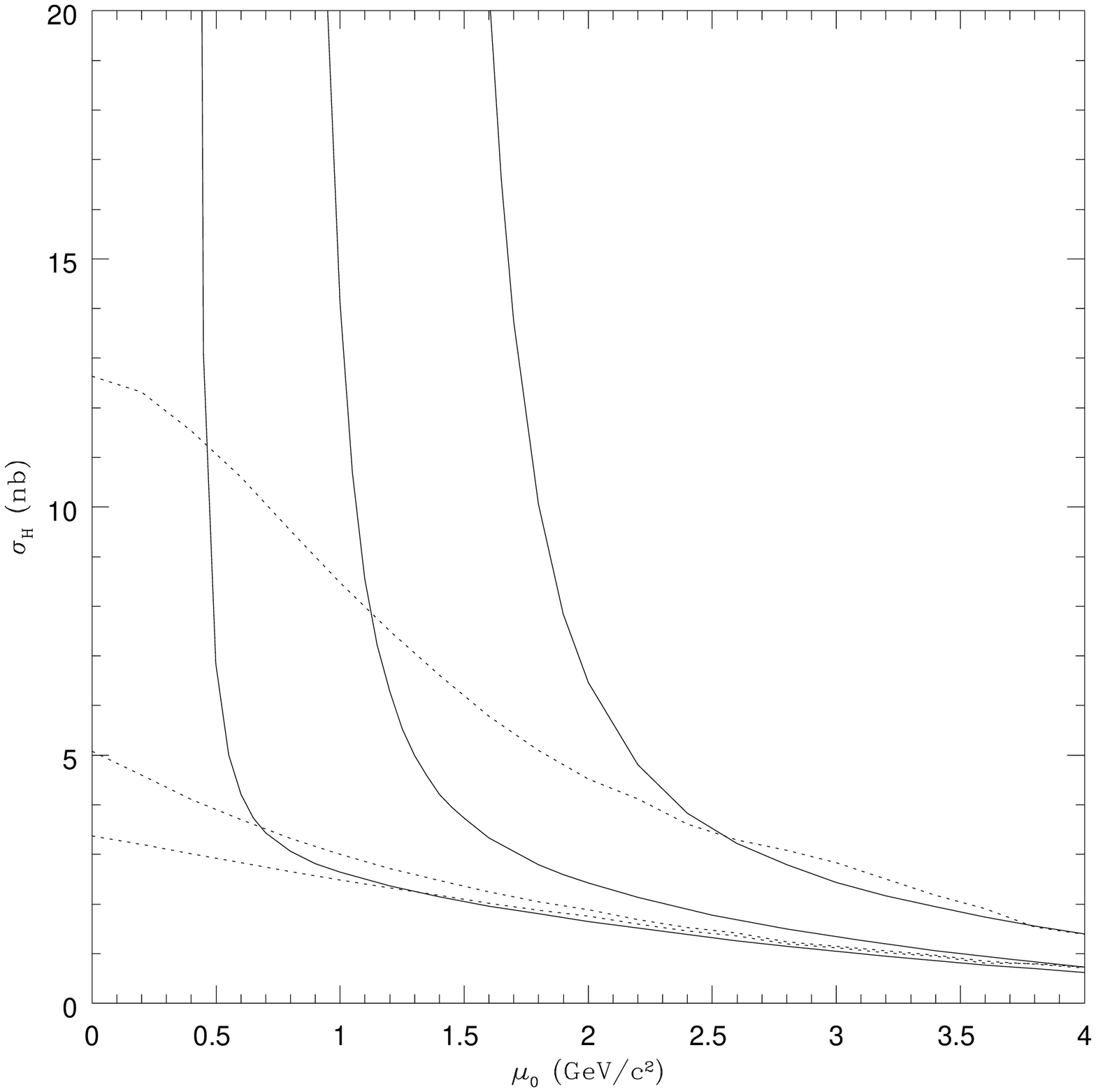,height=4.05in,width=5.05in,clip=}}
\caption[The $\mu_0$ dependence of the resummed cross section for 
$b$-quark production at HERA-B]
{The $\mu_0$ dependence of the resummed cross section for 
$b$-quark production at HERA-B with $m_b=4.75$ GeV$/c^2$
for the $q \bar q$ channel in the DIS scheme and in the $\overline{\rm MS}$
scheme, and for the $gg$ channel in the $\overline{\rm MS}$ scheme.
Plotted are $\sigma_{q\bar q}^{\rm res}$ (lower solid line DIS scheme,
middle solid line $\overline{\rm MS}$ scheme) and $\sigma_{gg}^{\rm res}$ 
(upper solid line).  Also we plot the sum 
$\sigma^{(0)}+\sigma^{(1)}\mid _{\rm app}+\sigma^{(2)}\mid _{\rm app}$
(lower dotted line for $q \bar q$ in the DIS scheme,
middle dotted line for $q \bar q$ in the $\overline{\rm MS}$ scheme,
and upper dotted line for $gg$).}
\label{fig. 2.19}
\end{figure}
  
In fig. 2.19 we examine the $\mu_0$ dependence of the 
resummed cross section.
We also show, for comparison, the $\mu_0$ dependence of
$\sigma^{(0)}+\sigma^{(1)}\mid _{\rm app}+\sigma^{(2)}\mid _{\rm app}$ 
with the same cut $s_4>s_0$.
The effect of the resummation shows in the difference between the two curves
for each channel.
At small $\mu_0$, $\sigma^{\rm res}$ diverges signalling the 
divergence of the running coupling constant. 
There is a region for each channel where the higher-order terms are numerically
important.
At large values of $\mu_0$ the two lines for each channel 
are practically the same.
For the $q \bar q$ channel in the DIS scheme the resummation is successful
in the sense that there is a relatively large region of $\mu_0$ where 
resummation is well behaved before we encounter the divergence. 
This region is reduced for the $q\bar q$ channel in the $\overline{\rm MS}$
scheme.
For the $gg$ channel, however, this region is even smaller.

From these curves we choose what we think are reasonable values for $\mu_0$.
We choose $\mu_0=0.6$ GeV/$c^2$ for the $q \bar q$ channel in the DIS scheme
($\mu_0/m_b \approx 13 \%$) and $\mu_0=1.7$ GeV/$c^2$ for the $gg$ channel
($\mu_0/m_b \approx  36 \%$). The values we chose 
for the $q \bar q$ and $gg$ channels are such that 
the resummed cross sections are slightly larger than the
sums $\sigma^{(0)}+\sigma^{(1)}\mid _{\rm app}+\sigma^{(2)}\mid _{\rm app}$.
Note that these $\mu_0$ values are not exactly
the same as those used in section 2.2,
where $\mu_0/m_t = 10 \%$ and $\mu_0/m_t = 25 \%$
for the $q \bar q$ and $gg$ channels respectively, 
which predicted the mass dependence of the top quark cross section. 
The $\mu_0$ parameters there were again
chosen via the criterion that the higher order terms in the perturbation theory
should not be too large. 

It is illuminating to compare fig. 2.19 with the 
corresponding plot for the top quark case fig. 2.11.
There one can infer that if we take the slightly larger $\mu_0$ values
given above there is very little change in the top quark cross section.
The reason is that in this case the $gg$ channel makes only a small 
contribution and the $\mu_0$ dependence in the $q\bar q$ channel reflects the 
small variation of the running coupling constant at a scale 
$\mu = 175$ GeV/$c^2$.
As the running coupling constant varies more rapidly at a scale
$\mu = 4.75$ GeV/$c^2\,$, the $\mu_0$ parameters should be taken from 
measurements at the lower scale and then used in the prediction
of the top quark cross section. This emphasizes the importance of 
the proposed measurement at HERA-B. It is clear from fig. 2.19 that we
cannot choose $\mu_0/m_b = 25 \%$ for the $gg$ channel for 
bottom quark production
but we can choose $\mu_0/m_t = 36 \%$ for the $gg$ channel for
top quark production, with very little change in the value of the
top quark cross section.
Both sets of parameters yield cross sections which are within the error bars 
of the recent CDF \cite{ObsCDF2} and D0 \cite{ObsD02}
experimental results for the top quark cross section.
Therefore our cut off parameters do have experimental justification.
We would also like to point out that an application of the principal value 
resummation method has been recently completed by Berger and Contopanagos
\cite{BC2} leading to 
essentially the same mass dependence of the top cross section as reported
in \cite{lsn22}, which again justifies our choice for $\mu_0$.
Finally note that we could just as easily have chosen to work in the
$\overline{\rm MS}$ scheme for both channels by changing
$\mu_0$ in the $q \bar q$ channel
to $\mu_0 \approx 1.3 $ GeV/$c^2$. The reason the DIS scheme is preferred
is simply because it has a larger radius of convergence. 

Using the values of $\mu_0$ that we chose from the previous graphs, 
we proceed to plot the resummed cross section  for $b$-quark
production at fixed-target $pp$ experiments versus beam momentum. 
\begin{figure}
\centerline{\psfig{file=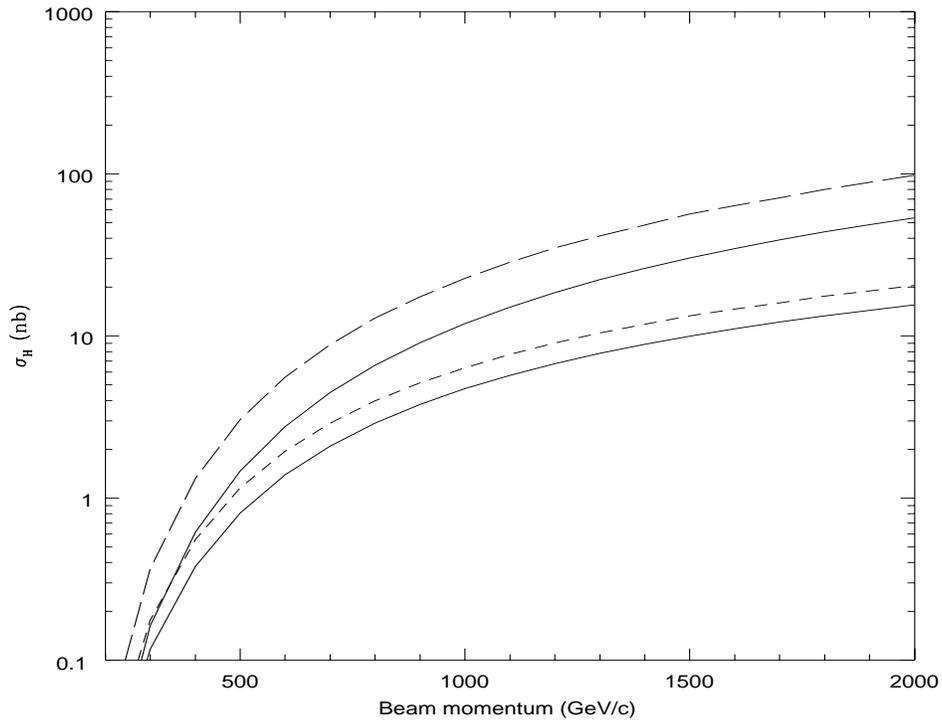,height=4.05in,width=5.05in,clip=}}
\caption[Resummed and NLO cross sections versus beam momentum
for $b$-quark production at fixed-target $pp$ experiments]
{Resummed and NLO cross sections versus beam momentum
for $b$-quark production at fixed-target $pp$ experiments for $m_b=4.75$
GeV$/c^2$. Plotted are the resummed cross sections for the
$q \bar q$ channel in the DIS scheme for $\mu_0=0.6$ GeV/$c^2$ 
(short-dashed line),
and for the $gg$ channel in the $\overline{\rm MS}$ scheme for
$\mu_0=1.7$ GeV/$c^2$ (long-dashed line); and the $O(\alpha_s^3)$
cross sections for the $gg$ channel in the $\overline{\rm MS}$ scheme
and the $q \bar q$ channel in the DIS scheme (upper and lower
solid lines, respectively).}
\label{fig. 2.20}
\end{figure}
We present the results in fig. 2.20 for the $q \bar q$ and $gg$ channels. 
For comparison the exact NLO results are shown as well. The 
resummed cross sections were calculated with the cut $s_4>s_0$ while no
such cut was imposed on the NLO result.
\begin{figure}
\centerline{\psfig{file=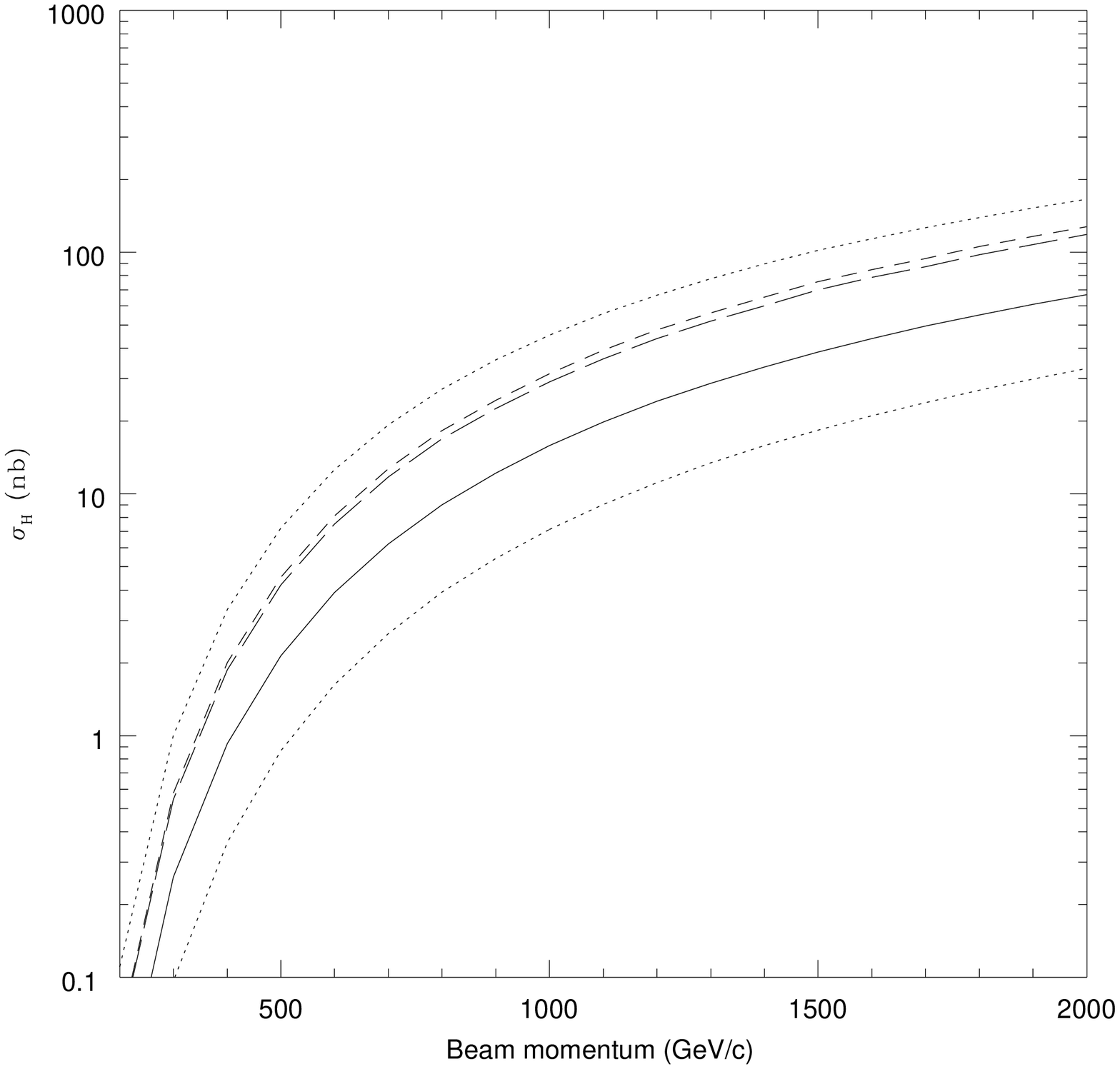,height=4.05in,width=5.05in,clip=}}
\caption[Resummed, improved, and NLO cross sections versus beam momentum
for $b$-quark production at fixed-target $pp$ experiments]
{Resummed, improved, and NLO cross sections versus beam momentum
for $b$-quark production at fixed-target $pp$ experiments for $m_b=4.75$
GeV$/c^2$. Plotted are the total resummed cross section  
(long-dashed line), the total improved cross section (short-dashed line),
and the total $O(\alpha_s^3)$ cross section ($\mu=m_b$ solid line, 
$\mu=m_b/2$ upper dotted line, $\mu=2m_b$ lower dotted line).}
\label{fig. 2.21}
\end{figure}
Then, in fig. 2.21 we plot the resummed cross section and the 
improved total cross section (2.2.6) (where we have
also taken into account the small negative contributions of the $qg$ and
$\bar q g$ channels) versus beam momentum and, for comparison, 
the total exact NLO cross section
for the three choices $\mu=m_b/2$, $m_b$, and $2m_b$. 
The total NLO cross section for $b$-quark production at HERA-B 
(beam energy 820 GeV) is 28.8 nb for
$\mu=m_b/2$; 9.6 nb for $\mu=m_b$; and 4.2 nb for $\mu=2m_b$. 
The resummed cross section is 18 nb. The improved total cross section for 
$b$-quark production at HERA-B is 19.4 nb.

\mysection{Conclusions}
We have presented NLO and resummed results for the cross sections 
for top quark production at the Fermilab Tevatron 
and for bottom quark production at HERA-B 
and at fixed-target $pp$ experiments in general. 
In both cases we found
that the threshold region gives the main contribution
to the NLO cross sections. Approximations for the soft gluon contributions
in that region have been compared with the exact results.  
The resummation of the leading S+V logarithms 
produces an enhancement of the NLO results.

For top quark production at the Fermilab Tevatron we saw that the
$q {\bar q}$ channel is dominant. We found that the resummation for this
channel is relatively well behaved and that the scale dependence 
of the NLO cross section is relatively flat. 
The total NLO cross section for top quark production
with $m_t=175$ GeV/$c^2$ at the Fermilab Tevatron with 
$\sqrt{S}=1.8$ TeV is 4.8 pb.
The upper and lower values of the improved cross section 
for top quark production
at the Fermilab Tevatron are 5.8 pb and 5.1 pb, respectively.

For bottom quark production at fixed-target $pp$ experiments
it was shown that the $gg$ channel is dominant. 
The leading S+V approximation is not very good in the $gg$ channel in the
$\overline {\rm MS}$ scheme in the kinematic region
that is important for bottom quark production at HERA-B. The addition
of subleading S+V terms and Coulomb terms improves the approximation
considerably. The resummation is not as successful as in the $q {\bar q}$
channel and the scale dependence of the NLO cross section is much bigger.
The total NLO cross section for $b$-quark production at HERA-B 
(beam energy 820 GeV) with $m_b=4.75$ GeV$/c^2$ is 9.6 nb (for $\mu=m_b$). 
The improved total cross section for 
$b$-quark production at HERA-B is 19.4 nb.   
%

\chapter{Top and bottom quark inclusive differential distributions}

The inclusive transverse momentum and rapidity distributions 
for top quark production at the Fermilab
Tevatron and bottom quark production at HERA-B are presented  both in  
order $\alpha_s^3$ in QCD and using the
resummation of the leading soft gluon corrections
in all orders of QCD perturbation theory.
The resummed results are uniformly larger than the
$O(\alpha_s^3)$ results for both distributions.

\mysection{Introduction}
At the Tevatron, the top quark is mainly
produced through $t\bar{t}$ pair production from the light mass
quarks and gluons in the colliding proton and
antiproton. Both the top quark and the top antiquark
then decay to $(W,b)$ pairs, and 
each $W$ boson can decay either hadronically or leptonically.
The $b$-quark becomes an on-mass-shell $B$-hadron which 
subsequently decays into leptons and (charmed) hadrons.
A large effort is being made to reconstruct the top quark mass from the 
measured particles in the decay, which is complicated by the fact
that the neutrinos are never detected. Also there are 
additional jets so it is not clear which ones to choose to 
recombine \cite{{OS3},{OTS3}}. The best channel for this mass 
reconstruction is where both $W$ bosons decay 
leptonically, one to a $(e,\nu_e)$ pair, the other to a 
$(\mu,\nu_{\mu})$ pair (a dilepton event) because the backgrounds
in this channel are small. 
When only  a single lepton is detected then it is necessary
to identify the $b$ quark in the decay to remove large backgrounds
from the production of $W +$ jets \cite{Berends3}. 
In all cases the reconstruction of the particles in the final state involves
both the details of the production of the top quark-antiquark
pair as well as the knowledge of their fragmentation
and decay products. 

In the analysis of the decay distributions one needs
knowledge of the inclusive differential distributions
of the heavy quarks in transverse momentum $p_T$ and rapidity $Y$.
These distributions are known in NLO \cite{{nde23},{bnmss3}}.
We present an analysis of the resummation effects on the
inclusive transverse momentum distribution of the top quark
assuming that it has a mass of $175$ GeV$/c^2$.
Also we discuss here how 
the resummation effects modify the rapidity distribution
of the top quark. Since there have been suggestions of using
the mass and angular distributions in top quark production to
look for physics beyond the standard model \cite{Lane3} it is very
important to know the normal QCD predictions for these quantities.

We also present a corresponding analysis for the $p_T$ and $Y$
distributions for bottom quark production at the HERA-B experiment. 
\mysection{Soft gluon approximation to the
inclusive distributions}
The partonic processes under discussion will be denoted by
\begin{equation}
i(k_1) + j(k_2) \rightarrow Q(p_1) + \bar Q(p_2) + g(k_3),
\end{equation}
where $i,j = g, q, \bar q$. The kinematical variables
\begin{equation}
s = ( k_1+k_2)^2 \quad , \quad t_1 = (k_2-p_2)^2 - m^2 \quad , \quad
u_1 = (k_1- p_2)^2 - m^2\quad ,
\end{equation}
are introduced in the calculation of the corrections
to the single particle inclusive differential distributions
of the heavy (anti)quark. We do not distinguish in the text
between the heavy quark and heavy antiquark since the distributions
are essentially identical in our calculations.
Here $s$ is the square of the 
parton-parton c.m. energy and the heavy quark transverse
momentum is given by $p_T= (t_1u_1/s-m^2)^{1/2}$.
The rapidity variable is defined by
$ \exp (2y)  = u_1/t_1$. 
Also as before we define $s_4=s+t_1+u_1$.

The transverse momentum $p_T$ of the heavy quark is related 
to our previous variables by
\begin{equation}
t_1 = - \frac{1}{2}\Big\{ s - s_4 -[(s - s_4)^2 - 4s m_T^2]^{1/2}\Big\}\,,
\end{equation}
\begin{equation}
u_1 = - \frac{1}{2}\Big\{ s - s_4 +[(s - s_4)^2 - 4s m_T^2]^{1/2}\Big\}\,,
\end{equation}
with $m_T^2 = m^2 + p_T^2.$ The double differential cross section is
therefore
\begin{equation}
s^2 \frac{d^2\sigma_{ij}(s, t_1, u_1)}{dt_1 \: du_1} = 
s[(s - s_4)^2 - 4s m_T^2]^{1/2}
\frac{d^2\sigma_{ij}(s, s_4, p_T^2)}{dp_T^2ds_4} \, ,
\end{equation}
with the boundaries
\begin{equation}
0 < p_T^2 < \frac{s}{4} - m^2\quad , \quad 0 < s_4 < s-2m_T s^{1/2}\,.
\end{equation}
The $O(\alpha_s^k)$ contribution to the inclusive transverse momentum
distribution $d\sigma_{ij}/dp_T^2$ is given by
\begin{eqnarray}
\frac{d\sigma_{ij}^{(k)}(s,p_T^2)}{dp_T^2} &=&
\frac{2}{s} \alpha_s^k(\mu^2) \sum_{l=0}^{2k-1} a_l(\mu^2) 
\int_0^{s-2m_Ts^{1/2}}\, ds_4
 \nonumber \\&&
\times\Big\{ \frac{1}{s_4} \ln^l\Big(\frac{s_4}{m^2}\Big) \theta(s_4 - \Delta)
+ \frac{1}{l+1} \ln^{l+1} \Big(\frac{\Delta}{m^2}\Big) \delta(s_4)\Big\}
 \nonumber \\ &&
 \times \frac{1}{[(s-s_4)^2 - 4sm_T^2]^{1/2}} \sigma^B_{ij}(s,s_4,p_T^2)
 \, ,
\end{eqnarray}
where we have inserted an extra factor of 2 so 
that $\int dp_T^2 \: d\sigma/dp_T^2 = \sigma_{\rm tot}$. 
After some algebra
we can rewrite this result as
\begin{eqnarray}
\frac{d\sigma_{ij}^{(k)}(s,p_T^2)}{dp_T^2} &=&
\alpha_s^k(\mu^2) \sum_{l=0}^{2k-1} a_l(\mu^2) 
\Big[\int_0^{s-2m_Ts^{1/2}}\, ds_4 \frac{1}{s_4} \ln^l\frac{s_4}{m^2}
 \nonumber \\ &&
\times\Big\{ \frac{d\bar\sigma_{ij}^{(0)}(s,s_4,p_T^2)}{dp_T^2} 
 - \frac{d\bar\sigma_{ij}^{(0)}(s,0,p_T^2)}{dp_T^2} \Big\} 
  \nonumber \\ &&  
+ \frac{1}{l+1} \ln^{l+1}\Big(\frac{s - 2m_T s^{1/2}}{m^2}\Big) 
\frac{d\bar\sigma_{ij}^{(0)}(s,0,p_T^2)}{dp_T^2} \Big]\,,
\end{eqnarray}
with the definition
\begin{equation}
\frac{d\bar\sigma_{ij}^{(0)}(s,s_4,p_T^2)}{dp_T^2} =
\frac{2}{s[(s-s_4)^2 - 4s m_T^2]^{1/2}} \sigma_{ij}^B(s,s_4,p_T^2)\,,
\end{equation}
where $d\bar\sigma^{(0)}_{ij}(s,0,p_T^2)/dp_T^2 \equiv 
d\sigma^{(0)}_{ij}(s,p_T^2)/dp_T^2 $ again represents the 
Born differential $p_T$ distribution. For the $q\bar q$ and $gg$
subprocesses we have the explicit results
\begin{eqnarray}
\frac{d\bar\sigma_{q\bar q}^{(0)}(s,s_4,p_T^2)}{dp_T^2} &=&
2\pi \alpha_s^2(\mu^2) K_{q\bar q} N C_F \frac{1}{s}
\frac{1}{[(s-s_4)^2 -4sm_T^2]^{1/2}}
\nonumber \\ &&
\times \Big[\frac{(s-s_4)^2 - 2sp_T^2}{s^2}\Big] \,,
\end{eqnarray}
and 
\begin{eqnarray}
\frac{d\bar\sigma_{gg}^{(0)}(s,s_4,p_T^2)}{dp_T^2} &=&
4\pi \alpha_s^2(\mu^2) K_{gg} N C_F \frac{1}{s}
\frac{1}{[(s-s_4)^2 -4sm_T^2]^{1/2}}
\nonumber \\ &&
\times\Big[C_F - C_A \frac{m_T^2}{s}\Big]
\nonumber \\ &&
\times \Big[\frac{(s-s_4)^2 - 2sm_T^2}{sm_T^2} +
\frac{4m^2}{m_T^2} \Big( 1 -\frac{m^2}{m_T^2}\Big)\Big] \,.
\end{eqnarray}
Since the above formulas are symmetric 
under the interchange $t_1 \leftrightarrow u_1$
the heavy quark and heavy antiquark inclusive $p_T$ distributions are
identical. Note that (3.2.8) is basically the integral of a plus 
distribution together with a surface term.

The corresponding formula to (3.2.8) for the rapidity $y$
of the heavy quark is obtained by using
\begin{equation}
t_1 = - \frac{(s-s_4)}{2}(1 - \tanh y)\,,
\end{equation}
\begin{equation}
u_1 = - \frac{(s-s_4)}{2}(1 + \tanh y)\,.
\end{equation}
The double differential cross section is
therefore
\begin{equation}
s^2 \frac{d^2\sigma_{ij}(s, t_1, u_1)}{dt_1 \: du_1} 
=
2 s^2 \frac{\cosh^2y}{s-s_4}
\frac{d^2\sigma_{ij}(s, s_4, y)}{dy \: ds_4}\,,
\end{equation}
with the boundaries
\begin{equation}
- \frac{1}{2}\ln \Big( \frac{1+\beta}{1-\beta}\Big) < y < 
  \frac{1}{2}\ln \Big( \frac{1+\beta}{1-\beta}\Big)   
\quad , \quad 0 < s_4 < s-2ms^{1/2}\cosh y\,,
\end{equation}
where $\beta^2 = 1 -4m^2/s$.
The $O(\alpha_s^k)$ contribution to the inclusive rapidity 
distribution $d\sigma_{ij}/dy$ is given by
\begin{eqnarray}
\frac{d\sigma_{ij}^{(k)}(s,y)}{dy} &=&
\alpha_s^k(\mu^2) \sum_{l=0}^{2k-1} a_l(\mu^2) 
\int_0^{s-2ms^{1/2}\cosh y}\, ds_4
 \nonumber \\&&
\times \Big\{ \frac{1}{s_4} \ln^l\Big(\frac{s_4}{m^2}\Big) \theta(s_4 - \Delta)
+ \frac{1}{l+1} \ln^{l+1}\Big( \frac{\Delta}{m^2}\Big) \delta(s_4)\Big\}
 \nonumber \\ &&
 \times \Big(\frac{s-s_4}{2s^2\cosh^2 y}\Big) \sigma^B_{ij}(s,s_4,y)
 \,.
\end{eqnarray}
After some algebra we can rewrite this result as
\begin{eqnarray}
\frac{d\sigma_{ij}^{(k)}(s,y)}{dy} &=&
\alpha_s^k(\mu^2) \sum_{l=0}^{2k-1} a_l(\mu^2) 
\Big[\int_0^{s-2ms^{1/2}\cosh y}\, ds_4 \frac{1}{s_4} 
\ln^l\Big(\frac{s_4}{m^2}\Big)
 \nonumber \\ &&
\times\Big\{ \frac{d\bar\sigma_{ij}^{(0)}(s,s_4,y)}{dy} 
 - \frac{d\bar\sigma_{ij}^{(0)}(s,0,y)}{dy} \Big\} 
  \nonumber \\ &&  
+ \frac{1}{l+1} \ln^{l+1}\Big(\frac{s - 2ms^{1/2}\cosh y}{m^2}\Big) 
\frac{d\bar\sigma_{ij}^{(0)}(s,0,y)}{dy} \Big]\,,
\end{eqnarray}
with the definition
\begin{equation}
\frac{d\bar\sigma_{ij}^{(0)}(s,s_4,y)}{dy} =
\frac{s-s_4}{2s^2 \cosh^2 y} \, \sigma_{ij}^B(s,s_4,y)\,,
\end{equation}
where $d\bar\sigma^{(0)}_{ij}(s,0,y)/dy \equiv 
d\sigma^{(0)}_{ij}(s,y)/dy $ again represents the 
Born differential $y$ distribution. For the $q\bar q$ and $gg$
subprocesses we have the explicit formulas
\begin{eqnarray}
\frac{d\bar \sigma_{q\bar q}^{(0)}(s,s_4,y)}{dy} &=&
\pi\alpha_s^2(\mu^2) K_{q\bar q} N C_F 
\frac{s-s_4}{2s^2\cosh^2 y}
\nonumber \\ &&
\times \Big[\frac{(s-s_4)^2}{2s^2\cosh^2 y}\Big(
\cosh^2 y + \sinh^2 y\Big) + \frac{2m^2}{s}\Big] \,,
\end{eqnarray}
and 
\begin{eqnarray}
\frac{d\bar \sigma_{gg}^{(0)}(s,s_4,y)}{dy} &=&
4\pi \alpha_s^2(\mu^2) K_{gg} N C_F 
\frac{s-s_4}{2s^2 \cosh^2 y }
\nonumber \\ &&
\times\Big[C_F - C_A \frac{(s-s_4)^2}{4s^2 \cosh^2 y }\Big]
\times \Big[\cosh^2 y + \sinh^2 y  
\nonumber \\ &&
+ \frac{8m^2s \cosh^2 y}{(s-s_4)^2} \Big( 1 -\frac{4m^2s\cosh^2 y}{(s-s_4)^2}
\Big)\Big] \,.
\end{eqnarray}
Since the above formulas are symmetric under 
the interchange $t_1 \leftrightarrow u_1$
the heavy quark and heavy antiquark inclusive $y$ distributions are
identical. Also (3.2.17) is again of the form of a plus distribution
together with a surface term. Finally, we note that the terms in
(3.2.8) and (3.2.17) are all finite.

\mysection{Resummation procedure in parton-parton collisions}
We now consider the resummation of the 
order $\alpha_s^k$ contributions to the $p_T$ distribution. We have
\begin{eqnarray}
\frac{d\sigma_{ij}(s,p_T^2)}{dp_T^2} &=&
\sum_{k=0}^{\infty} 
\frac{d\sigma_{ij}^{(k)}(s,p_T^2)}{dp_T^2}  
 \nonumber \\ &&
\! \! \! \! \! \! =\int_{s_0}^{s-2m_Ts^{1/2}}\, ds_4 
\frac{df(s_4/m^2, m^2/\mu^2)}{ds_4}
 \nonumber \\ &&
\times\Big\{ \frac{d\bar\sigma_{ij}^{(0)}(s,s_4,p_T^2)}{dp_T^2}
 - \frac{d\bar\sigma_{ij}^{(0)}(s,0,p_T^2)}{dp_T^2} \Big\}
  \nonumber \\ &&
+ f\Big( \frac{s-2m_Ts^{1/2}}{m^2}, \frac{m^2}{\mu^2}\Big) 
\frac{d\sigma_{ij}^{(0)}(s,p_T^2)}{dp_T^2} \,. 
\end{eqnarray}
Note that, as in chapter 2, 
we now have cut off the lower limit of the $s_4$ integration
at $s_4=s_0$ because $\bar\alpha_s$ diverges as $s_4 \rightarrow 0$.
This parameter $s_0$ must satisfy the conditions
$0 < s_0 <s-2m_Ts^{1/2}$ and $s_0/m^2 <<1$.
The derivative of 
$f(s_4/m^2,m^2/\mu^2)$ is obtained from
(2.1.24). It is equal to
\begin{eqnarray}
\frac{df(s_4/m^2,m^2/\mu^2)}{ds_4} &=&
\frac{1}{s_4} \Big\{ 2A \frac{C_{ij}}{\pi} \bar\alpha_s(\frac{s_4}{m^2},
m^2) \ln\frac{s_4}{m^2}+ \eta \Big\} 
\nonumber \\ &&
\times \exp\Big\{
 A \frac{C_{ij}}{\pi} \bar\alpha_s(\frac{s_4}{m^2},m^2)\ln^2\frac{s_4}{m^2}
\Big\} \frac{[s_4/m^2]^\eta}{\Gamma(1+\eta)}
\nonumber \\ &&
\times \exp(-\eta\gamma_E)\, , 
\end{eqnarray}
where we have neglected terms which are higher order in
$\bar\alpha_s$.

The analogous formula for the rapidity distribution is
\begin{eqnarray}
\frac{d\sigma_{ij}(s,y)}{dy} &=&
\sum_{k=0}^{\infty} 
\frac{d\sigma_{ij}^{(k)}(s,y)}{dy} 
 \nonumber \\ &&
\! \!\! \! \! \! =\int_{s_0}^{s-2ms^{1/2}\cosh y}\, ds_4 
\frac{df(s_4/m^2, m^2/\mu^2)}{ds_4}
 \nonumber \\ &&
\times\Big\{ \frac{d\bar\sigma_{ij}^{(0)}(s,s_4,y)}{dy}
 - \frac{d\bar\sigma_{ij}^{(0)}(s,0,y)}{dy} \Big\}
  \nonumber \\ &&
+ f\Big( \frac{s-2ms^{1/2}\cosh y}{m^2}, \frac{m^2}{\mu^2}\Big) 
\frac{d\sigma_{ij}^{(0)}(s,y)}{dy} \,, 
\end{eqnarray}
with the conditions $0<s_0<s-2ms^{1/2}\cosh y$ and $s_0/m^2 << 1$.

\mysection{Top quark differential distributions}

Since the $p_T$ distribution in hadron-hadron collisions is
not altered by the Lorentz transformation
along the collision axis from the parton-parton c.m. frame,
we can write an analogous formula to (2.2.1) for the heavy-quark
inclusive differential distribution in $p_T^2$
\begin{equation}
\frac{d\sigma^{(k)}_H(S,m^2,p_T^2)}{dp_T^2} = \sum_{ij}\int_{4m_T^2/S}^1
\,d\tau \,\Phi_{ij}(\tau,\mu^2)\, \frac{d\sigma_{ij}^{(k)}
(\tau S,m^2,p_T^2,\mu^2)}{dp_T^2}\,.
\end{equation}
In the case of the all-order 
resummed expression the lower boundary in (3.4.1) 
has to be modified according to the condition
$s_0 < s - 2m_Ts^{1/2}$ (see section 3.3).
The all-order resummed differential distribution in $p_T^2$ is
\begin{equation}
\frac{d\sigma^{\rm res}_H(S,m^2,p_T^2)}{dp_T^2} = \sum_{ij}\int_{\tau_0}^1
\,d\tau \, \Phi_{ij}(\tau,\mu^2) \frac{d\sigma_{ij}(\tau S,m^2,p_T^2,\mu^2)}
{dp_T^2}\,,
\end{equation}
with $d\sigma_{ij}/dp_T^2$ given in (3.3.1)
and 
\begin{equation}
\tau_0 = \frac{(m_T+(m_T^2+s_0)^{1/2})^2}{S}\,.
\end{equation}
The corresponding formula to (3.4.1) for the heavy quark inclusive
differential distribution in $Y$ is
\begin{equation}
\frac{d\sigma^{(k)}_H(S,m^2,Y)}{dY} = \sum_{ij}\int_{4m^2\cosh^2 y/S}^1
\,d\tau \,\Phi_{ij}(\tau,\mu^2)\, \frac{d\sigma_{ij}^{(k)}
(\tau S,m^2,y,\mu^2)}{dy}\,.
\end{equation}
Order by order in perturbation theory the heavy quark rapidity plots
in the parton-parton c.m. frame show peaks away from $y=0$ (see fig. 7
in \cite{bnmss3}). However, upon folding with the partonic densities the
heavy quark rapidities in the hadron-hadron c.m. frame peak near $Y=0$.
Therefore we will assume that the plots for the resummed rapidity distribution
show a similar feature. 
The all-order resummed differential distribution in $Y$ is therefore 
taken to be
\begin{equation}
\frac{d\sigma^{\rm res}_H(S,m^2,Y)}{dY} = \sum_{ij}\int_{\tau_0}^1
\,d\tau \,\Phi_{ij}(\tau,\mu^2)\, \frac{d\sigma_{ij}(\tau S,m^2,y,\mu^2)}
{dy}\,,
\end{equation}
with $d\sigma_{ij}/dy$ given in (3.3.3)
and 
\begin{equation}
\tau_0 = \frac{(m\cosh y +(m^2\cosh^2 y+s_0)^{1/2})^2}{S}\,.
\end{equation}
The hadronic heavy quark rapidity $Y$ is related to the 
partonic heavy quark rapidity $y$
by
\begin{equation}
Y=y+\frac{1}{2}\ln\frac{x_1}{x_2}\,.
\end{equation}

We now specialize to top quark production at the Fermilab Tevatron
where $\sqrt{S}=1.8$ TeV and choose the top quark mass to be $m_t=175$
GeV$/c^2$.
In the presentation of our results for the exact, approximate
and resummed hadronic cross sections 
we use the same (MRSD$\_ ' \:$) parametrization for the parton distributions
as we did in the previous chapter.

Since we know the exact $O(\alpha_s^3)$ result, we can make an even 
better estimate of the differential distributions by calculating
the perturbation theory improved $p_T$ and $Y$ distributions.
We define the improved $p_T$ distribution by
\begin {equation}
\frac{d\sigma_H^{\rm imp}}{dp_T}=\frac{d\sigma_H^{\rm res}}{dp_T}
+\frac{d\sigma_H^{(1)}}{dp_T}\mid _{\rm exact}
-\frac{d\sigma_H^{(1)}}{dp_T}\mid _{\rm app}\,,
\end{equation}
and the improved $Y$ distribution by
\begin {equation}
\frac{d\sigma_H^{\rm imp}}{dY}=\frac{d\sigma_H^{\rm res}}{dY}
+\frac{d\sigma_H^{(1)}}{dY}\mid _{\rm exact}
-\frac{d\sigma_H^{(1)}}{dY}\mid _{\rm app}\,,
\end{equation}
to exploit the fact that $d\sigma_H^{(1)}/dp_T\mid _{\rm exact}$ and  
$d\sigma_H^{(1)}/dY\mid _{\rm exact}
$ are known and $d\sigma_H^{(1)}/dp_T\mid _{\rm app}$ 
and $d\sigma_H^{(1)}/dY\mid _{\rm app}$ are included in 
$d\sigma_H^{\rm res}/dp_T$ and $d\sigma_H^{\rm res}/dY$ respectively.
We note that here $d\sigma^{(n)}$ denotes the $O(\alpha_s^{n+2})$
contribution to the differential cross section. Moreover , 
$d\sigma^{(n)}\mid _{\rm exact}$ denotes the exact 
calculated differential cross section,
and $d\sigma^{(n)}\mid _{\rm app}$ the approximate one where only the 
leading soft gluon corrections are taken into account. 

First we present the differential $p_T$ distributions at $\sqrt{S}=1.8$ TeV
for a top quark mass $m_t = 175$ GeV$/c^2$.
For these plots the mass factorization scale is not everywhere equal to $m_t$.
We chose $\mu=m_t$ in $s_0$, $f_k(s_4/m^2\,,m^2/\mu^2)$ and $\bar{\alpha}_s$, 
but $\mu=m_T$ in the MRSD$\_ ' \:$ parton distribution
functions and the running coupling constant $\alpha_s(\mu^2)$.
It should be noted, however, that it makes little difference if we choose
$\mu=m_t$ everywhere. 
The difference in the cross section is only a few percent 
so that the changes due to scale dependence are insignificant compared 
with the changes due to higher order resummation.
We begin with the results for the $q\bar{q}$ channel in the DIS scheme.
\begin{figure}
\centerline{\psfig{file=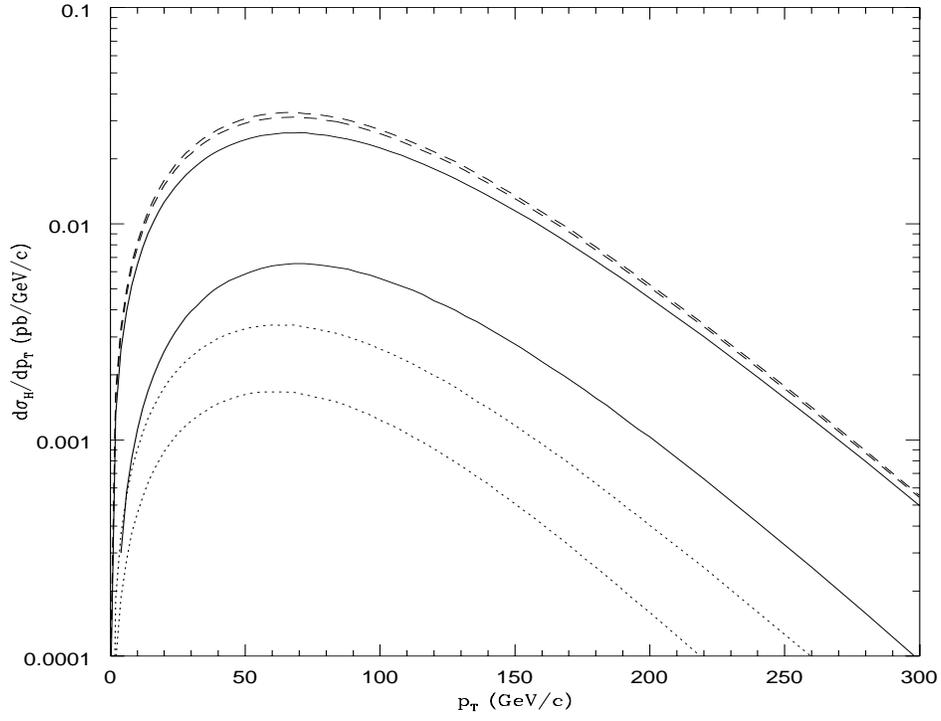,height=4.05in,width=5.05in,clip=}}
\caption[The top quark $p_T$ distributions for
the $q\bar{q}$ channel]
{The top quark $p_T$ distributions $d\sigma_H^{(k)}/dp_T$ for
the $q\bar{q}$ channel in the DIS scheme for a top quark mass 
$m_t=175$ GeV$/c^2$. Plotted are $d\sigma_H^{(0)}/dp_T$ (upper solid line),
$d\sigma_H^{(1)}/dp_T\mid _{\rm exact}$ (lower solid line),
$d\sigma_H^{(1)}/dp_T\mid _{\rm app}$ (upper dotted line),
$d\sigma_H^{(2)}/dp_T\mid _{\rm app}$ (lower dotted line),
and $d\sigma_H^{\rm res}/dp_T$ ($\mu_0=0.05\:m_t$ upper dashed line
and $\mu_0=0.1\:m_t$ lower dashed line).}
\label{fig. 3.1}
\end{figure}
In fig. 3.1 we show
the Born term $d\sigma_H^{(0)}/dp_T$, the first order exact result 
$d\sigma_H^{(1)}/dp_T\mid _{\rm exact}$, the first order approximation 
$d\sigma_H^{(1)}/dp_T\mid _{\rm app}$, the second 
order approximation $d\sigma_H^{(2)}/dp_T\mid _{\rm app}$,
and the resummed result $d\sigma_H^{\rm res}/dp_T$ for $\mu_0=0.05\:m_t$ 
and for $\mu_0=0.1\:m_t$.
These are the same values for $\mu_0$ that we used in chapter 2.
As we decrease $\mu_0$ the differential cross 
sections increase. 

We continue with the results for the $gg$ channel in 
the $\overline{\rm MS}$ scheme.
\begin{figure}
\centerline{\psfig{file=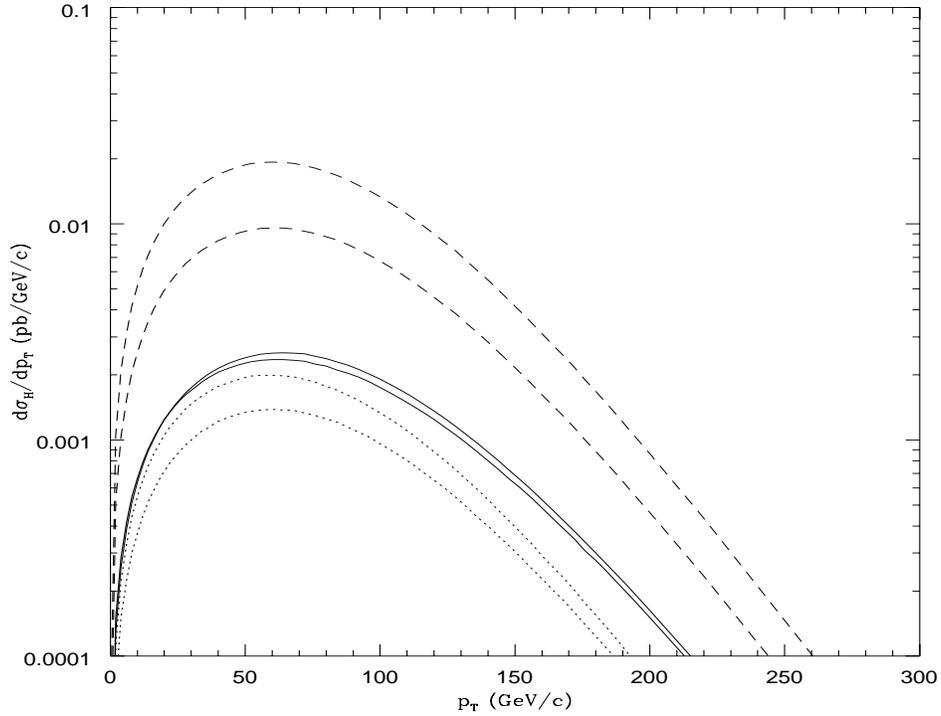,height=4.05in,width=5.05in,clip=}}
\caption[The top quark $p_T$ distributions for
the $gg$ channel]
{The top quark $p_T$ distributions $d\sigma_H^{(k)}/dp_T$ for
the $gg$ channel in the $\overline{\rm MS}$ scheme for a top quark mass 
$m_t=175$ GeV$/c^2$. Plotted are $d\sigma_H^{(0)}/dp_T$ (upper solid line
at large $p_T$),
$d\sigma_H^{(1)}/dp_T\mid _{\rm exact}$ (lower solid line at large $p_T$),
$d\sigma_H^{(1)}/dp_T\mid _{\rm app}$ (lower dotted line),
$d\sigma_H^{(2)}/dp_T\mid _{\rm app}$ (upper dotted line),
and $d\sigma_H^{\rm res}/dp_T$ ($\mu_0=0.2\:m_t$ upper dashed line
and $\mu_0=0.25\:m_t$ lower dashed line).}
\label{fig. 3.2}
\end{figure}
The corresponding plot is given in figure 3.2. 
In this case the values of $\mu_0$ have been chosen to be
$\mu_0=0.2\:m_t$ and $\mu_0=0.25\:m_t$, again as in chapter 2. 
The first and second order corrections in the $gg$ channel in the 
$\overline{\rm MS}$ scheme are larger than the respective ones in the 
$q\bar{q}$ channel in the DIS scheme. In fact, for the range of $p_T$ 
values shown the second-order approximate correction is larger than the
first-order approximation. Hence, the relative difference in magnitude
 between the improved $d\sigma_H^{\rm imp}/dp_T$ and 
the exact $O(\alpha_s^3)$ results 
is significantly larger than that for the $q\bar{q}$ channel 
in the DIS scheme. 

We finish our discussion of the differential $p_T$ distributions with the
results of adding the $q\bar{q}$ and $gg$ channels. The plot appears
in figure 3.3. We also show the total improved and $O(\alpha_s^3)$
distributions in fig. 3.4.
It is evident that resummation produces an 
enhancement of the exact $O(\alpha_s^3)$ result, with very little change 
in shape.
\begin{figure}
\centerline{\psfig{file=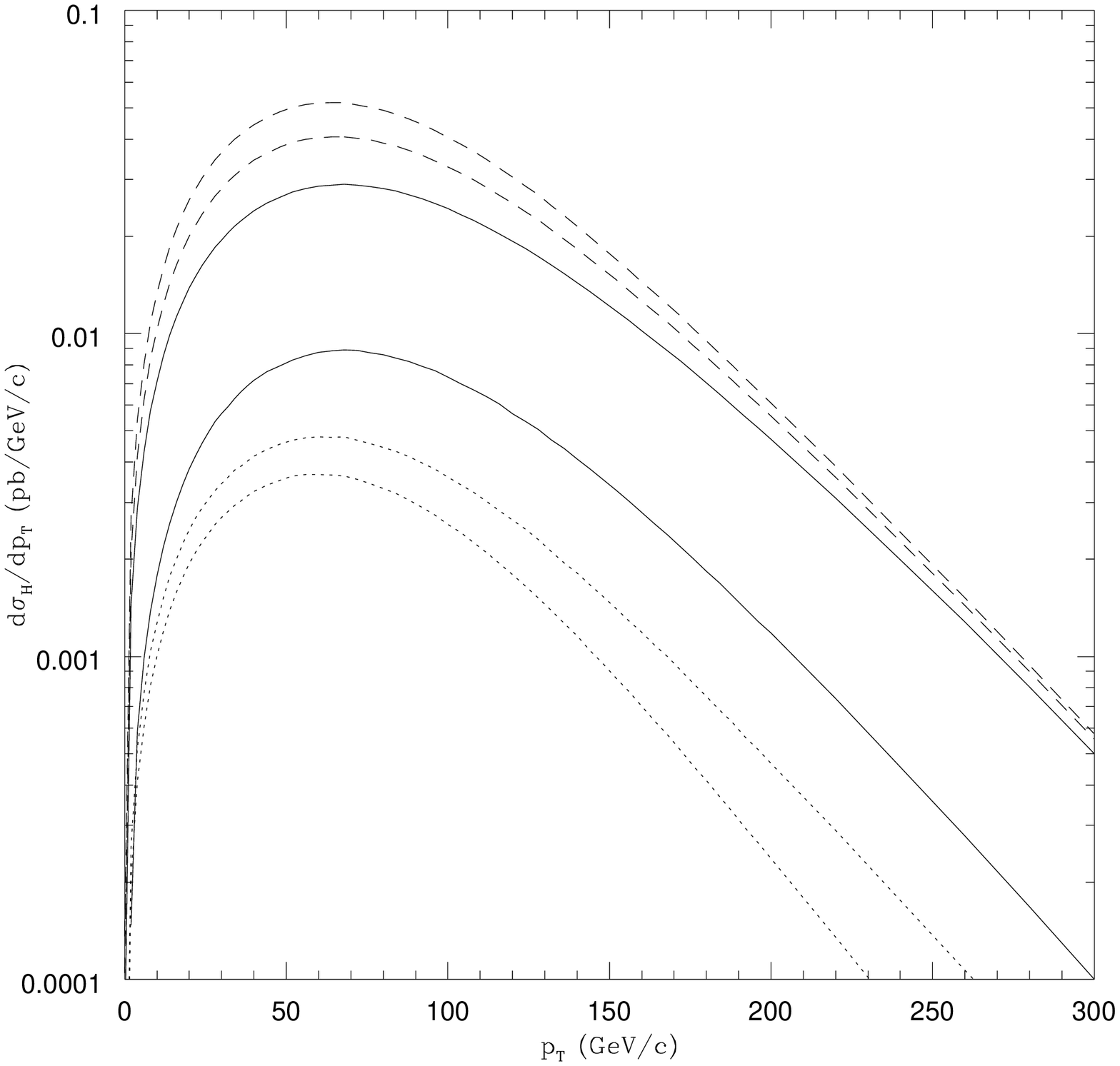,height=4.05in,width=5.05in,clip=}}
\caption[The top quark $p_T$ distributions $d\sigma_H^{(k)}/dp_T$ for
the sum of the $q\bar{q}$ and $gg$ channels]
{The top quark $p_T$ distributions $d\sigma_H^{(k)}/dp_T$ for
the sum of the $q\bar{q}$ and $gg$ channels for a top quark mass 
$m_t=175$ GeV$/c^2$. Plotted are $d\sigma_H^{(0)}/dp_T$ (upper solid line),
$d\sigma_H^{(1)}/dp_T\mid _{\rm exact}$ (lower solid line),
$d\sigma_H^{(1)}/dp_T\mid _{\rm app}$ (upper dotted line),
$d\sigma_H^{(2)}/dp_T\mid _{\rm app}$ (lower dotted line),
and $d\sigma_H^{\rm res}/dp_T$ (upper
and lower dashed lines).}
\label{fig. 3.3}
\end{figure}

\begin{figure}
\centerline{\psfig{file=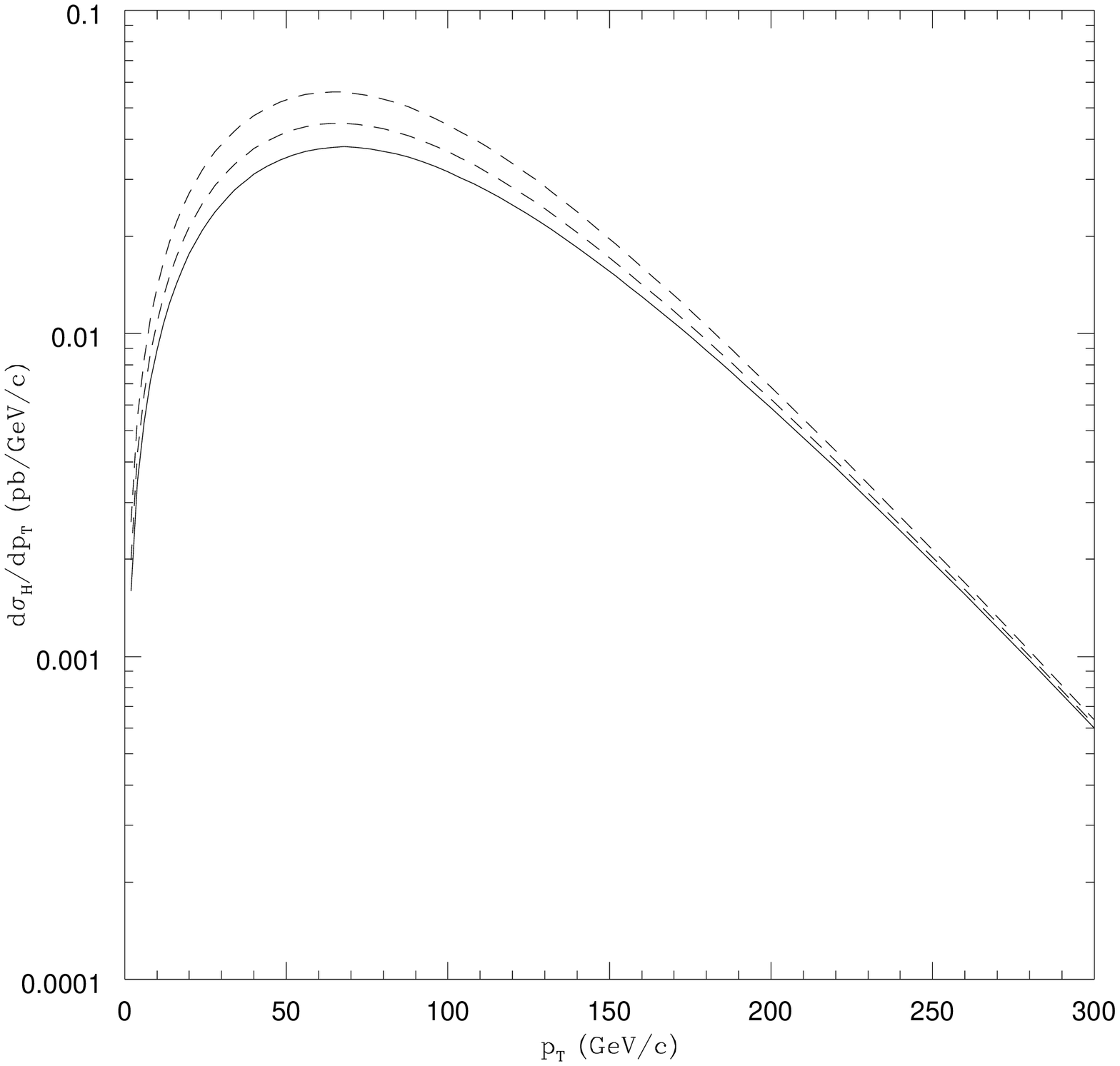,height=4.05in,width=5.05in,clip=}}
\caption[The top quark $p_T$ distributions $d\sigma_H/dp_T$ for the
sum of the $q\bar{q}$ and $gg$ channels]
{The top quark $p_T$ distributions  $d\sigma_H/dp_T$ for the
sum of the $q\bar{q}$ and $gg$ channels for a top quark mass 
$m_t=175$ GeV$/c^2$.
Plotted are $d\sigma_H^{(0)}/dp_T+d\sigma_H^{(1)}/dp_T\mid _{\rm exact}$
(solid line) and $d\sigma_H^{\rm imp}/dp_T$
(upper and lower dashed lines).} 
\label{fig. 3.4}
\end{figure}

Now we turn to a discussion of the differential $Y$ 
distributions at $\sqrt{S}=1.8$ TeV
for a top quark mass $m_t = 175$ GeV$/c^2$.
In this case we set the factorization mass scale equal to $m_t$ everywhere.
We begin with the results for the $q\bar{q}$ channel in the DIS scheme.
\begin{figure}
\centerline{\psfig{file=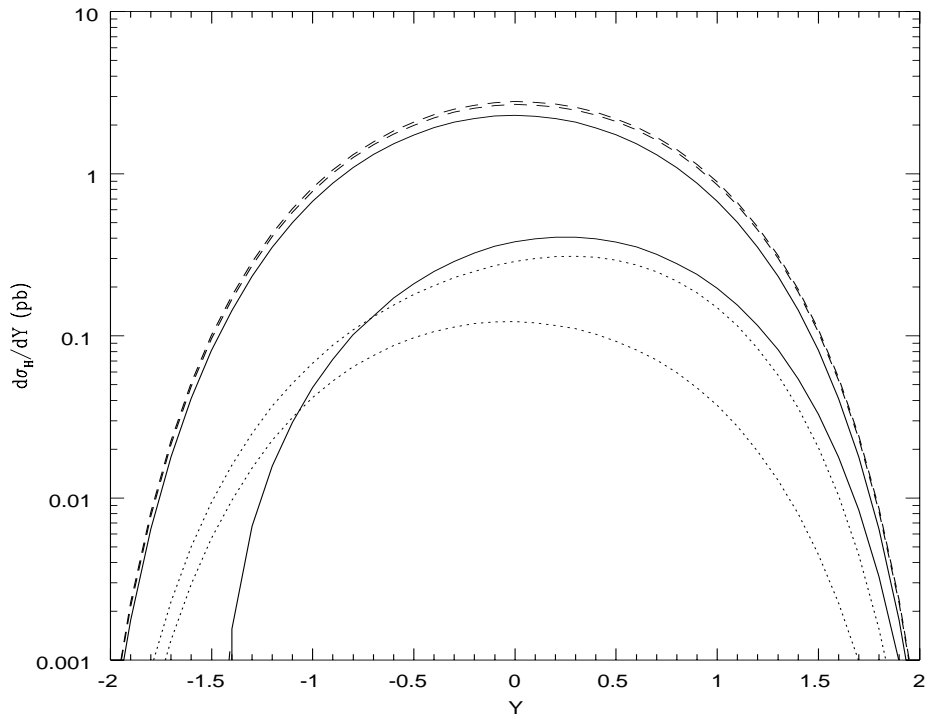,height=4.05in,width=5.05in,clip=}}
\caption[The top quark $Y$ distributions for
the $q\bar{q}$ channel]
{The top quark $Y$ distributions $d\sigma_H^{(k)}/dY$ for
the $q\bar{q}$ channel in the DIS scheme for a top quark mass 
$m_t=175$ GeV$/c^2$. Plotted are $d\sigma_H^{(0)}/dY$ (upper solid line),
$d\sigma_H^{(1)}/dY\mid _{\rm exact}$ (lower solid line),
$d\sigma_H^{(1)}/dY\mid _{\rm app}$ (upper dotted line),
$d\sigma_H^{(2)}/dY\mid _{\rm app}$ (lower dotted line),
and $d\sigma_H^{\rm res}/dY$ ($\mu_0=0.05\:m_t$ upper dashed line
and $\mu_0=0.1\:m_t$ lower dashed line).}
\label{fig. 3.5}
\end{figure}
In fig. 3.5 we show
the Born term $d\sigma_H^{(0)}/dY$, the first order exact result 
$d\sigma_H^{(1)}/dY\mid _{\rm exact}$, the first order 
\linebreak
approximation $d\sigma_H^{(1)}/dY\mid _{\rm app}$, the second 
order approximation $d\sigma_H^{(2)}/dY\mid _{\rm app}$,
and the resummed result $d\sigma_H^{\rm res}/dY$ for $\mu_0=0.05\:m_t$ 
and $\mu_0=0.1\:m_t$. 

\begin{figure}
\centerline{\psfig{file=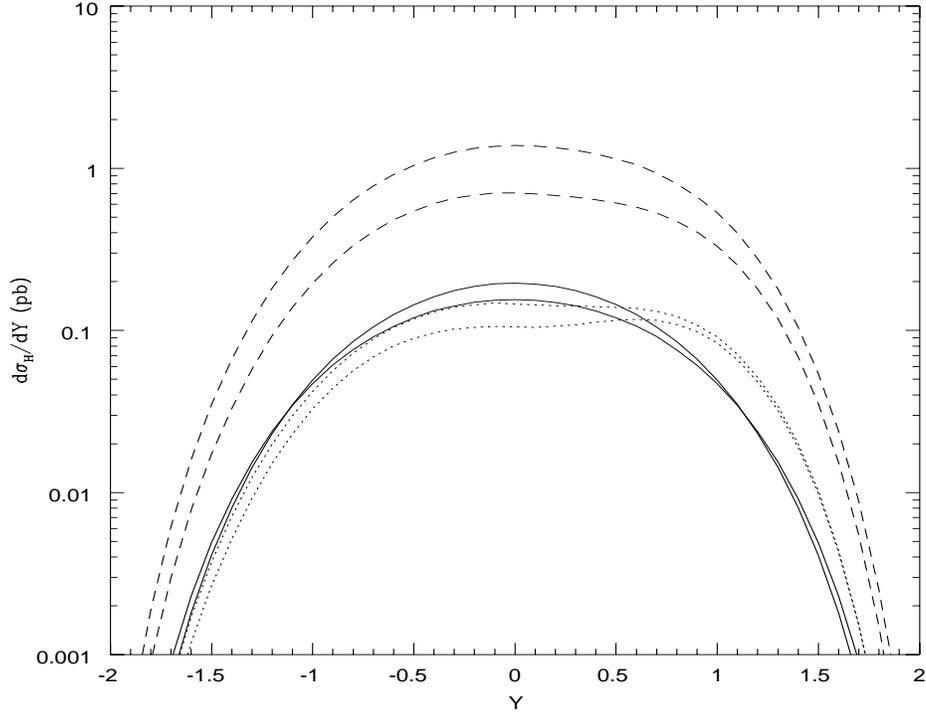,height=4.05in,width=5.05in,clip=}}
\caption[The top quark $Y$ distributions for
the $gg$ channel]
{The top quark $Y$ distributions $d\sigma_H^{(k)}/dY$ for
the $gg$ channel in the $\overline{\rm MS}$ scheme for a top quark mass 
$m_t=175$ GeV$/c^2$. Plotted are $d\sigma_H^{(0)}/dY$ (upper solid line 
at $Y=0$),
$d\sigma_H^{(1)}/dY\mid _{\rm exact}$ (lower solid line at $Y=0$),
$d\sigma_H^{(1)}/dY\mid _{\rm app}$ (lower dotted line),
$d\sigma_H^{(2)}/dY\mid _{\rm app}$ (upper dotted line),
and $d\sigma_H^{\rm res}/dY$ ($\mu_0=0.2\:m_t$ upper dashed line
and $\mu_0=0.25\:m_t$ lower dashed line).}
\label{fig.3.6}
\end{figure}

We continue with the results for the $gg$ channel 
in the $\overline{\rm MS}$ scheme.
The corresponding plot is given in figure 3.6. Here, the values 
of $\mu_0$ are $\mu_0=0.2\:m_t$ and $\mu_0=0.25\:m_t$. As in the case of 
the $p_T$ distributions, the first and second order 
corrections in this channel are larger than the respective ones in the 
$q\bar{q}$ channel in the DIS scheme.  For the range of $Y$ 
values shown the second-order approximate correction is larger than the
first-order approximation. Again, as in the $p_T$ distributions,
the relative difference in magnitude between the improved
$d\sigma_H^{\rm imp}/dY$ and the exact $O(\alpha_s^3)$ results 
is significantly larger than that in the $q\bar{q}$ channel
in the DIS scheme. 

Finally, we conclude our discussion of the differential 
$Y$ distributions by showing the
results of adding the $q\bar{q}$ and $gg$ channels. The plots appear
in figures 3.7 and 3.8. Again, it is evident that resummation produces 
a non-negligible modification of the exact $O(\alpha_s^3)$ result.
However, the shape of the distribution is unchanged.
\begin{figure}
\centerline{\psfig{file=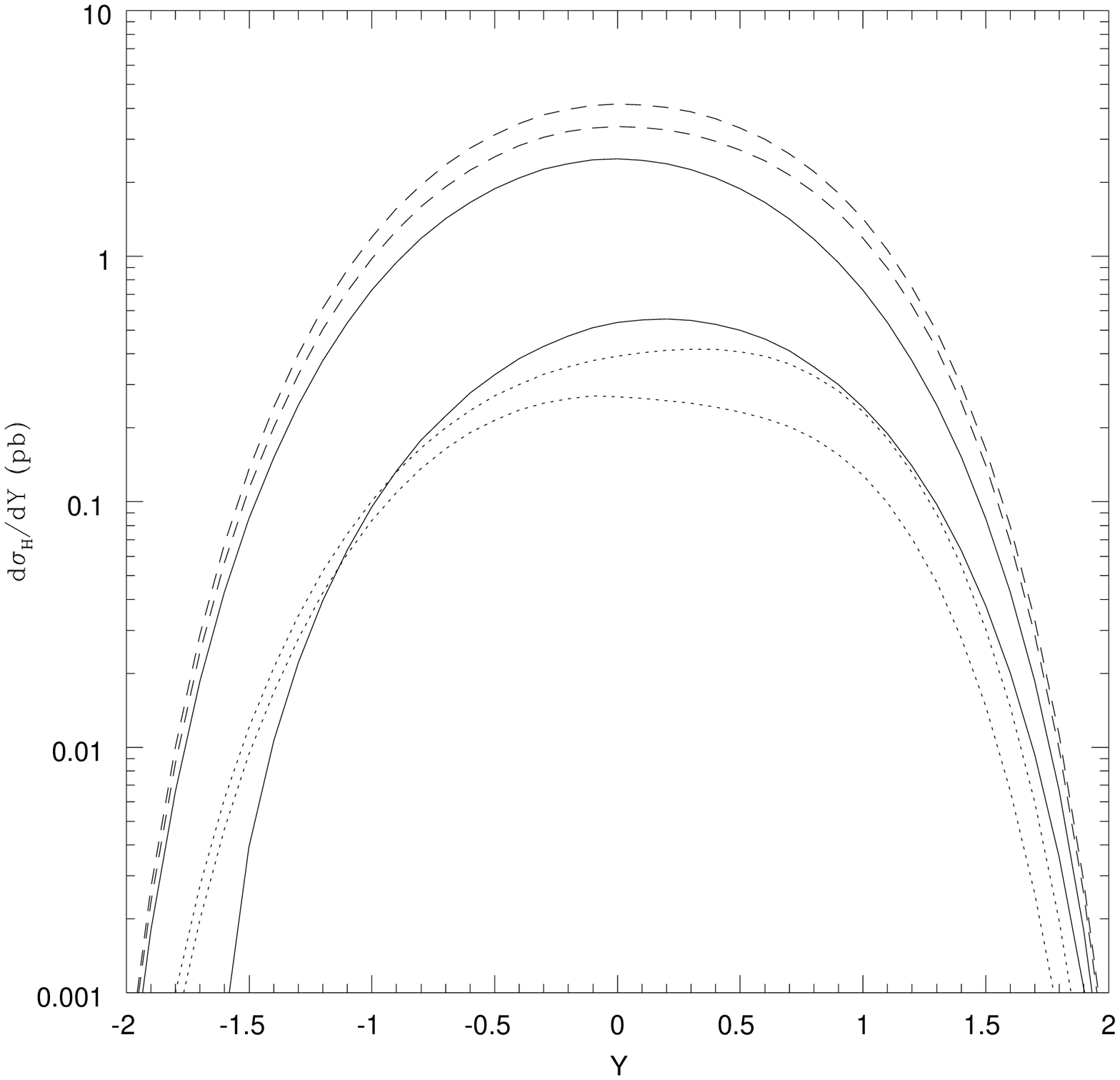,height=4.05in,width=5.05in,clip=}}
\caption[The top quark $Y$ distributions $d\sigma_H^{(k)}/dY$ for
the sum of the $q\bar{q}$ and $gg$ channels]
{The top quark $Y$ distributions $d\sigma_H^{(k)}/dY$ for
the sum of the $q\bar{q}$ and $gg$ channels  for a top quark mass 
$m_t=175$ GeV$/c^2$. Plotted are $d\sigma_H^{(0)}/dY$ (upper solid line),
$d\sigma_H^{(1)}/dY\mid _{\rm exact}$ (lower solid line),
$d\sigma_H^{(1)}/dY\mid _{\rm app}$ (upper dotted line),
$d\sigma_H^{(2)}/dY\mid _{\rm app}$ (lower dotted line),
and $d\sigma_H^{\rm res}/dY$ (upper 
and lower dashed lines).}
\label{fig. 3.7}
\end{figure}
\begin{figure}
\centerline{\psfig{file=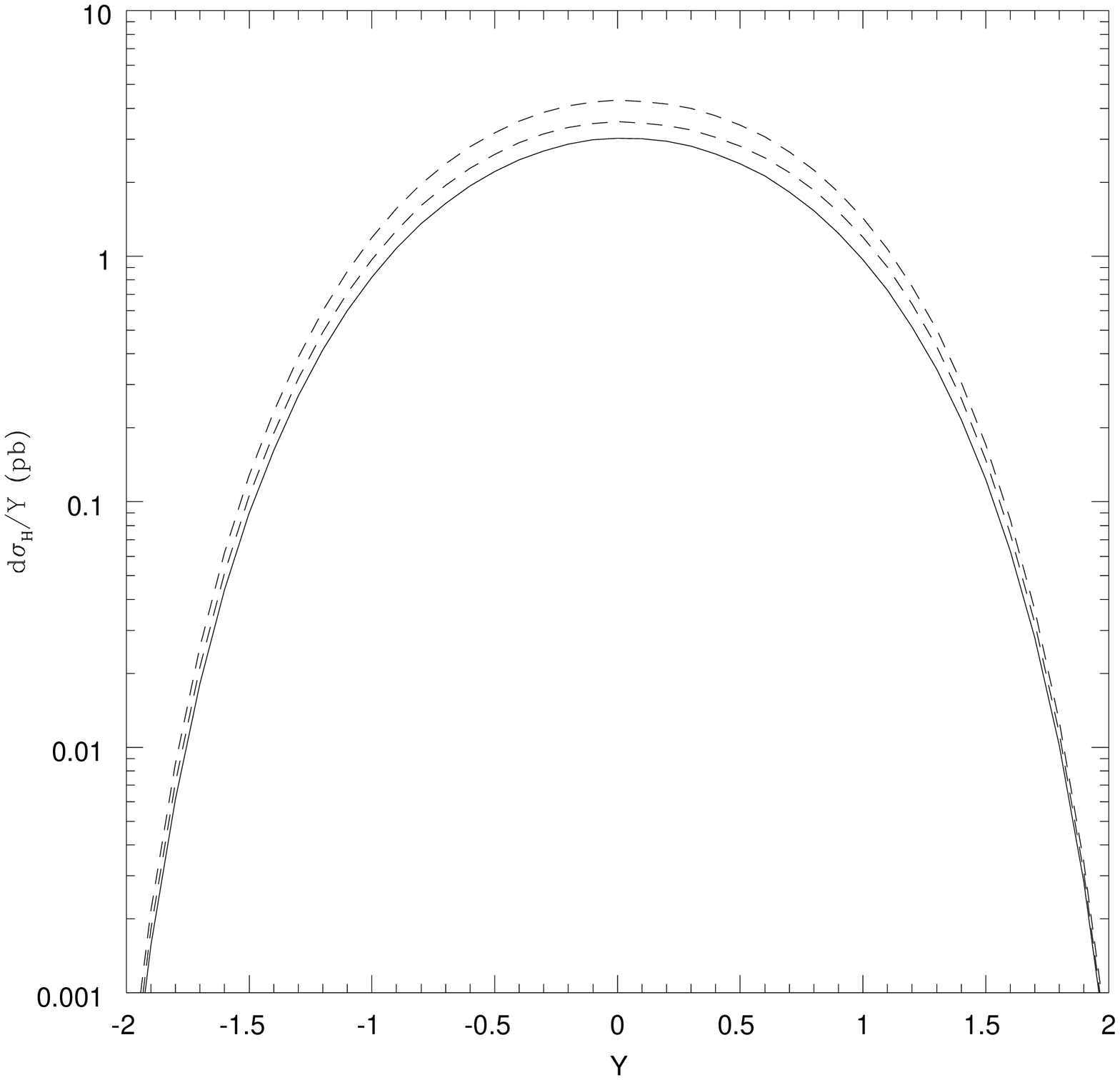,height=4.05in,width=5.05in,clip=}}
\caption[The top quark $Y$ distributions $d\sigma_H/dY$ for the
sum of the $q\bar{q}$ and $gg$ channels]
{The top quark $Y$ distributions  $d\sigma_H/dY$ for the
sum of the $q\bar{q}$ and $gg$ channels for a top quark mass 
$m_t=175$ GeV$/c^2$.
Plotted are $d\sigma_H^{(0)}/dY+d\sigma_H^{(1)}/dY\mid _{\rm exact}$
(solid line) and $d\sigma_H^{\rm imp}/dY$
(upper and lower dashed lines).} 
\label{fig. 3.8}
\end{figure}

\mysection{Bottom quark differential distributions}
In this section we present some results on the inclusive transverse momentum
and rapidity distributions of the bottom quark at HERA-B. 

We begin with the $p_T$ distributions.
For these plots the mass factorization scale is not everywhere equal to $m_b$.
We chose $\mu=m_b$ in $s_0$, $f_k(s_4/m^2\,,m^2/\mu^2)$ and $\bar{\alpha}_s$, 
but $\mu=m_T$ in the MRSD$\_ ' \:$ parton distribution
functions and the running coupling constant $\alpha_s(\mu^2)$.
\begin{figure}
\centerline{\psfig{file=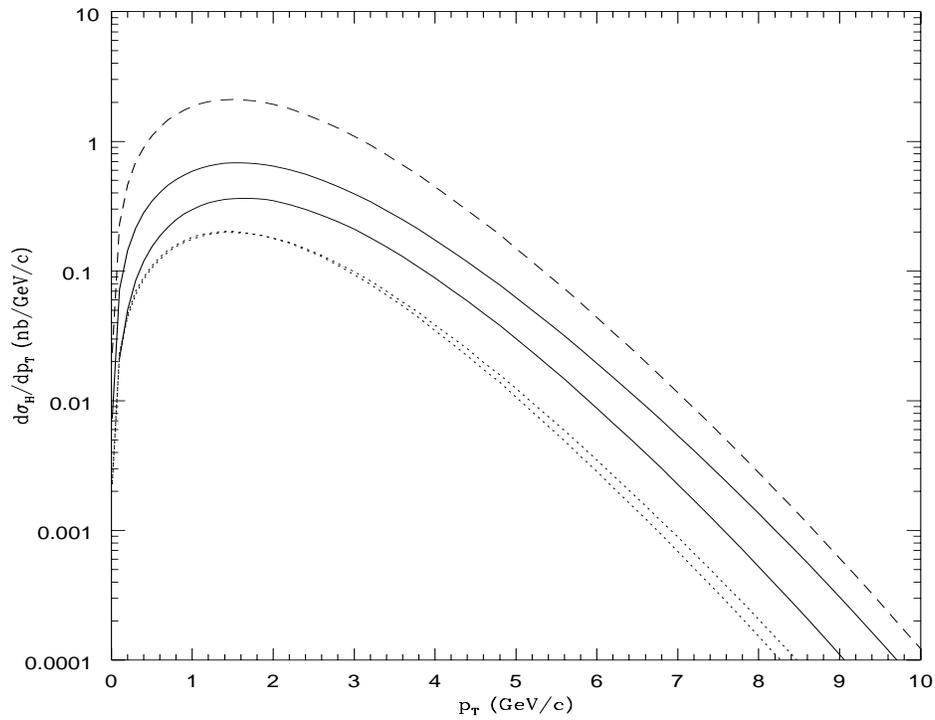,height=4.05in,width=5.05in,clip=}}
\caption[The bottom quark $p_T$ distributions at HERA-B for
the $q\bar{q}$ channel]
{The bottom quark $p_T$ distributions $d\sigma_H^{(k)}/dp_T$ 
at HERA-B for
the $q\bar{q}$ channel in the DIS scheme for 
$m_b=4.75$ GeV$/c^2$. 
Plotted are $d\sigma_H^{(0)}/dp_T$ (upper solid line),
$d\sigma_H^{(1)}/dp_T\mid _{\rm exact}$ (lower solid line),
$d\sigma_H^{(1)}/dp_T\mid _{\rm app}$ (upper dotted line),
$d\sigma_H^{(2)}/dp_T\mid _{\rm app}$ (lower dotted line),
and $d\sigma_H^{\rm res}/dp_T$ for $\mu_0=0.6$ GeV/$c^2$ (dashed line).}
\label{fig. 3.9}
\end{figure}
In fig. 3.9, we give the results for the $q \bar q$ channel in the DIS scheme.
We plot the Born term $d\sigma_H^{(0)}/dp_T$, the first order exact result 
$d\sigma_H^{(1)}/dp_T\mid _{\rm exact}$, the first order approximation 
$d\sigma_H^{(1)}/dp_T\mid _{\rm app}$, the second 
order approximation $d\sigma_H^{(2)}/dp_T\mid _{\rm app}$,
and the resummed result $d\sigma_H^{\rm res}/dp_T$ for $\mu_0=0.6$ GeV/$c^2$.
This is the same value for $\mu_0$ that we used in chapter 2 
for the total cross section. 
If we decrease $\mu_0$ the differential cross 
sections will increase. 
The resummed distribution
was calculated with the cut $s_4>s_0$ while no such cut was imposed 
on the phase space for the individual terms in the perturbation series.
We continue with the results for the $gg$ channel in 
the $\overline{\rm MS}$ scheme.
\begin{figure}
\centerline{\psfig{file=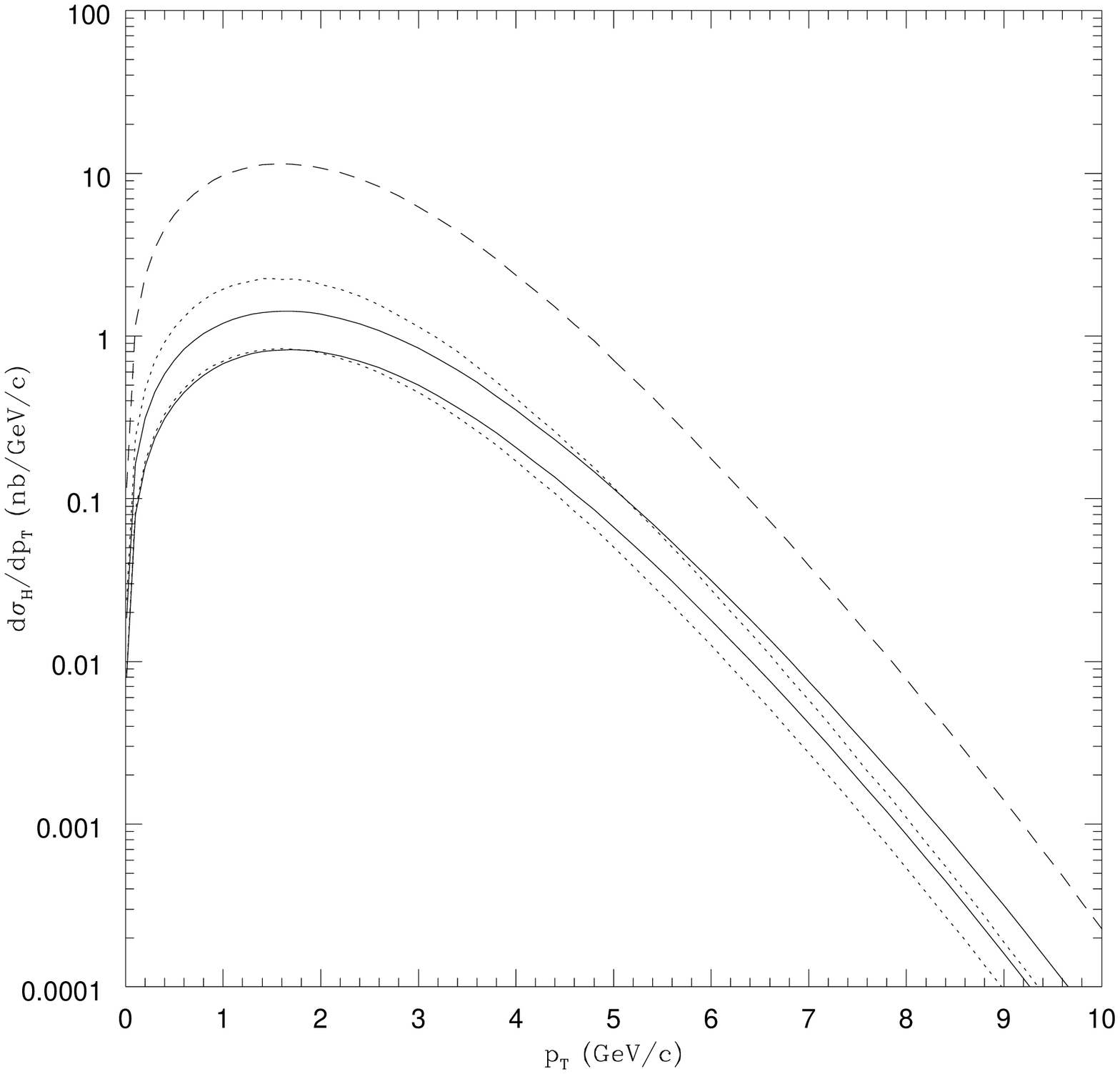,height=4.05in,width=5.05in,clip=}}
\caption[The bottom quark $p_T$ distributions at HERA-B for
the $gg$ channel]
{The bottom quark $p_T$ distributions $d\sigma_H^{(k)}/dp_T$ 
at HERA-B for
the $gg$ channel in the $\overline {\rm MS}$ scheme for 
$m_b=4.75$ GeV$/c^2$. 
Plotted are $d\sigma_H^{(0)}/dp_T$ (lower solid line),
$d\sigma_H^{(1)}/dp_T\mid _{\rm exact}$ (upper solid line),
$d\sigma_H^{(1)}/dp_T\mid _{\rm app}$ (lower dotted line),
$d\sigma_H^{(2)}/dp_T\mid _{\rm app}$ (upper dotted line),
and $d\sigma_H^{\rm res}/dp_T$ for $\mu_0=1.7$ GeV/$c^2$ (dashed line).}
\label{fig. 3.10}
\end{figure}
The corresponding plot is given in fig. 3.10.
We note that the corrections in this channel are large. In fact the exact
first-order correction is larger than the Born term and the approximate 
second-order correction is larger than the approximate first-order correction.
In this case the value of $\mu_0$ has been chosen to be
$\mu_0=1.7$ GeV/$c^2$ as in chapter 2. 
\begin{figure}
\centerline{\psfig{file=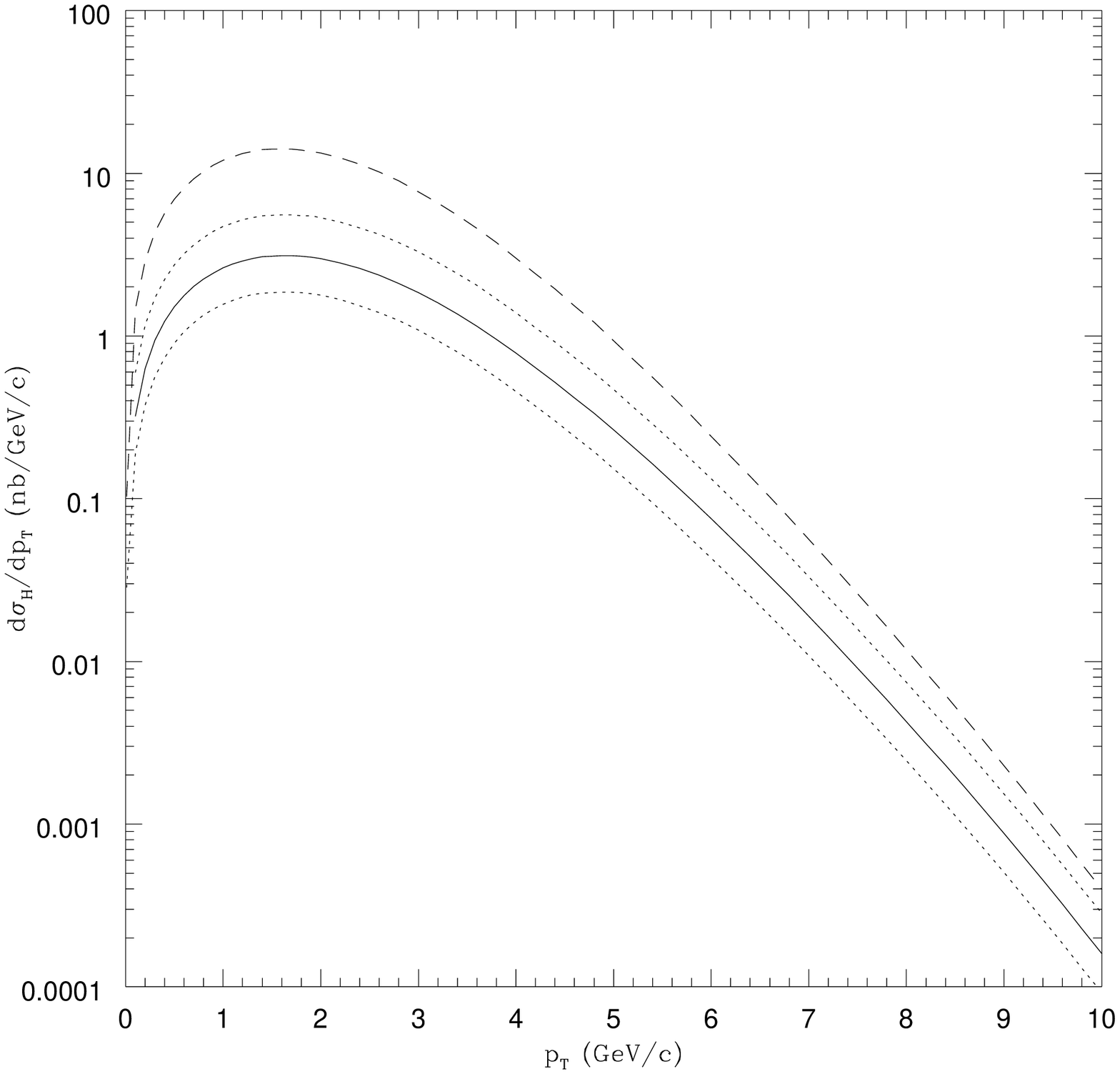,height=4.05in,width=5.05in,clip=}}
\caption[The bottom quark $p_T$ distributions at HERA-B for the sum
of all channels]
{The bottom quark $p_T$ distributions  $d\sigma_H/dp_T$ 
at HERA-B for the sum
of all channels for  $m_b=4.75$ GeV$/c^2$.
Plotted are $d\sigma_H^{(0)}/dp_T+d\sigma_H^{(1)}/dp_T\mid _{\rm exact}$
($\mu=m_b$ solid line, $\mu=m_b/2$ upper dotted line, $\mu=2m_b$ 
lower dotted line) 
and $d\sigma_H^{\rm imp}/dp_T$ (dashed line).}
\label{fig. 3.11}
\end{figure}
In fig. 3.11 we plot the improved $p_T$ distribution for the sum 
of all channels,
where we have included the small negative contributions of the $qg$
and $\bar q g$ channels.
For comparison we also show the total exact NLO results for 
$\mu=m_b/2$, $m_b$, and 
$2m_b$. The improved $p_T$ distribution is uniformly above the exact
$O(\alpha_s^3)$ results. 
We see that the effect of the resummation exceeds the uncertainty
due to scale dependence.

We finish with a discussion of the $Y$ distributions.
In this case we set the factorization mass scale equal to $m_b$ everywhere.
We begin with the $q \bar q$ channel. 
\begin{figure}
\centerline{\psfig{file=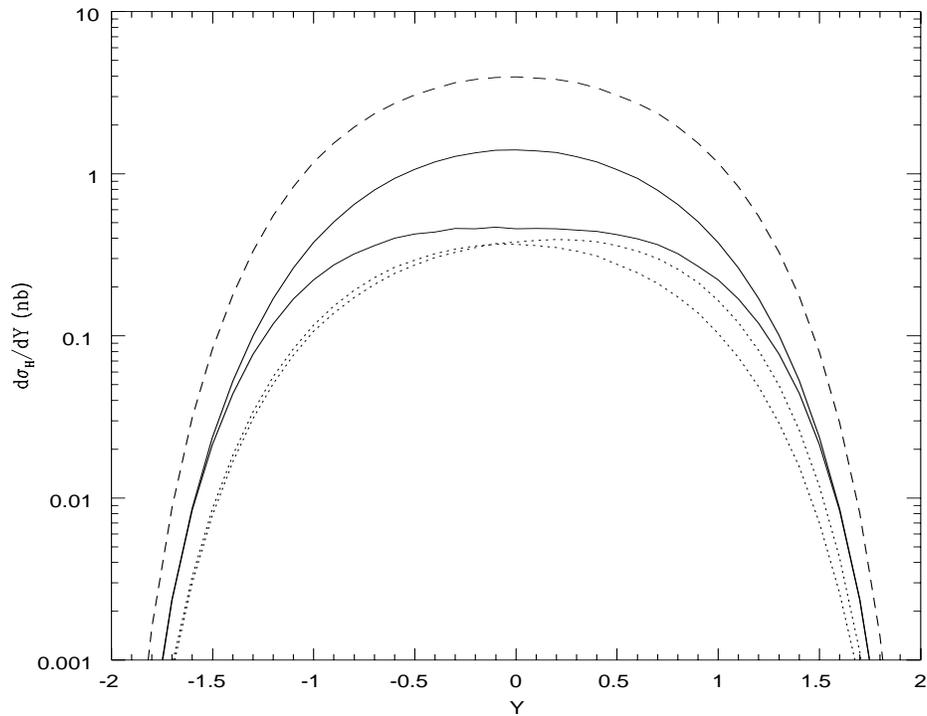,height=4.05in,width=5.05in,clip=}}
\caption[The bottom quark $Y$ distributions at HERA-B for
the $q\bar{q}$ channel]
{The bottom quark $Y$ distributions $d\sigma_H^{(k)}/dY$ 
at HERA-B for
the $q\bar{q}$ channel in the DIS scheme for $m_b=4.75$ GeV$/c^2$. 
Plotted are $d\sigma_H^{(0)}/dY$ (upper solid line),
$d\sigma_H^{(1)}/dY\mid _{\rm exact}$ (lower solid line),
$d\sigma_H^{(1)}/dY\mid _{\rm app}$ (upper dotted line at positive Y),
$d\sigma_H^{(2)}/dY\mid _{\rm app}$ (lower dotted line at positive Y),
and $d\sigma_H^{\rm res}/dY$ for $\mu_0=0.6$ GeV/$c^2$ (dashed line).}
\label{fig. 3.12}
\end{figure}
In fig. 3.12 we show
the Born term $d\sigma_H^{(0)}/dY$, the first order exact result 
$d\sigma_H^{(1)}/dY\mid _{\rm exact}$, the first order approximation 
$d\sigma_H^{(1)}/dY\mid _{\rm app}$, the second 
order approximation $d\sigma_H^{(2)}/dY\mid _{\rm app}$,
and the resummed result $d\sigma_H^{\rm res}/dY$ for $\mu_0=0.6$ GeV/$c^2$.
Again, the resummed distribution was calculated with the cut $s_4>s_0$ 
while no such cut was imposed 
on the phase space for the individual terms in the perturbation series.
We continue with the results for the $gg$ channel 
in the $\overline{\rm MS}$ scheme.
\begin{figure}
\centerline{\psfig{file=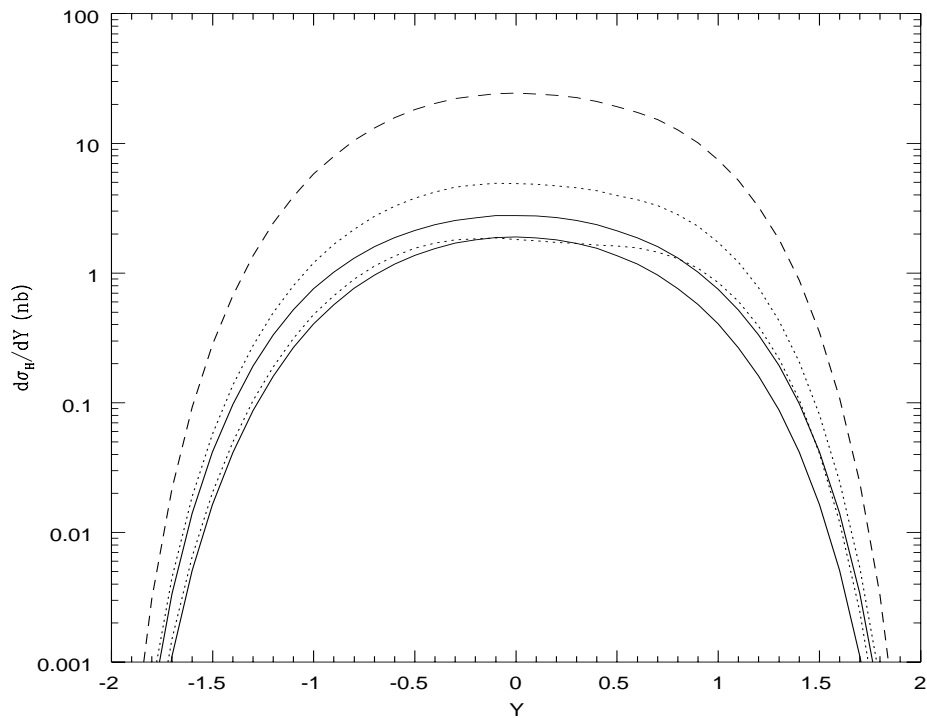,height=4.05in,width=5.05in,clip=}}
\caption[The bottom quark $Y$ distributions at HERA-B for
the $gg$ channel]
{The bottom quark $Y$ distributions $d\sigma_H^{(k)}/dY$ 
at HERA-B for
the $gg$ channel in the $\overline {\rm MS}$ scheme for $m_b=4.75$ GeV$/c^2$. 
Plotted are $d\sigma_H^{(0)}/dY$ (lower solid line),
$d\sigma_H^{(1)}/dY\mid _{\rm exact}$ (upper solid line),
$d\sigma_H^{(1)}/dY\mid _{\rm app}$ (lower dotted line),
$d\sigma_H^{(2)}/dY\mid _{\rm app}$ (upper dotted line),
and $d\sigma_H^{\rm res}/dY$ for $\mu_0=1.7$ GeV/$c^2$ (dashed line).}
\label{fig. 3.13}
\end{figure}
The corresponding plot is given in fig. 3.13. Here, the value 
of $\mu_0$ is $\mu_0=1.7$ GeV/$c^2$.
The corrections in this channel are large as was the case for the $p_T$
distributions.
\begin{figure}
\centerline{\psfig{file=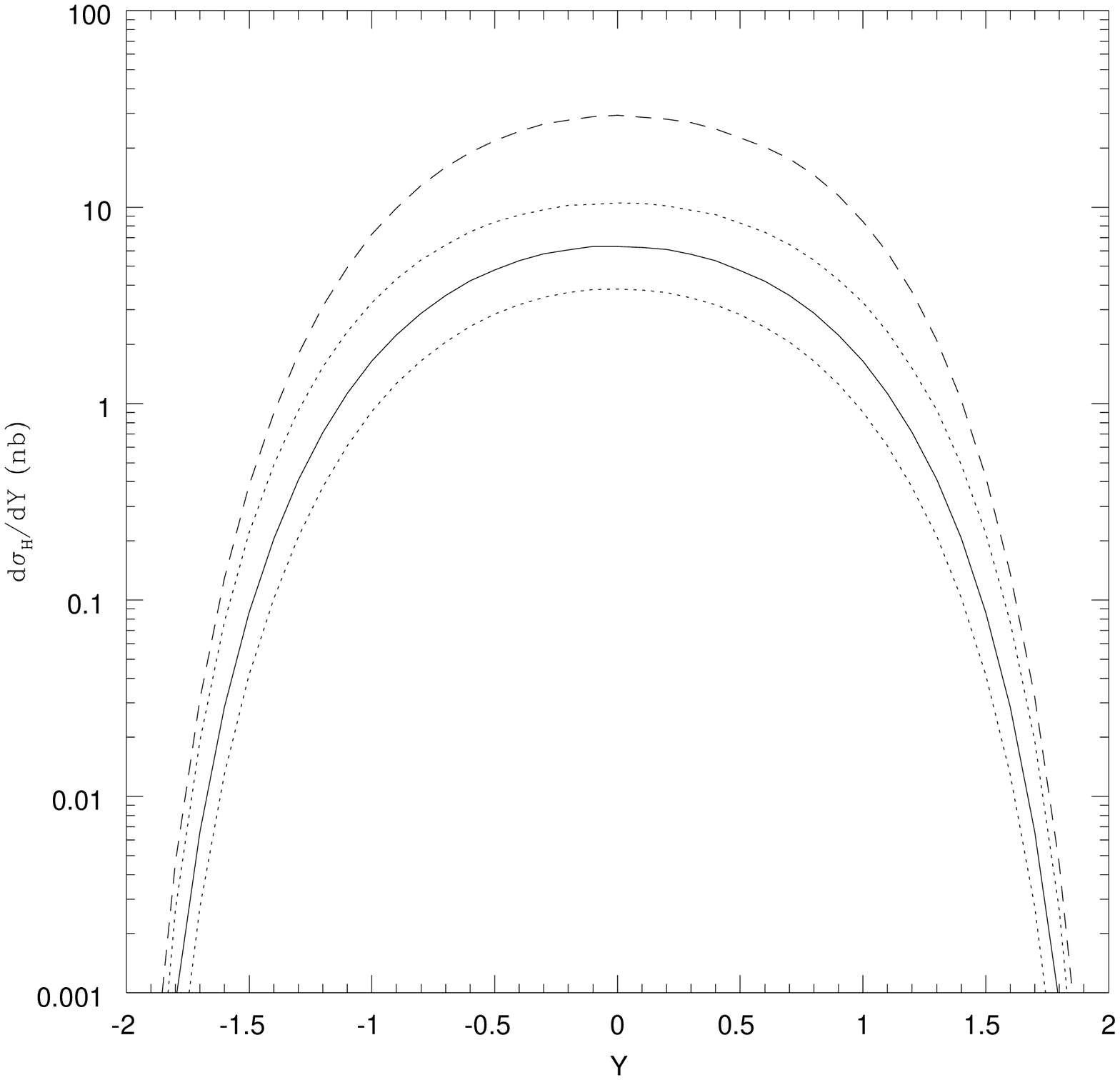,height=4.05in,width=5.05in,clip=}}
\caption[The bottom quark $Y$ distributions  
at HERA-B for the sum of all channels]
{The bottom quark $Y$ distributions  $d\sigma_H/dY$ 
at HERA-B for the sum of all channels for $m_b=4.75$ GeV$/c^2$.
Plotted are $d\sigma_H^{(0)}/dY+d\sigma_H^{(1)}/dY\mid _{\rm exact}$
($\mu=m_b$ solid line, $\mu=m_b/2$ upper dotted line, $\mu=2m_b$ 
lower dotted line) 
and $d\sigma_H^{\rm imp}/dY$(dashed line).}
\label{fig. 3.14}
\end{figure}
In fig. 3.14 we plot the improved $Y$ distribution for the sum 
of all channels,
where we have included the small negative contributions of the $qg$
and $\bar q g$ channels.
For comparison we also show the total exact NLO results
for $\mu=m_b/2$, $m_b$, and 
$2m_b$. The improved $Y$ distribution is uniformly above the 
exact $O(\alpha_s^3)$ results. 
Again, we see that the effect of the resummation exceeds the uncertainty
due to scale dependence.

\mysection{Conclusions}

We have shown that the resummation of soft gluon radiation 
produces a small difference between the perturbation theory 
improved distributions 
in $p_T$ and $Y$ and the exact $O(\alpha_s^3)$ distributions 
in $p_T$ and $Y$ for
the $q\bar{q}$ reaction in the DIS scheme for the values of $\mu_0$ chosen.
However, for the $gg$ channel in the $\overline{\rm MS}$ scheme the
resummation produces a large difference. The difference between the
resummed and the exact $O(\alpha_s^3)$ distributions depends 
on the mass factorization 
scheme (DIS or $\overline{\rm MS}$), the factorization scale $\mu$, 
as well as the 
specific reaction under consideration ($q\bar{q}$ or $gg$).
For top quark production at the Fermilab Tevatron
with $m_t=175$ GeV$/c^2$ the $gg$ channel is not 
as important numerically as 
the $q\bar{q}$ channel. However, since the corrections for the $gg$ channel
are quite large, resummation produces a non-negligible difference between
the perturbation theory improved and the exact 
$O(\alpha_s^3)$ distributions when adding the 
two channels. However, the shapes of the distributions are essentially
unchanged.
For bottom quark production at HERA-B 
with $m_b=4.75$ GeV$/c^2$ the $gg$ channel is
dominant so the enhancement of the NLO distributions is much bigger.
%

\chapter{Resummation of singular distributions 
in QCD hard scattering} 
We discuss the resummation of distributions that are singular
at the elastic limit of partonic phase space
(partonic threshold) in
QCD hard-scattering cross sections, such as  heavy quark production.
We show how nonleading soft logarithms exponentiate 
in a manner that depends on the
color structure within the underlying hard scattering.  This result
generalizes the resummation of threshold singularities for the
Drell-Yan process, in which the hard scattering proceeds through
color-singlet annihilation.
We illustrate our results for the
case of heavy quark production by light quark annihilation
and gluon fusion, and also for light quark production
through gluon fusion.

\mysection{General formalism}
In hard scattering cross sections factorized according to perturbative 
QCD the calculable short-distance function includes distributions that 
are singular when the total invariant mass of the partons reaches the
minimal value necessary to produce the observed final state. Such singular
distributions can give substantial QCD corrections to any order in $\alpha_s$. 

Expressions that resum these distributions to the short-distance 
function of Drell-Yan cross sections to arbitrary logarithmic accuracy
have been known for some time \cite{oldDY}.   
It has also been observed that leading distributions, 
and hence  leading logarithms in moment space, are the same for
many hard QCD cross sections. We used this fact as the
basis for our estimates of heavy quark 
production cross sections and differential distributions 
in the previous chapters.  
In chapter 2 we pointed out the inadequacy of the leading log approximation,
particularly for the $gg$ channel.
In this chapter, we shall extend our analysis
to the level of next-to-leading logarithms. 
We shall exhibit a method by which nonleading distributions 
may be treated, and will illustrate
this method in the cases of heavy quark production through 
light quark annihilation and gluon fusion, and light quark production
through gluon fusion.

We consider the inclusive cross section for the production of 
one or more particles, with total invariant
mass $Q$.  Examples include states produced by
QCD, such as heavy quark pairs or
high-$p_T$ jets, in addition to massive 
electroweak vector bosons, virtual or real, as in the Drell-Yan
process.  

To be specific,  
we shall discuss the summation of (``plus")
distributions, which are singular  for $z=1$, where
\begin{equation}
z={Q^2\over s},
\label{def}
\end{equation}
for the production of a heavy quark pair of total invariant
mass $Q$,
with $s$ the invariant mass squared
of the incoming partons that initiate the hard scattering.  
We shall refer to $z=1$ as  ``partonic threshold" 
\footnote{We emphasize that by partonic threshold, we
refer to c.m.\ total energy of the incoming partons
for a fixed final state; heavy
quarks, for example, are not necessarily produced at rest.},
or more accurately the ``elastic limit.''  
We assume that the cross section is defined so that
there are no uncancelled collinear divergences in the final state.  

The main complications relative to Drell-Yan \cite{oldDY} involve
the exchange of color in the hard scattering, and the presence of
final-state interactions.  In fact, these effects only
modify partonic threshold singularities at next-to-leading logarithm,  and we
give below explicit exponentiated
moment-space expressions which take them into account
at this level.  At the next level of accuracy, we shall see that resummation
requires ordered exponentials, in terms of calculable anomalous dimensions.

The properties of QCD that make this organization possible are the 
factorization of soft gluons from high-energy partons in perturbation 
theory \cite{fact}, and the
exponentiation of soft gluon effects \cite{expon}.  
Factorization is represented by
fig. 4.1 for the annihilation of a light quark pair to form a pair of
heavy quarks.  In this figure, momentum configurations that contribute singular
behavior near partonic threshold
are shown in a cut diagram notation \cite{fact}.
As shown, it is possible to factorize soft gluons from
the ``jets" of virtual and real particles that are on-shell and
parallel to the 
incoming, energetic light quarks, as well as from the outgoing heavy
quarks.
Soft-gluon factorization 
from incoming light-like partons is 
a result of relativistic limit \cite{fact},
while factorization from heavy quarks, even when they
are nonrelativistic is familiar from heavy-quark effective
theory. Once soft gluons are factored from them, the jets
may be identified with the parton distributions of the
initial state hadrons. The
hard interactions, labelled $H_I$ and $H^*_J$ in fig 4.1, corresponding
to contributions from the amplitude and its complex conjugate, respectively,
are labelled by the overall color exchange in each.
A general argument of how the exponentiation of Sudakov logarithms follows
from the factorization of soft and hard parts and jets is given in 
\cite{CLS}.

\begin{figure}
\centerline{
\psfig{file=figz.ps,height=2.5in,width=4.05in,clip=}}
\centerline{
\psfig{file=figz.ps,height=2.5in,width=4.05in,clip=}}
\caption[Cut diagram illustrating momentum configurations that give rise to
threshold enhancements in heavy quark production]
{Cut diagram illustrating momentum configurations that give rise to
threshold enhancements in heavy quark production:  (a) General
factorization theorem.  Away from partonic threshold all singularities
in the ``short-distance" subdiagram $H/S$ cancel;
(b) Expanded view of $H/S$ near threshold, showing the factorization
of soft gluons onto eikonal (Wilson) lines from incoming and outgoing
partons in the hard subprocess.  $H_I$ and $H^*_J$ represent the
remaining, truly short-distance, hard scattering.}
\label{fig. 4.1}
\end{figure}

For example,
with the quark-antiquark process shown, 
the choice of color structure is simple,
and may, for instance, be chosen as singlet or octet.  To make these choices
explicit, we label the colors of the incoming pair $i$ and $j$ for
the quark and antiquark respectively, and of
the outgoing (massive) pair $k$ and $l$ for the quark and antiquark.  
The hard scattering is then of the generic form
\begin{equation}
H_1=h_1(Q^2/\mu^2,\alpha_s(\mu^2))\, \delta_{ji}\delta_{lk}\, ,
\label{singletdef}
\end{equation}
for singlet structure (annihilation of color) in the $s$-channel.
For the $s$-channel octet, or more generally adjoint in color SU(N), 
we have, analogously
\begin{equation}
H_A=h_A(Q^2/\mu^2,\alpha_s(\mu^2))\, 
\sum_{c=1}^{N^2-1}\left [ T_c^{(F)}\right]_{ji}\; \left [ T_c^{(F)}\right]_{lk}\, ,
\label{octetdef}
\end{equation}
with $T_c^{(F)}$ the generators in the fundamental representation.
The functions $h_I$ are, as indicated, infrared safe, that is, free of 
both collinear and infrared divergences, even at partonic threshold.  

Taking into account possible choices of $H_I$ and $H^*_J$,
an expression that organizes all singular distributions
for heavy quark production is
\begin{eqnarray}
\frac{d\sigma_{h_1 h_2}}{dQ^2 \; d\cos\theta \; dy}&=&\sum_{ab} \; \sum_{IJ}
\int\frac{dx_a}{x_a}\frac{dx_b}{x_b} \phi_{a/h_1}(x_a,Q^2) 
\phi_{b/h_2}(x_b,Q^2)
\nonumber \\ &&
\times \delta\left(y-\frac{1}{2}\ln\frac{x_a}{x_b}\right) 
\ \Omega_{ab}^{(IJ)}\left(\frac{Q^2}{x_a x_b S},y,\theta,
\alpha_s(Q^2)\right) ,
\nonumber\\
\end{eqnarray}
where $y$ is the pair rapidity and $\theta$ is the scattering angle
in the pair center of mass frame.
The indices $I$ and $J$ label color tensors,
such as the singlet (\ref{singletdef}) and octet (\ref{octetdef}), with which
we contract
the color indices of the incoming and outgoing partons
that participate in the hard scattering. 
The variable $S$ is the invariant mass squared of the incoming hadrons. 
The functions $\phi_{a/h}$ are parton densities,
evaluated at scale $Q^2$.  
The function $\Omega$ contains all singular behavior
in the threshold limit, $z\rightarrow 1$.
$\Omega$ depends on the scheme in which we perform factorization,
the usual choices being $\overline{\rm MS}$ and DIS.  
Note that the resummation may be carried out at fixed $y$, 
so long as $y$ is not close to
the edge of phase space \cite{LaenenSterman}.

The color structure of the hard scattering influences contributions to
nonleading infrared behavior.  Not all soft gluons, however, are sensitive
to the color structure of the hard scattering.  Gluons that are both soft
and collinear to the incoming partons,
factorize into the parton distributions of the incoming hadrons. It is at
the level of nonleading purely soft gluons with central rapidities that color 
dependence appears,
in the resummation of soft gluon effects. Each choice of color structure
has, as a result, its own exponentiation for soft gluons \cite{BottsSt}.
Then, to next-to-leading-logarithm (NLL) 
it is possible to pick a color basis 
in which moments with respect to $z$ exponentiate, 
\begin{eqnarray}
\tilde{\Omega}_{ab}^{(IJ)}(n,y,\theta,Q^2)&=&\int_0^1 dz z^{n-1} 
\Omega_{ab}^{(IJ)}(z,y,\theta,\alpha_s(Q^2))
\nonumber \\ 
&=&H_{ab}^{(IJ)}(y,\theta,Q^2)
e^{E_{IJ}(n,\theta,Q^2)} \, ,
\label{omegaofn}
\end{eqnarray}
where the color-dependent exponents are given by
\begin{eqnarray}
E_{IJ}(n,\theta,Q^2)&=&-\int_0^1\frac{dz}{1-z}(z^{n-1}-1)
 \biggl [  \int_0^z {dy\over 1-y}\, g_1^{(ab)}[\alpha_s((1-y)(1-z)Q^2)]
\nonumber \\ &&
      +\, g_2^{(ab)}[\alpha_s((1-z) Q^2)]
+g_3^{(I)}[\alpha_s((1-z)^2 Q^2),\theta] \nonumber \\
&\ & \hbox{\hskip 1.0 true in}
+g_3^{(J)*}[\alpha_s((1-z)^2 Q^2),\theta]\, \biggr ]\, .
\nonumber \\
\label{Eofn}
\end{eqnarray}
The $g_i$ are finite functions of their arguments.
The $H_{ab}^{(IJ)}$ are infrared safe expansions in $\alpha_s(Q^2)$. 
$g_1^{(ab)}$ and $g_2^{(ab)}$ are universal among hard cross sections and color
structures for given incoming partons $a$ and $b$, but depend on whether these
partons are quarks or gluons.
On the other hand, $g_3^{(I)}$ summarizes
soft logarithms that depend directly on color exchange in the hard scattering,
and hence also on the identities and relative directions of the colliding 
partons (through $\theta$), both incoming and outgoing.

Just as in
the case of Drell-Yan, to reach the accuracy of NLL in the exponents, 
we need $g_1$ only to two loops, with leading logarithms coming entirely from
its one-loop approximation,
and $g_2$ and $g_3^{(I)}$ only to
a single loop.    More explicitly,
we take \cite{CSt}
\begin{equation}
g_1^{(ab)} = (C_a+C_b)\left ( {\alpha_s\over \pi} 
+\frac{1}{2} K \left({\alpha_s\over \pi}\right)^2\right )\, ,
\label{g1def}
\end{equation}
with $C_i=C_F\ (C_A)$ for an incoming quark (gluon), and
with $K$ given by
\begin{equation}
K= C_A\; \left ( {67\over 18}-{\pi^2\over 6 }\right ) - {5\over 9}n_f\, ,
\label{Kdef}
\end{equation}
where $n_f$ is the number of quark flavors.  
$g_2$ is given for quarks in the DIS scheme by
\begin{equation}
g_2^{(q {\bar q})}=-{3\over 2}C_F\; {\alpha_s\over\pi}\,  
\end{equation}
and it vanishes in the $\overline {\rm MS}$ scheme.
As pointed out in \cite{CT2}, one-loop contributions to
$g_3$ may always be absorbed into the one-loop contribution to $g_2$
and the two-loop contribution to $g_1$. Because
$g_3^{(I)}$ depends upon $I$, however, it is advantageous to keep
this nonfactoring process-dependence separate.
We shall describe how it is determined below.

First, let us sketch how these results come about \cite{CLS}.
After the normal factorization of parton distributions, soft gluons cancel
in inclusive hard scattering cross sections.  When restrictions are
placed on soft gluon emission, however, finite logarithmic
enhancements remain, and it is useful to 
separate soft partons from the hard scattering (which is then constrained to
be fully virtual).  
Soft gluons  may be factored from the hard scattering
into a set of Wilson lines, or ordered exponentials,
from which collinear singularities in the initial state are eliminated, either
by explicit subtractions or by a suitable choice of gauge \cite{fact}.
Assuming that the lowest-order process is two-to-two,
there will be two incoming {\em and} two outgoing Wilson lines.\footnote{In
Drell-Yan and other electroweak annihilation processes, there is a
pair of incoming lines only.} The result, illustrated in 
\linebreak
fig. 4.1b,
is of the form, $H_{ab}^{IJ}S_{IJ}$, summed over the same color basis as in
(4.1.4) above.

The resulting
hard scattering and soft-gluon functions both require renormalization,
which is organized by going to a basis in the space of color 
exchanges between the Wilson lines.  
The renormalization is carried out
by a counterterm matrix in this space of color structures. 
For incoming and outgoing lines of equal masses, such analyses have
been carried out to one loop in \cite{BottsSt}, \cite{SotSt}, and \cite{GK},
and to two loops in a related process in \cite{KK}.
For an underlying partonic process $a+b\rightarrow c+d$,
we then construct an anomalous
dimension matrix $\Gamma^{(ab\rightarrow cd)}_{S, IJ}$, where the indices
$I$ and $J$ vary over the various color exchanges possible in the
partonic process.  We write for the renormalization of $S$
\begin{equation}
S^{(0)}_{IJ}=\frac{1}{2}[Z_{S, II'} \delta_{JJ'}+Z_{S, JJ'} \delta_{II'}]
S_{I'J'},
\end{equation}
where $S^{(0)}$ denotes the unrenormalized quantity.
The soft function  $S_{IJ}$ then satisfies the
renormalization group equation \cite{BottsSt}
\begin{equation}
\left(\mu {\partial \over \partial \mu}+\beta(g){\partial \over \partial g}
\right)\,S_{IJ}
=-\left[\Gamma_{S, II'}\delta_{JJ'}+\delta_{II'}\Gamma_{S, JJ'}\right]
S_{I'J'}\, .
\label{gammaRG}
\end{equation}
In a minimal subtraction scheme with $\epsilon=n-4$
\begin{equation}
\Gamma_S (g)=-\frac{g}{2} \frac {\partial}{\partial g}{\rm Res}_{\epsilon 
\rightarrow 0} Z_S (g, \epsilon).
\end{equation}
This resummation of soft logarithms is analogous to
singlet evolution in deeply inelastic scattering, which involves the mixing 
of operators,
and hence of parton distributions.  The general solution, even in moment
space, is given in terms of ordered exponentials which, however, may
be diagonalized at leading logarithm.
For the resummation of soft logarithms in 
QCD cross sections, the same general pattern
holds, with mixing between hard color tensors.  Leading 
soft logarithms,
however, are next-to-leading overall in moment space, which 
allows the exponentiation (\ref{omegaofn}) at this level.

Of course, the analysis is simplest for 
external quarks, and most complicated for external gluons.  It is
also possible to imagine a similar analysis when there are more than
two partons in the final state.  This would be necesary
if we were to treat threshold corrections to ${\bar{\rm p}}{\rm p}
\rightarrow Q{\bar Q}+ {\rm jet}$, for instance, but we have not attempted 
to explore such processes in detail.

Given a choice of incoming and outgoing partons, 
next-to-leading logarithms
in the moment variable $n$ exponentiate as in (\ref{omegaofn}) in the color 
tensor
basis that diagonalizes $\Gamma^{(ab\rightarrow cd)}_{S, IJ}$,
with eigenvalues $\lambda_I$.  The
resulting soft function $g_3^{(I)}$ is then simply
\begin{equation}
g_3^{(I)}[\alpha_s((1-z)^2 Q^2),\theta]=-\lambda_I[\alpha_s((1-z)^2 Q^2),
\theta]\, ,
\end{equation}
where the eigenvalues are complex in general, and depend
on the directions of the incoming and outgoing partons, as shown.

\mysection{Applications to $q \bar{q} \rightarrow Q \bar{Q}$}
These considerations may be illustrated by heavy quark
production through light quark annihilation,
\begin{equation}
q(p_a)+{\bar q}(p_b) \rightarrow {\bar Q}(p_i) + Q(p_j)\, .
\end{equation}
In this case,
as in elastic scattering \cite{BottsSt,KK}, the anomalous
dimension matrix is only two-dimensional.  

As before we define the invariants 
\begin{equation}
s=(p_a+p_b)^2\, , \quad t_1=(p_a-p_i)^2-m^2\, , \quad u_1=(p_b-p_i)^2-m^2\, , 
\end{equation}
with $m$ the heavy quark mass, which satisfy
\begin{equation}
s+t_1+u_1=0
\end{equation}
at partonic threshold. 
We also define dimensionless vectors $v_i^{\mu}$ by
\begin{equation}
p_i^{\mu}={\cal Q}v_i^{\mu}
\end{equation}
which obey $v_i^2=0$ for the light incoming quarks and
$v_i^2=m^2/{\cal Q}^2$ for the outgoing heavy quarks.
Note that $\cal Q$ satisfies the kinematic relation
\begin{equation}
s=2{\cal Q}^2.
\end{equation}
We now calculate $\Gamma_S (g)$. 
The UV divergent $O(\alpha_s)$ contribution to $S$ is the sum of the 
graphs in fig. 4.2. The counterterms for $S$ are the ultraviolet divergent
coefficients times our basis color tensors:
\begin{eqnarray}
S_1&=&c_1 Z_{S, 11} + c_2 Z_{S, 21},
\\
S_2&=&c_1 Z_{S, 12} + c_2 Z_{S, 22}.
\end{eqnarray}

\begin{figure}
\centerline{
\psfig{file=figz.ps,height=5.05in,width=5.05in,clip=}}
\caption[UV divergent one-loop contributions to S for 
$q {\bar q} \rightarrow Q {\bar Q}$]
{UV divergent one-loop contributions to S for 
$q {\bar q} \rightarrow Q {\bar Q}$.}
\label{fig. 4.2}
\end{figure}

In our calculations we use the axial gauge gluon propagator
\begin{equation}
D^{\mu \nu}(k)=\frac{-i}{k^2+i\epsilon} N^{\mu \nu}(k), \quad
N^{\mu \nu}(k)=g^{\mu \nu}-\frac{n^{\mu}k^{\nu}+k^{\mu}n^{\nu}}{n \cdot k}
+n^2\frac{k^{\mu}k^{\nu}}{(n \cdot k)^2},
\end{equation}
with $n^{\mu}$ the gauge vector,
and eikonal rules for all external lines (fig. 4.3).
\begin{figure}
\centerline{
\psfig{file=figz.ps,height=2.5in,width=5.05in,clip=}}
\caption[Eikonal rules for
$q {\bar q} \rightarrow Q {\bar Q}$]
{Eikonal rules for 
$q {\bar q} \rightarrow Q {\bar Q}$. The gluon momentum flows out of the 
eikonal lines.} 
\label{fig. 4.3}
\end{figure}
If we denote a typical one-loop correction to $c_I$ as
$\omega^{(I)}(\delta_i v_i, \delta_j v_j, n, \Delta_i, \Delta_j)$,
where $\delta=+1 \,(-1)$ for the gluon momenta flowing in the same (opposite)
direction to the momentum of $v_i$ and $\Delta=+1 \, (-1)$ for $v_i$ 
corresponding to a quark (antiquark),
then we have:
\begin{eqnarray}
\omega^{(I)}&=&{\cal C}^{(I)} g^2
\int\frac{d^n q}{(2\pi)^n}\frac{-i}{q^2+i\epsilon}
\left\{\frac{\Delta_i\Delta_j v_i \cdot v_j}{(\delta_i v_i\cdot q+i\epsilon)
(\delta_j v_j\cdot q+i\epsilon)}\right.
\nonumber \\ &&
\left. -\frac{\Delta_i v_i \cdot n}{(\delta_i v_i\cdot q+i\epsilon)}
\frac{P}{(n \cdot q)} 
-\frac{\Delta_j v_j \cdot n}{(\delta_j v_j\cdot q+i\epsilon)}
\frac{P}{(n \cdot q)}
+n^2\frac{P}{(n \cdot q)^2}\right\},
\nonumber \\
\label{omega}
\end{eqnarray}
where ${\cal C}^{(I)}$ is a color tensor.
$P$ stands for principal value,
\begin{equation}
\frac{P}{(q \cdot n)^{\beta}}=\frac{1}{2}\left(\frac{1}{(q \cdot n+i\epsilon)
^{\beta}}+(-1)^{\beta}\frac{1}{(-q \cdot n+i\epsilon)^{\beta}}\right).
\end{equation}
We rewrite (\ref{omega}) as
\begin{eqnarray}
\omega^{(I)}&=&{\cal C}^{(I)}{\cal S}_{ij} 
\left[I_1(\delta_i v_i, \delta_j v_j)
-\frac{1}{2}I_2(\delta_i v_i, n)-\frac{1}{2}I_2(\delta_i v_i, -n)\right.
\nonumber \\ && \quad \quad
\left.-\frac{1}{2}I_3(\delta_j v_j, n)-\frac{1}{2}I_3(\delta_j v_j, -n)
+I_4(n^2)\right] \, ,
\end{eqnarray}
where the overall sign is given by
\begin{equation}
{\cal S}_{ij}=\Delta_i \: \Delta_j \: \delta_i \: \delta_j.
\end{equation}

We now evaluate the ultraviolet poles of the integrals.
For the integrals when both $v_i$ and $v_j$ refer to massive quarks we have
\begin{eqnarray}
I_1^{UV \;pole}&=&\frac{\alpha_s}{\pi}\frac{1}{\epsilon} L_{\beta} \, , \\ 
I_2^{UV \;pole}&=&-\frac{\alpha_s}{\pi}\frac{1}{\epsilon} L_i \, , \\
I_3^{UV \;pole}&=&-\frac{\alpha_s}{\pi}\frac{1}{\epsilon} L_j \, , \\
I_4^{UV \;pole}&=&-\frac{\alpha_s}{\pi}\frac{1}{\epsilon} \, ,
\end{eqnarray}
where the $L_\beta$ is the familiar velocity-dependent
eikonal function
\begin{equation}
L_{\beta}=\frac{1-2m^2/s}{\beta}\left(\ln\frac{1-\beta}{1+\beta}
+\pi i \right)\, ,
\end{equation}
with $\beta=\sqrt{1-4m^2/s}$.
The $L_i$ and $L_j$ are complicated functions of
the gauge vector $n$. We will see shortly that their
contributions are cancelled by the inclusion of self energies. 
Their explicit expressions are:
\begin{equation}
L_i=\frac{1}{2}[L_i(+n)+L_i(-n)] \, ,
\end{equation}
where
\begin{eqnarray}
L_i(\pm n)&=&\frac{1}{2}\frac{|v_i \cdot n|}{\sqrt{(v_i \cdot n)^2-2m^2n^2/s}}
\nonumber \\ &&
\left[\ln\left(\frac{\delta(\pm n) \, 2m^2/s-|v_i \cdot n| 
- \sqrt{(v_i \cdot n)^2-2m^2n^2/s}}
{\delta(\pm n) \, 2m^2/s-|v_i \cdot n| 
+ \sqrt{(v_i \cdot n)^2-2m^2n^2/s}}\right)\right.
\nonumber \\ && 
+\left.\ln\left(\frac{\delta(\pm n) \, n^2-|v_i \cdot n| 
- \sqrt{(v_i \cdot n)^2-2m^2n^2/s}}
{\delta(\pm n) \, n^2-|v_i \cdot n| 
+ \sqrt{(v_i \cdot n)^2-2m^2n^2/s}}\right)\right] 
\end{eqnarray}
with $\delta(n) \equiv |v_i \cdot n|/ (v_i \cdot n)$.

When $v_i$ refers to a massive quark and $v_j$ to a massless quark we have
\begin{eqnarray}
I_1^{UV \;pole}&=&\frac{\alpha_s}{2\pi}
\left\{\frac{1}{\epsilon^2}-\frac{1}{\epsilon}
\left[\gamma+2\ln\left(\frac{v_{ij} \cal{Q}}{m}\right)-\ln(4\pi) \right]
\right\}\, , \\ 
I_2^{UV \;pole}&=&-\frac{\alpha_s}{\pi}\frac{1}{\epsilon} L_i \, , \\
I_3^{UV \;pole}&=&\frac{\alpha_s}{2\pi}
\left\{\frac{1}{\epsilon^2}-\frac{1}{\epsilon}
\left[\gamma+\ln(\nu_j)-\ln(4\pi)\right]
\right\} \, , \\
I_4^{UV \;pole}&=&-\frac{\alpha_s}{\pi}\frac{1}{\epsilon} \, , 
\end{eqnarray}
where
\begin{equation}
\nu_a=\frac{(v_a \cdot n)^2}{|n|^2}\, ,
\end{equation}
and $v_{ij}=v_i \cdot v_j$. Note that the double poles cancel.

Finally when both $v_i$ and $v_j$ refer to massless quarks we have 
\cite{BottsSt}
\begin{eqnarray}
I_1^{UV \;pole}&=&\frac{\alpha_s}{\pi}
\left\{\frac{2}{\epsilon^2}-\frac{1}{\epsilon}
\left[\gamma+\ln\left(\frac{v_{ij}}{2}\right)
-\ln(4\pi) \right]\right\} \, , \\ 
I_2^{UV \;pole}&=&\frac{\alpha_s}{2\pi}
\left\{\frac{2}{\epsilon^2}-\frac{1}{\epsilon}
\left[\gamma+\ln(\nu_i)-\ln(4\pi)\right]\right\} \, , \\
I_3^{UV \;pole}&=&\frac{\alpha_s}{2\pi}
\left\{\frac{2}{\epsilon^2}-\frac{1}{\epsilon}
\left[\gamma+\ln(\nu_j)-\ln(4\pi)\right]\right\} \, , \\
I_4^{UV \;pole}&=&-\frac{\alpha_s}{\pi}\frac{1}{\epsilon}\, . 
\end{eqnarray}
Again, note that the double poles cancel.

Our calculations 
are most easily carried out in a color tensor basis
consisting of singlet exchange in the $s$ and $u$ channels,
\begin{equation}
c_1 = \delta_{ab}\delta_{ij}\, , \quad \quad
c_2 =  \delta_{aj}\delta_{bi}\, .
\label{alter}
\end{equation}
The color indices for the incoming quark and antiquark are $a$ and $b$,
respectively, and for the outgoing quark and antiquark $j$ and $i$, 
respectively.

In the basis (\ref{alter}) we find
\begin{eqnarray}
\Gamma_{S, 11}&=&\frac{\alpha_s}{\pi}C_F\left[\ln
\left(\frac{v_{ab}}{2}\right)-L_{\beta}
-\frac{1}{2}\ln(\nu_a\nu_b)-L_i-L_j+2-\pi i\right]
\nonumber \\ &&
-\frac{1}{N}\Gamma_{S, 21},
\nonumber \\
\Gamma_{S, 21}&=&\frac{\alpha_s}{2\pi}
\ln\left(\frac{v_{aj}v_{bi}}{v_{ai}v_{bj}}\right),
\nonumber \\ 
\Gamma_{S, 12}&=&\frac{\alpha_s}{2\pi}
\left[\ln\left(\frac{v_{ab}}{2}\right)-\ln(v_{ai}v_{bj})-L_{\beta}
+\ln\left(\frac{2m^2}{s}\right)-\pi i\right],
\nonumber \\
\Gamma_{S, 22}&=&\frac{\alpha_s}{\pi}C_F
\left[\ln(v_{aj}v_{bi})-\ln\left(\frac{2m^2}{s}\right)
-\frac{1}{2}\ln(\nu_a\nu_b)-L_i-L_j+2\right]
\nonumber \\ &&
-\frac{1}{N}\Gamma_{S, 12}.
\end{eqnarray}
The matrix depends, as expected, on the directions of the Wilson
lines, which may be reexpressed in terms of ratios of
kinematic invariants for the partonic scattering.  
We eliminate the gauge dependence of the heavy quarks by including the 
self-energy graphs in fig. 4.4.
The contribution of the self-energy graphs (in the diagonal
elements only) is the following:
\begin{equation}
\frac{\alpha_s}{\pi} C_F(L_i+L_j-2).
\end{equation}
\begin{figure}
\centerline{
\psfig{file=figz.ps,height=2.05in,width=5.05in,clip=}}
\caption[Heavy quark self-energy contributions to S for 
$q {\bar q} \rightarrow Q {\bar Q}$]
{Heavy quark self-energy contributions to S for 
$q {\bar q} \rightarrow Q {\bar Q}$.}
\label{fig. 4.4}
\end{figure}
Next we absorb the $k^{\mu}k^{\nu}$ contribution to $g_2$, since
it also appears in Drell Yan.
This gives a $-(\alpha_s/\pi)C_F$ 
in the diagonal elements.
We also have in the axial gauge $\nu_a=1/2$.
Then in terms of our invariants $s$, $t_1$, and $u_1$, the
anomalous dimension matrix becomes
\begin{eqnarray}
\Gamma_{S, 11}&=&\frac{\alpha_s}{\pi}\left\{C_F\left[\ln
\left(\frac{u_1^2}{t_1^2}\right)-L_{\beta}-1-\pi i\right]-\frac{C_A}{2}
\ln\left(\frac{u_1^2}{t_1^2}\right)\right\},
\nonumber \\
\Gamma_{S, 21}&=&\frac{\alpha_s}{2\pi}
\ln\left(\frac{u_1^2}{t_1^2}\right),
\nonumber \\ 
\Gamma_{S, 12}&=&\frac{\alpha_s}{2\pi}
\left[\ln\left(\frac{m^2 s}{t_1^2}\right)-L_{\beta}-\pi i\right],
\nonumber \\
\Gamma_{S, 22}&=&\frac{\alpha_s}{\pi}\left\{C_F
\left[\ln\left(\frac{u_1^2}{t_1^2}\right)-L_{\beta}-1-\pi i\right]\right.
\nonumber \\ &&
\left.+\frac{C_A}{2}\left[-\ln\left(\frac{m^2s}{t_1^2}\right)+L_{\beta}
+\pi i\right]\right\}\, .
\end{eqnarray}
For arbitrary
$\beta$ and fixed scattering angle, we must solve for the relevant  
diagonal basis of color
structure, and determine the eigenvalues.  
At ``absolute" threshold, $\beta=0$,
we find 
\begin{eqnarray}
\Gamma^{\rm th}_{S, 11}&=&-\frac{\alpha_s}{\pi}C_F \left(\pi i
+\frac{\pi i}{2 \beta}\right),
\nonumber \\ 
\Gamma^{\rm th}_{S, 21}&=&0,
\nonumber \\ 
\Gamma^{\rm th}_{S, 12}&=&\frac{\alpha_s}{2\pi}\left(1-\pi i
-\frac{\pi i}{2 \beta}\right),
\nonumber \\ 
\Gamma^{\rm th}_{S, 22}&=&\frac{\alpha_s}{\pi}\left[-C_F \left(\pi i
+\frac{\pi i}{2 \beta}\right)
-\frac{C_A}{2}\left(1-\pi i-\frac{\pi i}{2 \beta}\right)\right].
\end{eqnarray}
Notice that 
$\Gamma_S^{\rm th}$ is diagonalized in a basis of singlet and octet
exchange in the $s$ channel, 
\begin{equation}
c_{\rm singlet}=c_1, \quad \quad c_{\rm octet}=-\frac{1}{2N}c_1+\frac{c_2}{2},
\end{equation}
with eigenvalues,
\begin{equation}
\lambda_{\rm singlet}=\Gamma^{\rm th}_{S, 11} \, , \quad\quad 
\lambda_{\rm octet}=\Gamma^{\rm th}_{S, 22} \, .
\end{equation}

The general result in this $s$ channel singlet-octet basis becomes:
\begin{eqnarray}
\Gamma^{(1,8)}_{S, 11}&=&-\frac{\alpha_s}{\pi}C_F(L_{\beta}+1+\pi i),
\nonumber \\
\Gamma^{(1,8)}_{S, 21}&=&\frac{2\alpha_s}{\pi}
\ln\left(\frac{u_1}{t_1}\right),
\nonumber \\ 
\Gamma^{(1,8)}_{S, 12}&=&\frac{\alpha_s}{\pi}
\frac{C_F}{C_A} \ln\left(\frac{u_1}{t_1}\right),
\nonumber \\
\Gamma^{(1,8)}_{S, 22}&=&\frac{\alpha_s}{\pi}\left\{C_F
\left[4\ln\left(\frac{u_1}{t_1}\right)-L_{\beta}-1-\pi i\right]\right.
\nonumber \\ &&
\left.+\frac{C_A}{2}\left[-3\ln\left(\frac{u_1}{t_1}\right)
-\ln\left(\frac{m^2s}{t_1u_1}\right)+L_{\beta}+\pi i \right]\right\}\, .
\end{eqnarray}
$\Gamma_S$ is also diagonalized in this singlet-octet basis
when the parton-parton c.m. scattering angle is
$\theta=90^\circ$ (where $u_1=t_1$) with
eigenvalues
\begin{eqnarray}
\lambda_{\rm singlet}&=&-{\alpha_s\over \pi}C_F (L_{\beta}+1+\pi i) \, , 
\\  
\lambda_{\rm octet}&=&{\alpha_s\over \pi}\left[-C_F (L_{\beta}+1+\pi i)
+{C_A\over 2}(L_{\beta}-\ln\left(\frac{m^2s}{t_1^2}\right)+\pi i)\right]\, .
\nonumber\\
\end{eqnarray}

It is of interest, of course, to compare the one-loop expansion of
our results to known one-loop calculations, at the level of NLO.
We may give our result as a function of $z$, since the inverse
transforms are trivial.
They may be found in terms of the Born cross section,
the one-loop factoring contributions of $g_1^{(q{\bar q})}$ 
and $g_2^{(q{\bar q})}$,
and $\Gamma_{22}^{(1,8)}$.
In the DIS scheme the result is
\begin{eqnarray}
\sum_{IJ}\; {\Omega^{(IJ)}_{q{\bar q}}(z,u_1,t_1,s)}{^{(1)}}
&=&
\sigma_{\rm Born}
\frac{\alpha_s}{\pi} \frac{1}{1-z}\left\{C_F\left[2\ln(1-z)
 +\frac{3}{2} \right.\right.\nonumber \\
&\ & \quad \quad 
\left.\left.+8\ln\left(\frac{u_1}{t_1}\right) 
-2-2 {\rm Re} \, L_{\beta}+2\ln\left(\frac{s}{\mu^2}\right)\right. \right]
\nonumber \\
&\ & \quad 
\left.+C_A\left[-3\ln\left(\frac{u_1}{t_1}\right)+L_{\beta}
-\ln\left(\frac{m^2s}{t_1 u_1}\right)\right]\right\}\, .
\nonumber\\
\label{longeq}
\end{eqnarray}
The logarithm of $s/\mu^2$ describes the evolution of the parton distributions.
This result cannot be compared directly to
the one-loop results of \cite{mengetal} for arbitrary $\beta$,   
where the singular behavior is given in terms of the variable $s_4$,
with
\begin{equation}
s_4= (p_j+k)^2-m^2 \approx 2p_j\cdot k\, ,
\end{equation}
where $k=p_a+p_b-p_i-p_j$ is the momentum carried away by the gluon.  At
partonic threshold, both $s_4$ and $(1-z)$ vanish, 
but even for small $s_4$, angular
integrals over the gluon momentum with $s_4$ held fixed
are rather different than those with
$1-z \approx 2(p_i+p_j)\cdot k/s$ held fixed.  

Nevertheless, the cross sections become identical in the 
$\beta\rightarrow 0$ limit, where we may make a direct
comparison.
Near $\beta=0$,   our cross section becomes
\begin{eqnarray}
\sum_{IJ}\; {\Omega^{(IJ)}_{q{\bar q}}(z,u_1,t_1,s)}{^{(1)}}|_{\rm th}&=&
\sigma_{\rm Born} \frac{\alpha_s}{\pi} \frac{1}{1-z}\left\{C_F\left[2\ln(1-z) 
+\frac{3}{2}\right.\right.
\nonumber \\ &&
\quad\quad \quad \left.\left.
+2\ln\left(\frac{4m^2}{\mu^2}\right)\right]-C_A\right\} \, .
\end{eqnarray}
Near $s=4m^2$, we  may identify $2m^2(1-z)=s_4$, and this expression
becomes identical to the $\beta\rightarrow 0$
limit of eq.\ (30) of \cite{mengetal}.  It is also worth
noting that even for $\beta>0$, the two cross
sections remain remarkably close, differening only
at first nonleading logarithm in the abelian ($C_F^2$)
term,
due to the interplay of angular integrals 
with leading singularities for the differing treatments of
phase space.  

As for the Drell-Yan cross section, our analysis 
applies not only to absolute threshold for the production of the
heavy quarks ($\beta=0$), but also to partonic threshold
for the production of moving heavy quarks.
When $\beta$ nears unity, however,
the anomalous dimensions themselves develop (collinear) singularities
associated with the fragmentation of the heavy quarks, which
in principle may be factored into nonperturbative
fragmentation functions.  

Finally, we have checked that our anomalous dimension matrix for heavy
outgoing quarks (4.2.30) reduces in the limit $m \rightarrow 0$
to the anomalous dimension matrix for light
outgoing quarks, which is \cite{BottsSt}
\begin{eqnarray}
\Gamma_{S, 11}&=&\frac{\alpha_s}{\pi}C_F\left[\ln
\left(\frac{v_{ab}v_{ij}}{4}\right)
-\frac{1}{2}\ln(\nu_a\nu_b\nu_i\nu_j)+2-2\pi i\right]
\nonumber \\ &&
-\frac{1}{N}\Gamma^{(1)}_{S, 21},
\nonumber \\
\Gamma_{S, 21}&=&\frac{\alpha_s}{2\pi}
\ln\left(\frac{v_{aj}v_{bi}}{v_{ai}v_{bj}}\right),
\nonumber \\ 
\Gamma_{S, 12}&=&\frac{\alpha_s}{2\pi}
\left[\ln\left(\frac{v_{ab}v_{ij}}{v_{ai}v_{bj}}\right)-\-2\pi i\right],
\nonumber \\
\Gamma_{S, 22}&=&\frac{\alpha_s}{\pi}C_F
\left[\ln\left(\frac{v_{aj}v_{bi}}{4}\right)
-\frac{1}{2}\ln(\nu_a\nu_b\nu_i\nu_j)+2\right]
\nonumber \\ &&
-\frac{1}{N}\Gamma^{(1)}_{S, 12}.
\end{eqnarray}

\mysection{Applications to $gg \rightarrow Q {\bar Q}$ 
and $gg \rightarrow q \bar{q}$}
In this section we give the results for the anomalous dimension matrix
when the incoming partons are gluons and the outgoing partons are heavy quarks
\begin{equation}
g(p_a)+g(p_b) \rightarrow Q(p_i) + {\bar Q}(p_j)\, .
\end{equation}
For the sake of completeness we also give results for the case
when the outgoing quarks are light
\begin{equation}
g(p_a)+g(p_b) \rightarrow q(p_i) + {\bar q}(p_j)\, .
\end{equation}
In fig. 4.5 we show the UV divergent one-loop contributions to S for 
$gg \rightarrow Q {\bar Q}$ or $gg \rightarrow q {\bar q}$.

\begin{figure}
\centerline{
\psfig{file=figz.ps,height=5.05in,width=5.05in,clip=}}
\caption[UV divergent one-loop contributions to S for 
$gg \rightarrow Q {\bar Q}$ or $q {\bar q}$]
{UV divergent one-loop contributions to S for 
$gg \rightarrow Q {\bar Q}$ or $q {\bar q}$.}
\label{fig. 4.5}
\end{figure}

Our analysis is similar to the one in the previous section.
We use the same integrals for the calculation of the $\omega^{(I)}$.
The eikonal rules for incoming gluons are slightly modified and are given
in fig. 4.6.
\begin{figure}
\centerline{
\psfig{file=figz.ps,height=2.5in,width=5.05in,clip=}}
\caption[Eikonal rules for
$gg \rightarrow Q {\bar Q}$ or $q {\bar q}$]
{Eikonal rules for 
$gg \rightarrow Q {\bar Q}$ or $q {\bar q}$. 
The gluon momentum flows out of the 
eikonal lines.} 
\label{fig. 4.6}
\end{figure}
We choose the following basis for the color factors:
\begin{equation}
c_1=\delta^{ab}\,\delta_{ij}, \quad c_2=d^{abc}\,T^c_{ij},
\quad c_3=i f^{abc}\,T^c_{ij}.
\end{equation}
Again, the counterterms for $S$ are the ultraviolet divergent
coefficients times our basis color tensors:
\begin{eqnarray}
S_1&=&c_1 Z_{S, 11} + c_2 Z_{S, 21} + c_3 Z_{S, 31},
\\
S_2&=&c_1 Z_{S, 12} + c_2 Z_{S, 22} + c_3 Z_{S, 32},
\\
S_3&=&c_1 Z_{S, 13} + c_2 Z_{S, 23} + c_3 Z_{S, 33}.
\end{eqnarray}

Our results for the anomalous dimension matrix when the outgoing quarks are
heavy are:
\begin{eqnarray}
\Gamma_{S, 11}&=&\frac{\alpha_s}{\pi}\left\{C_F (-L_{\beta}-L_i-L_j+1)
+C_A\left[\ln
\left(\frac{v_{ab}}{2}\right)
-\frac{1}{2}\ln(\nu_a\nu_b)+1-\pi i\right]\right\},
\nonumber \\
\Gamma_{S, 21}&=&0,
\nonumber \\ 
\Gamma_{S, 31}&=&\frac{\alpha_s}{\pi}
\ln\left(\frac{v_{ai}v_{bj}}{v_{aj}v_{bi}}\right),
\nonumber \\ 
\Gamma_{S, 12}&=&0,
\nonumber \\ 
\Gamma_{S, 22}&=&\frac{\alpha_s}{\pi}\left\{C_F (-L_{\beta}-L_i-L_j+1)
+\frac{C_A}{2}\left[
\ln\left(\frac{v_{ab}}{2}\right)
+\frac{1}{2}\ln(v_{ai}v_{bj}v_{aj}v_{bi})\right.\right.
\nonumber \\ &&
\left.\left.+L_{\beta}-\ln\left(\frac{2m^2}{s}\right)
-\ln(\nu_a\nu_b)+2-\pi i\right]\right\},
\nonumber \\ 
\Gamma_{S, 32}&=&\frac{N^2-4}{4N}\Gamma_{S, 31},
\nonumber \\ 
\Gamma_{S, 13}&=&\frac{1}{2}\Gamma_{S,31},
\nonumber \\ 
\Gamma_{S, 23}&=&\frac{C_A}{4}\Gamma_{S,31},
\nonumber \\ 
\Gamma_{S, 33}&=&\Gamma_{S, 22}.
\end{eqnarray}

Again, we eliminate the gauge dependence of the heavy quarks by including the 
self-energy graphs in fig. 4.7.
The contribution of the self-energy graphs (in the diagonal
elements only) is, as before, 
\begin{equation}
\frac{\alpha_s}{\pi} C_F(L_i+L_j-2).
\end{equation}
\begin{figure}
\centerline{
\psfig{file=figz.ps,height=2.05in,width=5.05in,clip=}}
\caption[Heavy quark self-energy contributions to S for 
$gg \rightarrow Q {\bar Q}$]
{Heavy quark self-energy contributions to S for 
$gg \rightarrow Q {\bar Q}$.}
\label{fig. 4.7}
\end{figure}
In analogy to the previous section, we also have an additional
$-(\alpha_s/\pi)C_A$ in the diagonal elements. 
Then in terms of the invariants $s$, $t_1$, and $u_1$,
the anomalous dimension matrix becomes  
\begin{eqnarray}
\Gamma_{S,11}&=&\frac{\alpha_s}{\pi}[-C_F(L_{\beta}+1)-C_A\pi i],
\nonumber \\
\Gamma_{S,21}&=&0,
\nonumber \\
\Gamma_{S,31}&=&\frac{\alpha_s}{\pi}\ln\left(\frac{t_1^2}{u_1^2}\right),
\nonumber \\
\Gamma_{S,12}&=&0,
\nonumber \\
\Gamma_{S,22}&=&\frac{\alpha_s}{\pi}\left\{-C_F(L_{\beta}+1)
+\frac{C_A}{2}\left[\ln\left(\frac{t_1 u_1}{m^2 s}\right)+L_{\beta}-\pi i
\right]\right\},
\nonumber \\
\Gamma_{S,32}&=&\frac{N^2-4}{4N}\Gamma_{S,31},
\nonumber \\
\Gamma_{S,13}&=&\frac{1}{2}\Gamma_{S,31},
\nonumber \\
\Gamma_{S,23}&=&\frac{C_A}{4}\Gamma_{S,31},
\nonumber \\
\Gamma_{S,33}&=&\Gamma_{S,22}.
\end{eqnarray}

At threshold the anomalous dimension matrix becomes diagonal 
with eigenvalues
\begin{eqnarray} 
\Gamma^{\rm th}_{S, 11}&=&\frac{\alpha_s}{\pi}
\left[-C_F\frac{\pi i}{2\beta}-C_A\pi i\right],
\\ 
\Gamma^{\rm th}_{S, 22}&=&\frac{\alpha_s}{\pi}\left[
-C_F\frac{\pi i}{2\beta}
+\frac{C_A}{2}\left(-1-\pi i+\frac{\pi i}{2\beta}\right)\right],
\\
\Gamma^{\rm th}_{S, 33}&=&\Gamma_{S, 22}^{\rm th}.
\end{eqnarray}
We also note that the matrix is diagonalized at $\theta=90^{\circ}$.

Again, it is interesting to compare the one-loop expansion of our results
to the one-loop calculations in \cite{mengetal}. 
In this case our calculation is complicated by the fact
that the color decomposition is not trivial as it was for $q\bar{q}$.
We have to decompose the Born cross section into three terms according to 
our color tensor basis. 
After some algebra and putting  $C_F=4/3$ and $C_A=3$, our result becomes 
\begin{eqnarray}
\sum_{IJ}\; {\Omega^{(IJ)}_{gg}(z,u_1,t_1,s)}{^{(1)}}
&=&
\alpha_s^3 \frac{1}{1-z} K_{gg} B_{\rm QED}(s, t_1, u_1)
\left\{\frac{t_1 u_1}{s^2}\left[-288 \ln (1-z)\right.\right.
\nonumber \\ &&
\left.\left.-144 \ln\left(\frac{s}{\mu^2}\right)
-72\ln\left(\frac{t_1 u_1}{m^2 s}\right)
-8 L_{\beta}+64\right]\right.
\nonumber \\ &&
+128\ln(1-z)+64 \ln\left(\frac{s}{\mu^2}\right)
+28 \ln\left(\frac{t_1 \, u_1}{m^2 s}\right)
\nonumber \\ &&
\left. -\frac{4}{9}L_{\beta} -28-\frac{4}{9}\right\}\, ,
\end{eqnarray}
where
\begin{equation}
B_{\rm QED}(s, t_1, u_1)=\frac{t_1}{u_1}+\frac{u_1}{t_1}
+\frac{4m^2s}{t_1 \, u_1}\left(1-\frac{m^2s}{t_1 \, u_1}\right)\, .
\end{equation}
The logarithms of $s/\mu^2$ describe the evolution of the parton distributions.

As we discussed in the previous section,
our result cannot be compared directly to the one-loop results
of \cite{mengetal}, but as $\beta\rightarrow 0$ our expression
becomes identical to the $\beta\rightarrow 0$ limit of the sum of
eqs. (36-38) in \cite{mengetal}. Even for $\beta>0$, the two cross sections
remain remarkably close. 

Finally, the anomalous dimension matrix for the case when 
the outgoing quarks are light is given by
\begin{eqnarray}
\Gamma_{S, 11}&=&\frac{\alpha_s}{\pi}\left\{
C_F\left[\ln\left(\frac{v_{ij}}{2}\right)-\frac{1}{2}\ln(\nu_i\nu_j)
+1-\pi i\right]\right.
\nonumber \\ &&
\left.+C_A\left[\ln\left(\frac{v_{ab}}{2}\right)-\frac{1}{2}\ln(\nu_a\nu_b)
+1-\pi i\right]\right\}  ,
\nonumber \\ 
\Gamma_{S, 21}&=&0,
\nonumber \\ 
\Gamma_{S, 31}&=&\frac{\alpha_s}{\pi}
\ln\left(\frac{v_{ai}v_{bj}}{v_{aj}v_{bi}}\right),
\nonumber \\ 
\Gamma_{S, 12}&=&0,
\nonumber \\ 
\Gamma_{S, 22}&=&\frac{\alpha_s}{\pi}\left\{
C_F\left[\ln\left(\frac{v_{ij}}{2}\right)-\frac{1}{2}\ln(\nu_i\nu_j)
+1-\pi i\right]\right.
\nonumber \\ &&
\left.+C_A\left[\frac{1}{4}
\ln(v_{ai}v_{bj}v_{aj}v_{bi})+\frac{1}{2}\ln\left(\frac{v_{ab}}{v_{ij}}\right)
-\frac{1}{2}\ln(\nu_a\nu_b)-\ln 2+1\right]\right\} ,
\nonumber \\ 
\Gamma_{S, 32}&=&\frac{N^2-4}{4N}\Gamma_{S, 31},
\nonumber \\ 
\Gamma_{S, 13}&=&\frac{1}{2}\Gamma_{S, 31},
\nonumber \\ 
\Gamma_{S, 23}&=&\frac{C_A}{4}\Gamma_{S, 31},
\nonumber \\ 
\Gamma_{S, 33}&=&\Gamma_{S, 22}.
\end{eqnarray}
Again, we note that the matrix is diagonalized at $\theta=90^{\circ}$.

Finally, we have checked that our anomalous dimension matrix for heavy
outgoing quarks (4.3.7) reduces to the anomalous dimension matrix for light
outgoing quarks (4.3.15) in the limit $m \rightarrow 0$. 
\mysection{Conclusions}
We have illustrated the application of a general method for
resumming next-to-leading logarithms at partonic threshold
in QCD cross sections.  
We have given explicit results for heavy quark production
through light quark annihilation and gluon fusion, and for light 
quark production through gluon fusion.
Possible extensions include, of course, 
dijet and multijet production.  
We reserve estimates of the phenomenological
importance of these nonleading terms to future work, but we
hope that whether they give small contributions or large,
the method will improve the reliability of perturbative QCD calculations
of hard scattering cross sections.






\begin{thebibliography}{99}
\bibitem{ObsCDF1}(CDF Collaboration) F. Abe {\it et al.}, 
Phys. Rev. Lett. {\bf 74}, 2626 (1995).
\bibitem{ObsD01} (D0 Collaboration) S. Abachi {\it et al.},
Phys. Rev. Lett. {\bf 74}, 2632 (1995). 
\bibitem{Politzer} H. D. Politzer, Phys. Rev. Lett. {\bf 30}, 1346 (1973);
D. J. Gross and F. Wilczek, {\it ibid} {\bf 30}, 1343 (1973).
\bibitem{css1}J. C. Collins, D. E. Soper, and G. Sterman, in
{\it Perturbative Quantum Chromodynamics}, 
ed. A. H. Mueller (World Scientific, Singapore, 1989), p. 1.
\bibitem{gs1}G. Sterman, Nucl. Phys. {\bf B281}, 310 (1987);
S. Catani and L. Trentadue, Nucl. Phys. {\bf B327}, 323 (1989); 
{\bf B353}, 183 (1991). More recent references can be located by
consulting 
L. Alvero and H. Contopanagos, Nucl. Phys. {\bf B436}, 184 (1995); 
{\it ibid} {\bf B456}, 497 (1995);
P.J. Rijken and W.L. van Neerven, Phys. Rev. D {\bf 51}, 44 (1995).
\bibitem{LSN1} E. Laenen, J. Smith, and W. L. van Neerven,
Nucl. Phys. {\bf B369}, 543 (1992). 
\bibitem{lsn21} E. Laenen, J. Smith, and W. L. van Neerven,
Phys. Lett. B {\bf 321}, 254 (1994). 
\bibitem{betal1}W. Beenakker, H. Kuijf, W. L. van Neerven, and J. Smith,
Phys. Rev. D {\bf 40}, 54 (1989);
W. Beenakker, W. L. van Neerven, R. Meng, 
G. A. Schuler, and J. Smith,  Nucl. Phys. {\bf B351}, 507 (1991).
\bibitem{nde11} P. Nason, S. Dawson, and R. K. Ellis, Nucl. Phys. {\bf B303}, 
607 (1988).
\bibitem{Mueller1} A. H. Mueller, Nucl. Phys. {\bf B250}, 327 (1985);
M. Beneke, Phys. Lett. B {\bf 307}, 154 (1993);
G. Grunberg, Phys. Lett. B {\bf 349}, 469 (1995);
Yu. A. Simonov,
Pisma Zh. Eksp. Teor. Fiz. {\bf 57}, 513 (1993)
[JETP Lett. {\bf 57}, 525 (1993)].
\bibitem{ConAlv1} H. Contopanagos and G. Sterman, Nucl. Phys. 
{\bf B419}, 77 (1994); L. Alvero and H. Contopanagos, 
Nucl. Phys. {\bf B436}, 184 (1995).
\bibitem{KSb1} N. Kidonakis and J. Smith, Report No. ITP-SB-95-16, 
hep-ph/9506253, 1995.
\bibitem{KSh1} N. Kidonakis and J. Smith, Mod. Phys. Lett. A {\bf 11}, 
587 (1996).
\bibitem{NKJS1} N. Kidonakis and J. Smith, Phys. Rev. D {\bf 51}, 6092 (1995).
\bibitem{NKGS1} N. Kidonakis and G. Sterman, 
Report No. ITP-SB-96-7, hep-ph/9604234, 1996. 
\end{thebibliography}

\begin{thebibliography}{99}
\bibitem{bnmss2}W. Beenakker, W. L. van Neerven, R. Meng, 
G. A. Schuler, and J. Smith, Nucl. Phys. {\bf B351}, 507 (1991).
\bibitem{MSSN2}R. Meng, G. A. Schuler, J. Smith, and W. L. van Neerven,
Nucl. Phys. {\bf B339}, 325 (1990).
\bibitem{betal2}W. Beenakker, H. Kuijf, W. L. van Neerven, and J. Smith,
Phys. Rev. D {\bf 40}, 54 (1989).
\bibitem{LSN2} E. Laenen, J. Smith, and W. L. van Neerven,
Nucl. Phys. {\bf B369}, 543 (1992).
\bibitem{mrs2} A. D. Martin, R. G. Roberts, and W. J. Stirling, Phys. Lett.
B {\bf 306}, 145 (1993).
\bibitem{PDFLIB2} H. Plothow-Besch,  `PDFLIB: Nucleon, Pion
and Photon Parton Density Functions and $\alpha_s$ Calculations',
Users's Manual - Version 4.16, W5051 PDFLIB, 1994.01.11, CERN-PPE.
\bibitem{nde12} P. Nason, S. Dawson, and R. K. Ellis, Nucl. Phys. {\bf B303}, 
607 (1988).
\bibitem{lsn22} E. Laenen, J. Smith, and W. L. van Neerven,
Phys. Lett. B {\bf 321}, 254 (1994). 
\bibitem{BC2} E. Berger and H. Contopanagos, Phys. Lett. B
{\bf 361}, 115 (1995); ANL-HEP-PR-95-82, hep-ph/9603326, 1996.
\bibitem{CT2} S. Catani, M. L. Mangano, P. Nason, and L. Trentadue,
CERN-TH-96-86, hep-ph/9604351, 1996.
\bibitem{herab12}H. Aubrecht et al., {\it An Experiment to Study
CP violation in the B system Using an Internal Target at the HERA
Proton Ring}, Letter of Intent, DESY-PRC 92/04 (1992).
\bibitem{herab22}H. Aubrecht et al., {\it An Experiment to Study
CP violation in the B system Using an Internal Target at the HERA
Proton Ring}, Progress Report, DESY-PRC 93/04 (1993).
\bibitem{ObsCDF2}(CDF Collaboration) F. Abe {\it et al.}, 
Phys. Rev. Lett. {\bf 74}, 2626 (1995).
\bibitem{ObsD02} (D0 Collaboration) S. Abachi {\it et al.},
Phys. Rev. Lett. {\bf 74}, 2632 (1995). 
\end{thebibliography}

\begin{thebibliography}{99}
\bibitem{OS3} L. H. Orr and W. J. Stirling, Phys. Rev. D {\bf 51}, 
1077 (1995). 
\bibitem{OTS3} L. H. Orr, T. Stelzer, and W. J. Stirling, Phys. Rev. D 
{\bf 52}, 124 (1995). 
\bibitem{Berends3} F. A. Berends, J. B. Tausk, and W. T. Giele,
Phys. Rev. D {\bf 47}, 2746 (1993).
\bibitem{nde23}P. Nason, S. Dawson, and R. K. Ellis, Nucl. Phys.
{\bf B327}, 49 (1989); {\bf B335}, 260(E) (1990).
\bibitem{bnmss3}W. Beenakker, W. L. van Neerven, R. Meng, 
G. A. Schuler, and J. Smith, Nucl. Phys. {\bf B351}, 507 (1991).
\bibitem{Lane3} K. Lane, Phys. Rev. D {\bf 52}, 1546 (1995).

\end{thebibliography}

\begin{thebibliography}{99}

\bibitem{oldDY} G.\ Sterman, Nucl.\ Phys.\ {\bf B281}, 310 (1987);
 S.\ Catani and L.\ Trentadue, 
Nucl.\ Phys.\ {\bf B327}, 323 (1989).

\bibitem{fact} J. C.\ Collins, D. E.\ Soper, and G.\ Sterman, in 
{\it Perturbative Quantum Chromodynamics},
ed.\ A. H.\ Mueller (World Scientific, Singapore, 1989), p. 1.

\bibitem{expon}  J. G. M.\ Gatheral, Phys.\ Lett.\ B {\bf 133}, 90 (1983);
J. C. Collins, in {\it Perturbative Quantum Chromodynamics},
ed.\ A. H.\ Mueller (World Scientific, Singapore, 1989), p. 573.

\bibitem{CLS} H. Contopanagos, E. Laenen, and G. Sterman, 
Report No. ANL-HEP-25, hep-ph/9604313, 1996.

\bibitem{LaenenSterman} E.\ Laenen and G.\ Sterman, in 
{\it Proceedings of The Fermilab
Meeting, DPF 92}, 7th  meeting of the 
American Physical Society Division of Particles
and Fields (Batavia, IL, 1992), ed.\
C. H.\ Albright {\it et al} (World Scientific,
Singapore, 1993), p.\ 987.

\bibitem{BottsSt} J.\ Botts and G.\ Sterman, Nucl.\ Phys.\ {\bf B325}, 
62 (1989).

\bibitem{CSt} H.\ Contopanagos and G.\ Sterman, Nucl.\ Phys.\ {\bf B400},
211 (1993); {\bf B419}, 77 (1994).

\bibitem{CT2}  S.\ Catani and L.\ Trentadue, {\bf B353}, 183 (1991).

\bibitem{SotSt} M.\ Sotiropoulos and G.\
Sterman,  Nucl.\ Phys.\ {\bf B419}, 59 (1994).

\bibitem{GK} G. P. \ Korchemsky, Phys.\ Lett.\ B {\bf 325}, 459 (1994).

\bibitem{KK}  I. A.\ Korchemskaya, G. P.\ Korchemsky,
Nucl.\ Phys.\ {\bf B437}, 127 (1995). 

\bibitem{mengetal} R.\ Meng, G.A.\ Schuler, J.\ Smith and W. L.\ van Neerven,
Nucl.\ Phys.\ {\bf B339}, 325 (1990).

\end{thebibliography}
\end{document}